\documentclass{ws-ijmpa}
\usepackage[super,compress]{cite}
\usepackage{graphicx}
\usepackage{xspace}
\usepackage{hyperref}
\hypersetup{colorlinks=false, citebordercolor={0 1 0},  linkbordercolor={1 0 0}, urlbordercolor={0 1 1} }
\newcommand{\DZero}    {D0\xspace} 
\newcommand{\EtAl}{\textit{et al.}}
\newcommand{\VPY}[3]{\textbf{#1}, #2 (#3)} 
\newcommand{\CiteJournal}[4]{#1 \VPY{#2}{#3}{#4}}
\newcommand{\DZeroAuth}[1][V.~M.~Abazov]{\DZero{} Collaboration (#1 \EtAl{})}

\newcommand{\PRDname}{Phys. Rev.~D}
\newcommand{\PRLname}{Phys. Rev. Lett.}
\newcommand{\PLBname}{Phys. Lett.~B}
\newcommand{\JHEPname}{J.~High~Energy Phys.}

\newcommand{\PRD}[3]{\CiteJournal{\PRDname}{#1}{#2}{#3}}
\newcommand{\PRL}[3]{\CiteJournal{\PRLname}{#1}{#2}{#3}}
\newcommand{\PLB}[3]{\CiteJournal{\PLBname}{#1}{#2}{#3}}
\newcommand{\JHEP}[3]{\CiteJournal{\JHEPname}{#1}{#2}{#3}}

\newcommand{\pT}{p_T}
\newcommand{\Mtj}{M_{\text{3jet}}}
\newcommand{\sherpa}{{\sc sherpa}}
\newcommand{\pythia}{{\sc pythia}}

\newcommand{\Dphi}{\Delta \phi\,{}_{\text{dijet}}}
\newcommand{\ptmin}{p_{T\text{min}}}
\newcommand{\ptmax}{p_{T\text{max}}}
\newcommand{\Rdphi}{R_{\Delta \phi}}
\newcommand{\Dphimax}{\Delta \phi_{\rm max}}
\newcommand{\Rdr}{R_{\Delta R}}
\newcommand{\DR}{\Delta R}
\newcommand{\ptnbrmin}{p_{T \rm min}^{\rm nbr}}
\newcommand{\as}{\alpha_s}

\newcommand{\asmz}{\alpha_s(M_Z)}
\newcommand{\asq}{\alpha_s(Q)}
\newcommand{\aspt}{\alpha_s(p_T)}
\newcommand{\Rtt}{R_{\text{3/2}}}
\newcommand{\chijj}{\chi_{\text{dijet}}}
\newcommand{\Ptg}{p_{T}^{\gamma}}

\newcommand{\lt}{\!<\!}

\newcommand{\gb}{\ensuremath{\gamma+{b}}\xspace}
\newcommand{\gc}{\ensuremath{\gamma+{c}}\xspace}

\newcommand{\GeVc}{\ensuremath{\text{GeV}}\xspace}
\newcommand{\GeVcc}{\ensuremath{\text{GeV}}\xspace}
\def \ptgg  {$q_T^{\gamma\gamma}$}
\def \mgg  {$M_{\gamma\gamma}$}
\def \dphigg {$\Delta\phi_{\gamma\gamma}$}

\newcommand{\gpTHRj}{$\gamma+{\rm 3~jet}$\xspace}
\newcommand{\gpHFjj}{$\gamma+{b/c~\rm jet+2~jet}$\xspace}
\def\Pomeron{I\hspace{-1.2mm}P}

\begin{document}
\title{REVIEW OF PHYSICS RESULTS FROM THE TEVATRON:
        QCD PHYSICS}

\author{Christina Mesropian}

\address{The Rockefeller University\\
 Laboratory of Experimental High Energy Physics,\\
1230 York Avenue\\
New York, NY 10065, USA\\
christina.mesropian@rockefeller.edu}

\author{Dmitry Bandurin}

\address{University of Virginia, Department of Physics, \\
Charlottesville, Virginia 22904, USA\\
bandurin@fnal.gov}

\maketitle

\begin{history}
\received{Day Month Year}
\revised{Day Month Year}
\end{history}

\begin{abstract}
We present a summary of  results from studies of quantum chromodynamics
at the Fermilab Tevatron collider by the CDF and the D0 experiments. These include 
Run II results for the time period up to the end of Summer 2014. A brief description of Run I results
is also given.
This review covers a wide spectrum of topics, and includes measurements 
with jet and vector boson final states in the hard (perturbative) energy regime,
as well as studies of soft physics such as diffractive and elastic scatterings, 
underlying and minimum bias events, hadron fragmentation, and multiple parton
interactions.

\end{abstract}

\ccode{PACS numbers: 13.87.Ce, 13.85.Qk, 12.38.Qk, 14.70.Fm, 24.10.Ht, 12.40.Nn}

\newpage
\tableofcontents
\newpage

\section{Introduction}
Quantum chromodynamics (QCD) is the theory of interacting quarks and gluons, 
which are the fundamental constituents of hadrons. QCD achieved remarkable success in  describing 
the strong interaction processes at hadron colliders at short distances, i.e. large momentum transfers, 
 by applying  well developed perturbative 
techniques. However QCD still lacks good understanding of quark-gluon interactions at large distances, 
or low  moment transfers, mostly due to the mathematical complexity of the theory and 
the non-applicability of perturbative methods at this range.
 
QCD studies at the Tevatron contributed significantly to the major progress 
in understanding the  strong interactions.
In this section we describe QCD measurements
performed by the CDF and D0 collaborations at the Fermilab Run II Tevatron $p\bar{p}$ collider
using data collected at center-of-mass energy $\sqrt{s} = 1.96$ TeV.
They address various aspects of QCD theory,
providing rigorous tests of predictions for hadron colliders,
and guiding priorities to reduce uncertainties for the most problematic parts of the theory.

We start our review with a brief summary of Run I physics results, obtained at $\sqrt{s} = 0.63$ and $1.8$ TeV,
in Section \ref{sec:run1}.

In Section \ref{sec:jets} we present
the inclusive jet, dijet production and three-jet cross section measurements
which are used to test perturbative QCD (pQCD) calculations,
constrain parton distribution functions (PDFs), and extract a precise value of
the strong coupling constant $\alpha_s$. 
They are also used to search for new phenomena expected at high energies.

Section \ref{seq:gamma} describes measurements with photon final states.
Inclusive photon ($\gamma$) and $\gamma$+jet production cross-section measurements
provide information for tuning QCD theory predictions
and particularly can be used as a direct constraint for global fits to gluon and other PDFs.
The diphoton production cross-sections check the validity
of next-to-leading-order (NLO) pQCD predictions, soft-gluon resummation methods
implemented in theoretical calculations, and contributions from
the parton-to-photon fragmentation diagrams.

Events with $W/Z$+jets productions are used to measure many kinematic
distributions allowing extensive tests and tunes of predictions from
pQCD NLO and Monte Carlo  event generators. They are discussed in Section \ref{sec:Vjets}.
%

All previously mentioned measurements belong to 
the processes that  can be treated in the framework of perturbative QCD. 
The majority of hadron-hadron collision processes are related to the general unsolved problems of 
soft strong interactions, and their studies are discussed in Section \ref{sec:softQCD}.
The charged-particle transverse momenta ($p_T$) and multiplicity distributions
in inclusive minimum bias events are used to  tune non-perturbative
QCD models, including those describing multiple parton interactions (MPI).
Events with inclusive production of $\gamma$ and 2 or 3 jets 
are used to study increasingly important MPI phenomenon at high $p_T$,
measure an effective interaction cross section, 
allowing the prediction of rates of double parton interactions, and providing constraints for existing MPI models.
The study of characteristics of soft particle production enables us to differentiate between 
various  approaches describing hadronization. 
Elastic scattering $p\bar{p}\rightarrow p\bar{p}$ is an important
process that probes the structure of the proton. 
Study of diffractive processes is an important source in understanding many interesting 
aspects of QCD such as low-$x$ structure of the proton and the behavior of QCD in the high density
regime, and provides an ultimate approach in understanding non-perturbative QCD.

We summarize our results in Section \ref{sec:summary}.

\section{Summary of Run I results}
\label{sec:run1}

With the increased luminosity performance of the Tevatron in Run I, when each experiment collected around 20 pb$^{-1}$ of data in  1992-1993 (Run 1A) and 100 pb$^{-1}$ of data in 1994-1996 (Run 1B) with a  small data sample of 600 nb$^{-1}$ being collected at $\sqrt{s}$=630 GeV, the new era of precision $p\bar{p}$ QCD measurements began.

 The cross section measurements for inclusive jet and dijet 
production (see Section \ref{sec:jets}) were no longer limited by statistical uncertainties and resulting systematic uncertainties were comparable to uncertainties from  theoretical predictions. This improvement revealed significant flexibility
  in parton distribution functions, especially for large $x$ gluons, and motivated inclusion of the Tevatron jet data in the global PDF analyses to constrain gluon distributions\cite{jetreview}, thus making predictions more precise, particularly in processes where gluon-quark scattering dominates.

QCD predictions were tested further by comparing with the measurements of the ratio of inclusive jet cross sections at $\sqrt{s}$= 630 and 1800 GeV, dijet cross sections at large rapidity, and a set of photon and photon+jet final state measurements (see Section \ref{seq:gamma})  for both $\sqrt{s}$=1800 GeV and  $\sqrt{s}$=630 GeV.
The strong coupling constant, a free parameter of QCD, was measured from  inclusive jet production, and its {\it running} was tested on a wide range of  momentum transfers. 
The groundwork for extensive Run II studies of $W/Z +$jet final states (see Section \ref{sec:Vjets})
 was laid by measurements of the cross sections and the properties of  vector boson production in association with jets.

The  soft strong interactions were studied in detail by measuring charged particle distributions, 
developing new approaches for studies of the underlying event, and measuring  the effective cross-section in  events with the multi-parton interactions (Section \ref{sec:softQCD}).

Legacy of Run I diffractive measurements  were observations of rapidity gaps between two jets, 
many observations  regarding the
diffractive structure function of the pomeron, and the breakdown
of QCD factorization in hard diffraction between Tevatron and
HERA.

\section{Jet final states}
\label{sec:jets}

\subsection{Measurements of multijet cross sections}

\begin{figure}[htbp]
\hspace*{20mm}  \includegraphics[scale=0.3]{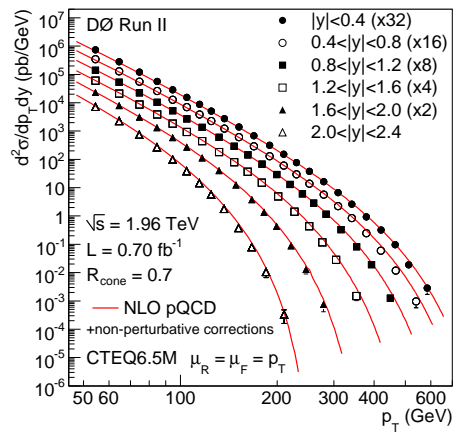}
\caption{The inclusive jet cross section as a function of jet
$p_T$ in six rapidity $|y|$ bins. The data points are multiplied by 2, 4,
8, 16, and 32 for the bins $1.6 < |y| < 2.0, 1.2 < |y| < 1.6,
0.8 < |y| < 1.2, 0.4 < |y| < 0.8$, and $|y| < 0.4$, respectively.}
\label{fig:incjets1_d0}
\end{figure}

\begin{figure}[htbp]
\hspace*{10mm}  \includegraphics[scale=0.55]{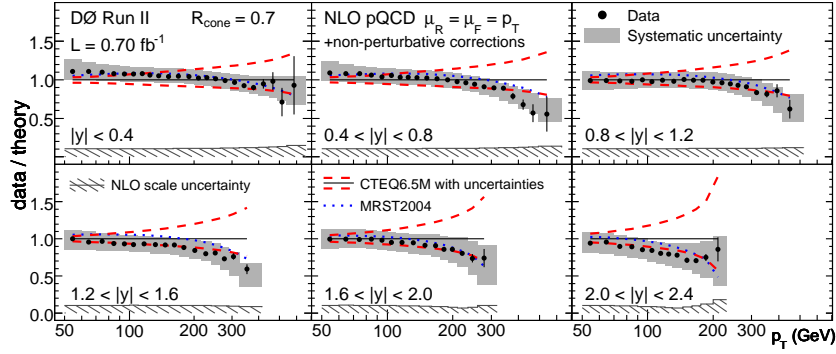}
\caption{Measured data divided by theory predictions. 
The data systematic uncertainties are displayed by the full shaded band. NLO pQCD calculations, 
with renormalization and factorization scales set to jet pT using the CTEQ6.5M PDFs and including 
non-perturbative corrections, are compared to the data. The CTEQ6.5 PDF uncertainties are shown as 
dashed lines and the predictions with MRST2004 PDFs as dotted lines. The theoretical uncertainty, 
determined by changing the renormalization and factorization scales between $p_T/2$ and $2p_T$, is shown
at the bottom of each figure.}
\label{fig:incjets2_d0}
\end{figure}

\begin{figure}[htbp]
\hspace*{0mm}  \includegraphics[scale=0.55]{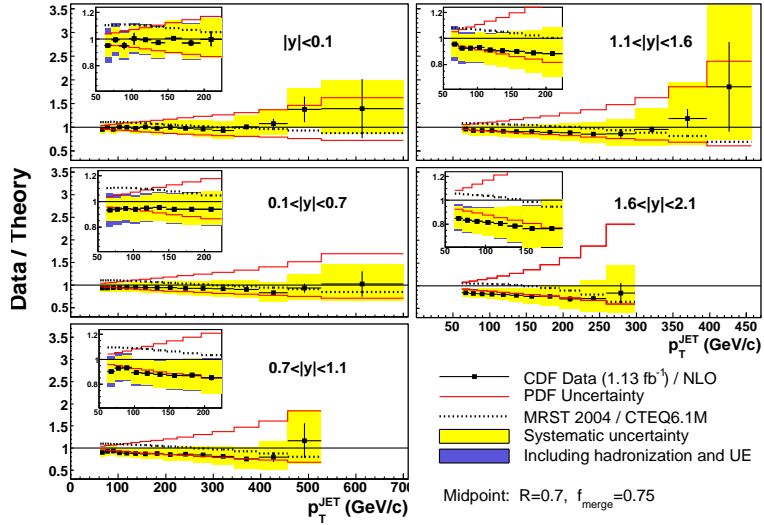}
\caption{The ratios of the measured inclusive jet cross sections at the
  hadron level with the Midpoint jet clustering algorithm to the NLO
  pQCD predictions (corrected to the hadron level) 
  in five rapidity regions.
  Also shown are the experimental systematic uncertainties on the measured cross section,
  the uncertainties in the hadronization and underlying event
  corrections added in quadrature with the experimental systematic uncertainties,
  and the PDF uncertainties on the theoretical predictions.
}
\label{fig:incjets_cdf}
\end{figure}

Stringent
tests of NLO pQCD were obtained from the study of final states
with high $E_T$ jets: inclusive jet and dijet differential cross sections, dijet
mass, dijet angular and multijet distributions. 
%
The Run II data provide a thorough testing of pQCD theory predictions at short distances 
through measurements of differential inclusive jet, dijet and three-jet cross
sections. 
Both experiments measured the inclusive jet cross sections as a function of jet
transverse momentum $p_T$ in several rapidity $y$ regions.
The D0 collaboration measured cross section using
jets found by the Midpoint cone algorithm \cite{ConeAlgo} with radius $R=0.7$
for transverse momenta from 50 GeV to 600 GeV and jet rapidities in the range -2.4 to 2.4.
Figure ~\ref{fig:incjets1_d0} shows the differential  cross section ($d^2\sigma/dp_T dy$)
measured by D0 collaboration \cite{incj_d0}, and
Fig.~\ref{fig:incjets2_d0}
shows a ratio of the measured cross section to NLO pQCD predictions. 
Fig.~\ref{fig:incjets_cdf}
shows a similar ratio for the jet cross sections measured by the CDF collaboration~\cite{incj_cdf} up to $|y|<2.1$.
(Similar measurements of the inclusive jet cross section have been made by 
the CDF collaboration using the $k_T$ jet clustering algorithm~\cite{incj_cdf_kt}.)
Both measurements are in agreement with pQCD predictions.
However, data with uncertainties smaller than those from theoretical calculations (mostly from PDF), 
favor a smaller
gluon content at high parton momentum fractions $x$ ($x>$0.2).
The jet measurements, being dominated by systematic uncertainties, 
are performed using data with 0.4--1 fb$^{-1}$ of integrated luminosity.

\begin{figure}[htbp]
\centering
 \includegraphics[scale=0.3]{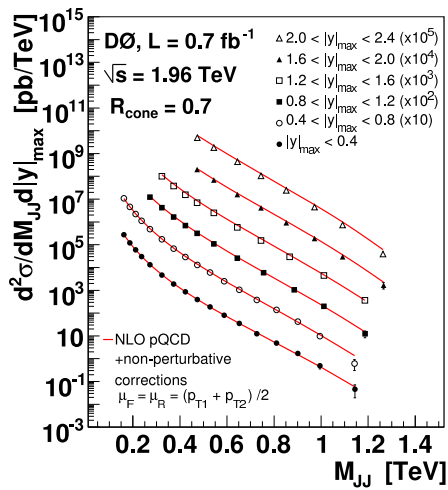}
\caption{The dijet production cross section as a function of invariant mass in intervals of
$|y|_{\rm max}$ compared to NLO predictions that include non-perturbative corrections.
Uncertainties shown are statistical only.}
\label{fig:dijet_d0}
\end{figure}

Figure~\ref{fig:dijet_d0} presents a measurement 
of the dijet production cross section as a function of
the dijet invariant mass and of the largest rapidity of
the two highest $p_T$ jets 
\cite{dijet_d0}.  
The data are described by NLO pQCD predictions using MSTW2008NLO~\cite{MSTW}
PDFs in all rapidity regions, and are not well described
by CTEQ6.6 PDF \cite{cteq6}, particularly at high jet rapidities.

\begin{figure}[htbp]
\hspace*{10mm}  \includegraphics[scale=0.7]{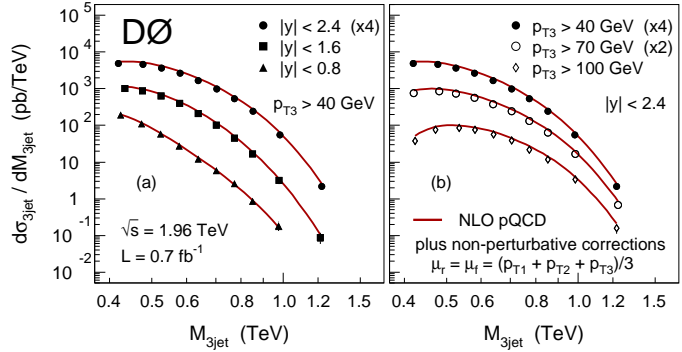}
\caption{The differential cross section $d\sigma_{\text{3jet}} / d\Mtj$
(a) in different rapidity regions and
(b) for different $p_{T3}$ requirements.
The solid lines represent the NLO pQCD matrix element calculations using MSTW2008NLO PDFs
and $\as(M_Z)=0.1202$ which are corrected for non-perturbative effects.}
\label{fig:3jet1_d0}
\end{figure}

\begin{figure}
\centering
\includegraphics[scale=0.75]{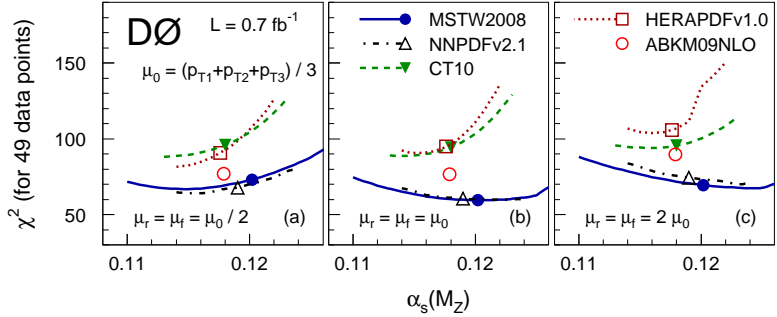}
\caption{
The $\chi^2$ values between theory and data,
as a function of the value of $\asmz$ used in the matrix elements
and PDFs.
The results are shown for different PDF parametrizations
and for different choices of the renormalization and
factorization scales.
The positions of the central $\asmz$ values in the different PDF sets
are indicated by the markers.}
\label{fig:3jet2_d0}
\end{figure}

The differential cross section in the three-jet invariant mass ($\Mtj$) is measured by the D0 collaboration
in five scenarios, spanning different rapidity regions and for different
requirements on the jet transverse momenta (see Fig.~\ref{fig:3jet1_d0})~\cite{3jet_d0}.
Jets are ordered in descending $\pT$ with the requirements 
$p_{T1} > 150$\,GeV and $p_{T3} > 40$\,GeV
(and no further requirement on $p_{T2}$).
The rapidities of the three leading $\pT$ jets are restricted to
$|y| < 0.8$, $|y| < 1.6$, or $|y| < 2.4$, in three different measurements.
Two additional measurements are made for $p_{T3} > 70$\,GeV
and $p_{T3} > 100$\,GeV, both requiring $|y| < 2.4$.
The data are compared to pQCD calculations at NLO
in $\alpha_s$ for different PDF parametrizations,
by computing $\chi^2$ values for different scale choices and
different $\asmz$ values (see Fig.~\ref{fig:3jet2_d0}).
The best description of the data is obtained for the MSTW2008NLO \cite{MSTW}
and NNPDFv2.1 \cite{NNPDF} PDF parametrizations which describe both
the normalization and the shape of the observed $\Mtj$ spectra.
The PDF parametrizations from ABKM09NLO \cite{ABKM} give a reasonable description
of the data, although with a slightly different shape of the $\Mtj$ spectrum.
The central results from the
CT10 \cite{CT10} and HERAPDFv1.0 PDF \cite{HERAPDF} sets predict a different $\Mtj$ shape
and are in poorer agreement with the data.

\begin{figure}[htbp]
\hspace*{-3mm}  \includegraphics[scale=0.85]{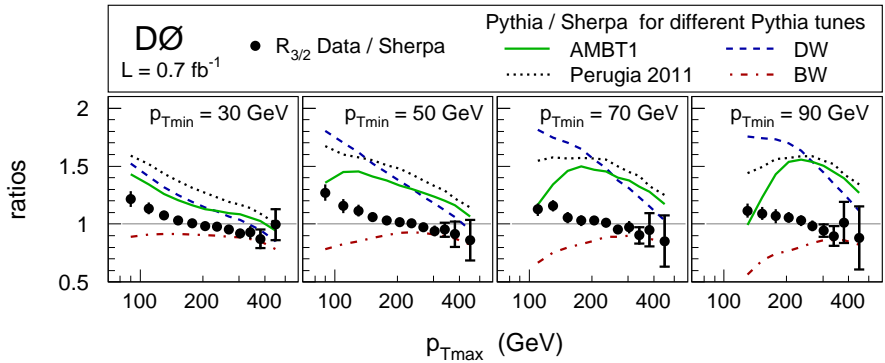}
\caption{
The measured $R_{3/2}$ results, normalized to the predictions of the {\sc sherpa}
Monte Carlo event generator. The inner
uncertainty bars represent the statistical uncertainties
while the total uncertainty bars represent the quadratic sums of statistical
and systematic uncertainties. Overlaid are the predictions from the {\sc pythia}
Monte Carlo event generator for four different tunes,
also normalized to the {\sc sherpa} predictions.}
\label{fig:r32_d0}
\end{figure}

The ratio  of three-jet to two-jet cross sections ($\Rtt$)
has also been measured \cite{r32}.
The ratio $\Rtt$ is presented in Fig.~\ref{fig:r32_d0} for the minimum jet $p_T$ ($\ptmin$) 
requirements of 30, 50, 70, and 90 GeV, as a function of the highest jet $p_T$ ($\ptmax$) in the range of 80--500~GeV.
The \sherpa\ event generator \cite{Sherpa} describes
the data within approximately $-10$\% to $+20$\%, but predicts a slightly different $\ptmax$\ dependence.
None of the \pythia\ MPI tunes
DW, BW, A, AMBT1, S Global, and Perugia 2011 \cite{PYT}
describes the data.
The data are well described by the pQCD predictions at NLO
in $\as$, corrected for non-perturbative effects estimated from hadronization 
and underlying event corrections using \pythia tunes DW and AMBT1.

\begin{figure}[htbp]
\center
\includegraphics[scale=0.65]{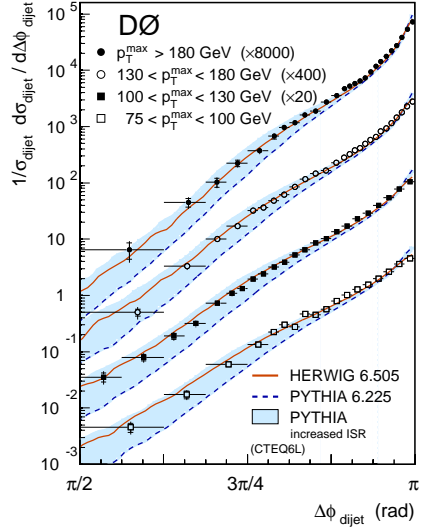}
\caption{The $\Dphi$ distributions in different $\ptmax$ ranges.
Results from {\sc herwig} and {\sc pythia} are overlaid on the data.
Data and predictions with $\ptmax > 100\,\text{GeV}$ are scaled by
successive factors of 20 for purposes of presentation.}
\label{fig:dphi1_d0}
\end{figure}

Multi-parton radiation is a complex aspect of pQCD theory and related phenomenology.
The proper description of radiative processes is crucial for a wide range of precision measurements as
well as for searches for new phenomena
where the influence of QCD radiation is unavoidable.
A clean and simple way to study radiative processes is to
examine azimuthal decorrelations in dijet events.
Results from {\sc herwig} (version 6.505) and {\sc pythia} (version 6.225) Monte Carlo generators,
both using default parameters and the CTEQ6L~\cite{cteq6} PDFs, are compared
to the $\Dphi$ measurement in the events with at least two jets \cite{azim_d0} in Fig.~\ref{fig:dphi1_d0}.
The minimum jet $p_T$, $\ptmin$, is 40 GeV while  $\ptmax$ is varied.
The data are described by
{\sc herwig} 
well over the entire $\Dphi$ range including the region around $\Dphi\approx\pi$. 
{\sc pythia} with default parameters describes the data poorly---the
distribution is too narrowly peaked at $\Dphi\approx\pi$ and lies
significantly below the data over most of the $\Dphi$ range.
The maximum $p_T$ in the initial-state parton shower is directly
related to the maximum virtuality that can be adjusted in {\sc pythia}.
The shaded bands indicate the range of variation
when the maximum allowed virtuality is increased from the
current default by a factor of four~\cite{parp67}.
These variations result in significant changes in the low $\Dphi$ region
clearly demonstrating the sensitivity of this measurement.
Consequently, global efforts to tune Monte Carlo event generators
should benefit from including these data.
%
NLO pQCD describes the data except for very large $\Dphi$ where the calculation does not provide a reliable prediction.

\begin{figure}[htbp]
\center
\includegraphics[scale=0.7]{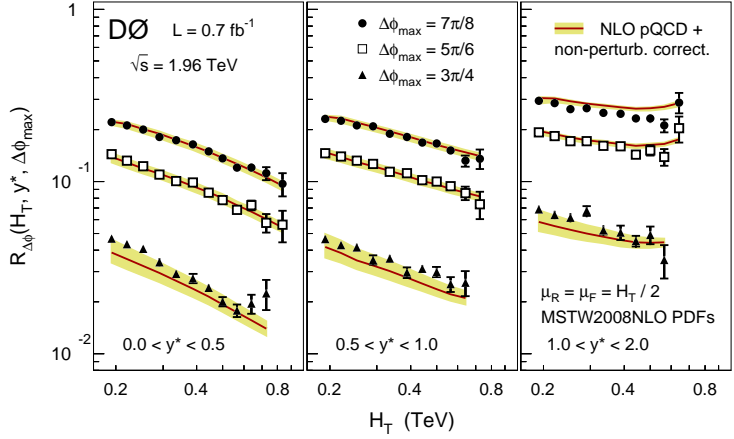}
\caption{The results for $\Rdphi$ as a function of $H_T$
in three different regions of $y^*$
and for three different $\Dphimax$ requirements.
The error bars indicate the statistical and systematic
uncertainties summed in quadrature.
The theoretical predictions are shown with their uncertainties.}
\label{fig:dphi2_d0}
\end{figure}

The combined rapidity and $p_T$ dependence of dijet azimuthal decorrelations has been also studied \cite{dphi_eta_d0}.
This measurement is based on a new quantity $\Rdphi$, 
defined as the fraction of the inclusive dijet
cross section with a decorrelation of $\Dphi < \Dphimax$.
The ratio $\Rdphi$ is measured
as a function of the total jet transverse 
momentum $H_T$, the rapidity $y^*=|y_{\rm jet1}-y_{\rm jet2}|$, and 
the maximal azimuthal decorrelation with $\Dphimax$, see Fig.~\ref{fig:dphi2_d0}.
For all values of $\Dphimax$ and at fixed $H_T$, dijet azimuthal
decorrelations increase with $y^*$, while they decrease with $H_T$ 
over most of the $H_T$ range at fixed $y^*$.
Predictions of NLO pQCD, corrected for non-perturbative effects,
give a good description of the data, except in the kinematic region
of large dijet rapidity intervals $y^* > 1$ and small decorrelations with
$\Dphimax = 7\pi/8$.

\subsection{Extraction of $\alpha_s$}


The D0 inclusive jet data has  been used to extract values of the strong coupling constant
$\alpha_s$ in the interval of $50 < p_T^{\rm jet} < 145$ GeV \cite{d0_as}.
The best fit over 22 data points leads 
to $\alpha_s(m_Z)=0.1161^{+0.0041}_{-0.0048}$ with improved accuracy as compared to the Run I CDF result~\cite{cdf_as},
$\alpha_s(m_Z)=0.1178^{+0.0122}_{-0.0121}$,
 and also in agreement with result from HERA jet data \cite{hera_as}.

\begin{figure}[htbp]
\center
\includegraphics[scale=0.67]{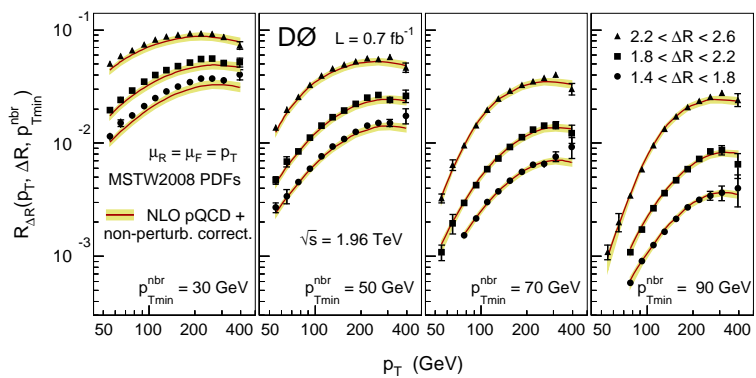}
\caption{The measurement of $\Rdr$ as a function of inclusive jet $p_T$
for three different intervals in $\DR$ and for four different
requirements of $\ptnbrmin$.
The inner uncertainty bars indicate the statistical uncertainties,
and the total uncertainty bars display the quadratic sum of
the statistical and systematic uncertainties.
The theory predictions are shown with their uncertainties.}
\label{fig:dr1_as}
\end{figure}

\begin{figure}[htbp]
\center
\includegraphics[scale=0.75]{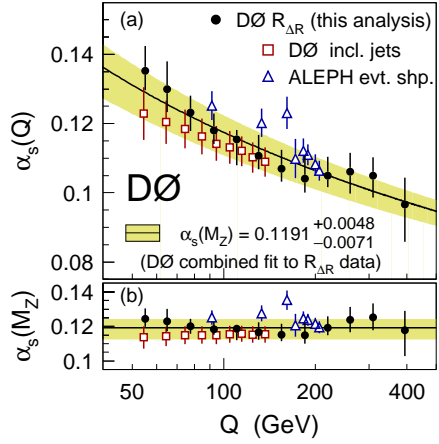}
\caption{The strong coupling $\as$ at large momentum transfers, $Q$,
presented as $\asq$ (a) and evolved to $M_Z$ using the RGE (b).
The uncertainty bars indicate the total uncertainty, including
the experimental and theoretical contributions.
The new $\as$ results from $\Rdr$ are compared to previous
results obtained
from inclusive jet cross section data~\cite{d0_as}
and from event shape data~\cite{Dissertori}.
The $\asmz$ result from the combined fit to all selected data points (b)
and the corresponding RGE prediction (a) are also shown.}
\label{fig:dr2_as}
\end{figure}

A new quantity $\Rdr$ which probes the angular correlations of jets has been introduced \cite{d0_rdr}.
It is defined as the number of neighboring jets above a given
transverse momentum threshold which accompany a given jet
within a given distance $\DR$ in the plane of rapidity and
azimuthal angle.
$\Rdr$ is measured as a function of inclusive jet $p_T$ in
different annular regions of $\DR$ between a jet and its neighboring jets
and for different requirements on the minimal transverse momentum
of the neighboring jet $\ptnbrmin$ (see Fig.~\ref{fig:dr1_as}).
The data for $p_T > 50\,$GeV are well-described by pQCD calculations
in NLO in $\as$ with non-perturbative corrections applied.
Results for $\aspt$ are extracted using the data with $\ptnbrmin \ge 50\,$GeV,
integrated over $\DR$.
The extracted $\aspt$ results from $\Rdr$ are, to good approximation,
independent of the PDFs and thus independent of assumptions on the renormalization group equation (RGE).
The results are in good agreement with previous results and consistent with
the RGE predictions  for the running of $\as$ for momentum transfers
up to 400\,GeV (see Fig.~\ref{fig:dr2_as}).
The combined $\asmz$ result, 
integrated over $\DR$ and $p_T$, is $\asmz = 0.1191^{+0.0048}_{-0.0071}$, in good agreement
with the world average value~\cite{PDG2012}.

\subsection{Jet substructure}

Studying the jet substructure allows for tuning parton showering and 
search for heavy resonances decaying hadronically and separated by a small angle.
It has been one of important topics of Run I jet program (see e.g. Ref.~\citen{jetstruct_run1}).

\begin{figure}
\center
{\includegraphics[width=8cm]{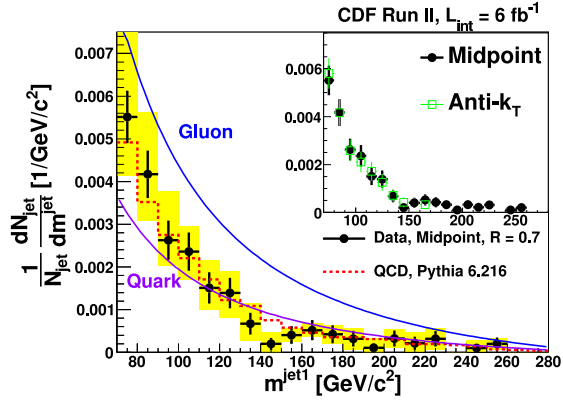} }
 \caption{  
\label{Fig:Jet1MassR07Compared}
The normalized jet mass distribution for jets with $p_T > 400$ GeV and $|\eta|\in (0.1,0.7)$.
The uncertainties shown are statistical (black lines) and systematic (yellow bars).
The theory predictions for the jet function for quarks and gluons are shown as solid curves
and  have an estimated uncertainty of $\sim30$\%.
We also show the {\sc pythia} Monte Carlo prediction (red dashed line).
The inset compares jets found by Midpoint (full black circles) and anti-$k_T$ (open green squares) algorithms.
}
\end{figure}

In Run II, the CDF collaboration studied structure of high $p_T$ jets by selecting only events with at least one jet
having $p_T>400$ GeV, $0.1<|y|<0.7$ and considering jets with cone sizes $R=0.4,0.7$ and $1.0$ \cite{cdf_jet_struct}.
The jet mass is calculated using 4-vectors of calorimeter towers in a jet. 
Special selection to remove the $t$-quark events have been applied.
Its mass distribution unfolded to the particle level is shown in Fig.~\ref{Fig:Jet1MassR07Compared}.  
The data are in agreement {\sc pythia} predictions and are located between  
the 
predictions  for quark and gluon jets. The data
confirm that the high mass jets are mostly caused by quark fragmentation.

\subsection{Jet shapes}

\begin{figure}
\vskip -5mm
\center
{\includegraphics[width=7cm]{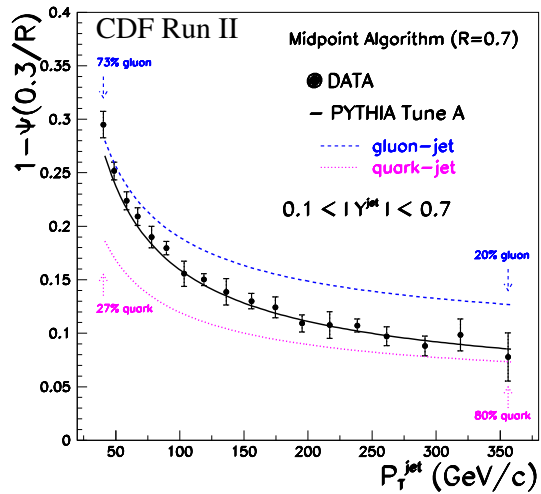} }
 \caption{  
\label{Fig:incjet_shapes}
The measured  $1 - \Psi(0.3/R)$ as a function of $p_T^{\rm jet}$
for jets with $0.1 < |y^{\rm jet}| < 0.7$ and $37 < p_T^{\rm jet} < 380 \ \GeVc$.
Error bars indicate the statistical and systematic uncertainties added in quadrature.
The predictions of {\sc pythia} Tune A (solid line) and the
separate predictions for quark-initiated jets (dotted line) and gluon-initiated jets (dashed line)
are shown for comparison. The arrows indicate the fraction of quark- and gluon-initiated jets at low and very high $P_T^{\rm jet}$, as predicted by {\sc pythia} Tune A.
}
\end{figure}

Jet shapes have been studied using inclusive jet production events 
in the kinematic region~$37 < p_{T}^{\rm jet} < 380 \ \GeVc$ and $0.1 < |y^{\rm jet}| < 0.7$
by the CDF experiment~\cite{InclJetShape}. 
Figure \ref{Fig:incjet_shapes} shows the measured  fractional total $p_T$ outside a cone of radius $r=0.3$
around the jet axis, $1 - \Psi(0.3/R)$, as a function of $p_T^{\rm jet}$. 
Here $\Psi$ is defined as 
\begin{equation}
\Psi(r) = \frac{1}{\rm N_{jet}} \sum_{\rm jets} \frac{p_T(0,r) }{p_T(0,R)}, \ \ \ \ 0 \leq  r \leq R.
\end{equation}

Jets become narrower as $p_{T}^{\rm jet}$ increases
which can be mainly attributed to the change in the quark- and gluon-jet mixture in the final state and the running
of the strong coupling with $p_T^{\rm jet}$.
{\sc pythia} Tune A Monte Carlo predictions, which includes  enhanced contributions from initial-state
gluon radiation and  secondary parton interactions between remnants, provides a good description of the data.
{\sc herwig} gives a reasonable description of the measured jet shapes but tends to
produce jets that are too narrow at low $p_{T}^{\rm jet}$ which can be attributed to the absence
of soft contributions from multiple parton interactions in {\sc herwig}. Jet shape measurements thus can be used to introduce strong
constraints on phenomenological models  describing soft-gluon radiation and the underlying event
in hadron-hadron interactions.
Similar studies with $b$-jets are also done \cite{bJetShape}.

\subsection{New phenomena searches}

\begin{figure}
\center
{\includegraphics[width=7cm]{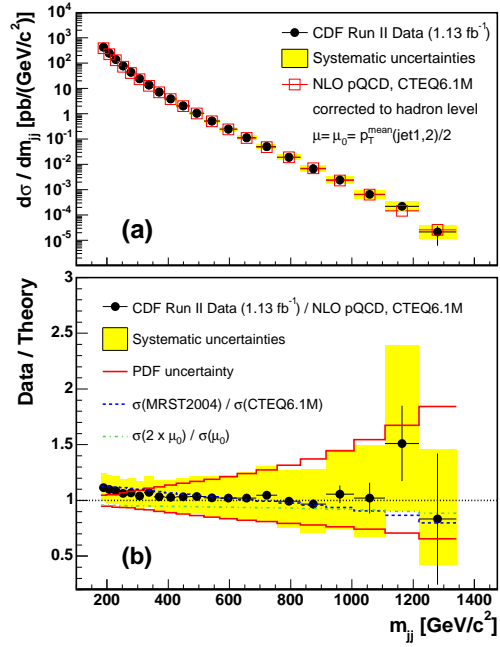} }
 \caption{  
\label{Fig:cdf_djm}
 (a) The measured dijet mass spectrum
    for both jets to have $|y|<1$
    compared to the NLO pQCD prediction obtained using the CTEQ6.1 PDFs.
    (b) The ratio of the data to the NLO pQCD prediction.
    The experimental systematic uncertainties, theoretical uncertainties
    from PDF, the ratio of MRST2004/CTEQ6.1, and
    the dependence on the choice of renormalization and factorization
    scales 
    are also shown.
    An additional 6\% uncertainty in the determination of the luminosity
    is not shown.
}
\end{figure}

The  CDF collaboration performed a search for new particles which decay into
dijets by measuring the dijet mass spectrum using
$p\bar p$ collision data from 1.1 fb$^{-1}$ of integrated luminosity~\cite{cdf-dijet}.
Since jets produced by new physics are expected to be produced more centrally than by Standard Model processes 
only events with two leading jets with $\mid y\mid\leq$1.0 are used. 
The measured dijet mass spectrum, see Fig.~\ref{Fig:cdf_djm} is found to be consistent with NLO
pQCD predictions based on recent PDFs and does not show evidence of a
mass resonance from new particle production.
Upper limits at the 95\% confidence level (CL) on new particle production cross sections  were set.
The mass exclusions for the excited quark, axigluon, flavor-universal coloron, $E_6$ diquark,
color-octet techni-$\rho$, $W'$, and $Z'$ for a specific
representative set of model parameters has also been determined.

The D0 collaboration measured~\cite{d0-angular-distribution} normalized  angular distributions in $\chijj=\exp{(|y_1 - y_2|)}$.
They are well-described by theory calculations at NLO in the strong coupling constant
and are used to set limits on quark compositeness, ADD large extra dimensions \cite{ArkaniHamed:1998rs,Atwood:1999qd},
and TeV$^{-1}$ extra dimensions models \cite{Dienes:1998vg}.


\section{Photon final states}
\label{seq:gamma}

\subsection{Inclusive photon production}

\begin{figure}
\center
\includegraphics[width=6.5cm]{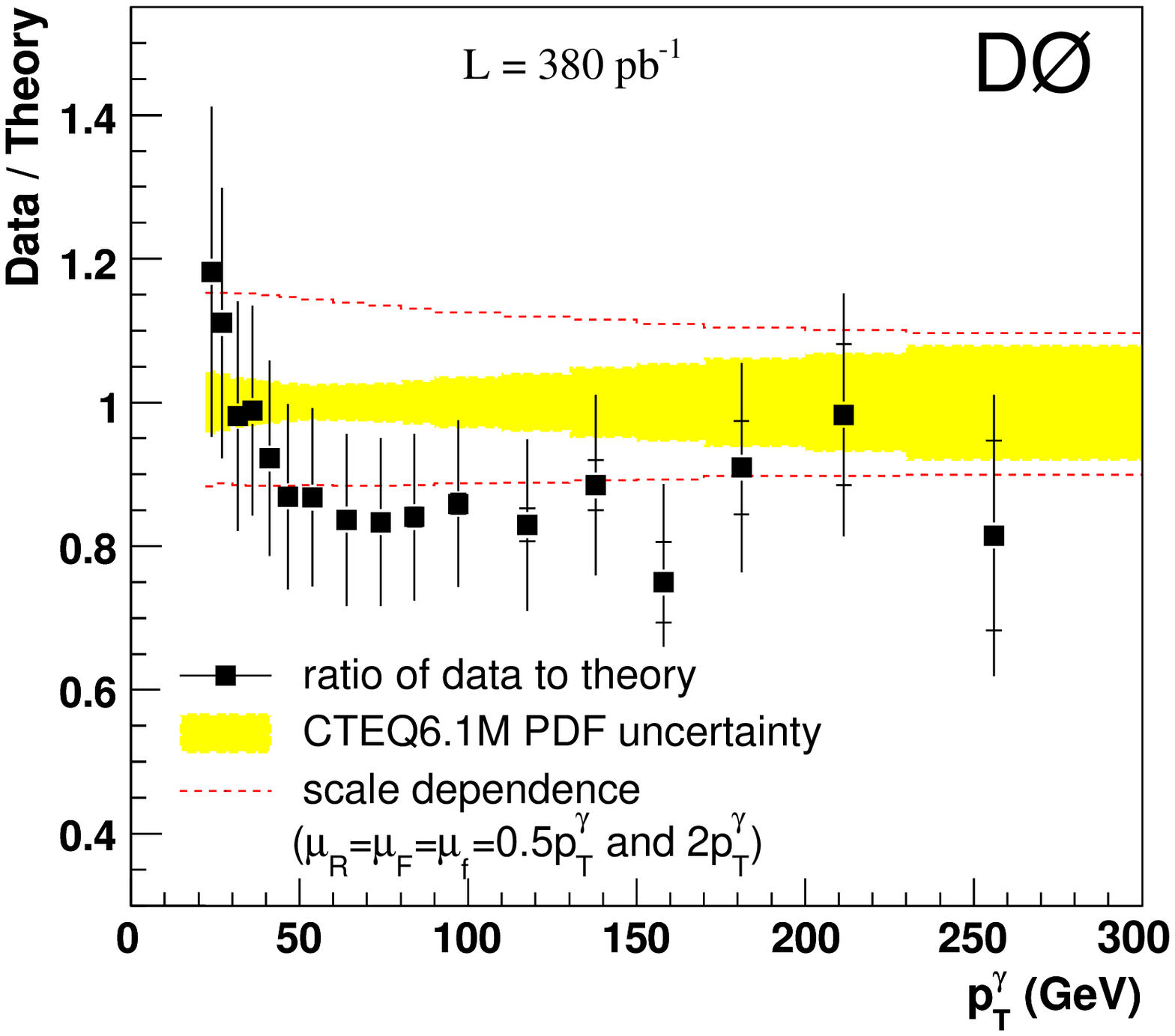} 
\includegraphics[width=6cm]{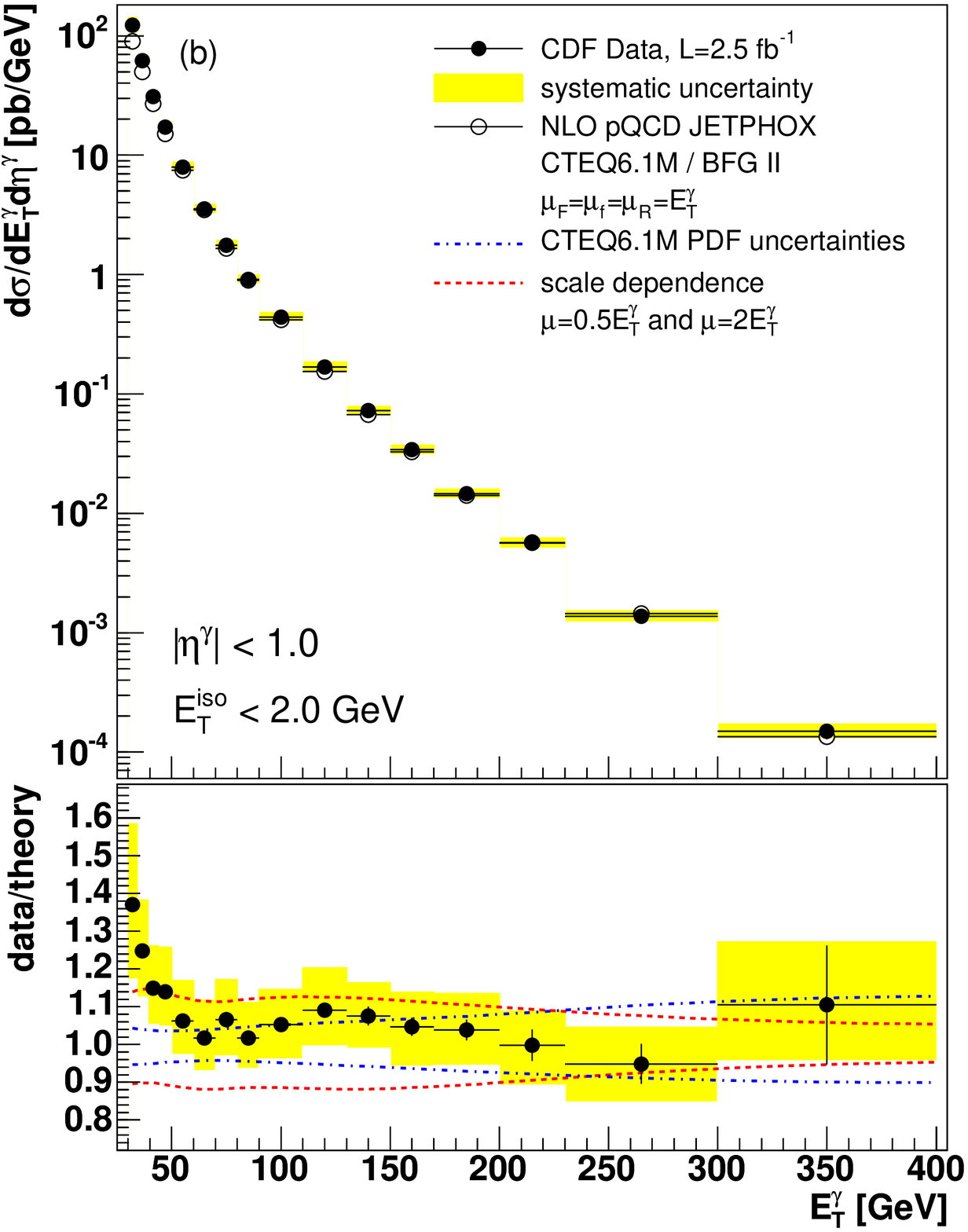}
 \caption{  
\label{Fig:incgam}
The ratio of the measured cross section to the theoretical predictions from
{\sc jetphox}. The  plot (a) is for D0 and the  plot (a) is for CDF measurements.
The full vertical lines correspond to the overall uncertainty
while the internal line indicates just the statistical uncertainty. Dashed lines 
represents the change in the cross section when varying the theoretical scales by
factors of two. The shaded region indicates the uncertainty in the cross section
estimated with CTEQ6.1M PDFs. 
}
\end{figure}

The high $p_T$ photons emerge directly from $p\bar{p}$ collisions and provide
a probe of the parton hard scattering process with a dominating contribution
from $qg$ initial state.
Being a direct probe of the parton dynamics, they are of a permanent interest in
high energy physics. A few cross section measurements were done in Run I (see Ref.~\citen{photons_run1}).
In Run II, the inclusive photon production cross sections 
have been measured by D0 and CDF collaborations with photons 
in the central rapidity region \cite{d0_incgam,cdf_incgam}. 
The results shown in Fig.~\ref{Fig:incgam} are in agreement within experimental 
uncertainties between the two experiments, 
and both indicate some tension between NLO pQCD and data at low $p_T$.

The D0 and CDF inclusive photon data together with ATLAS and CMS data 
\cite{atlas_incgam, cms_incgam} have been used to constrain 
the gluon PDF at low $x$ values \cite{gluonPDF_incgam}.

\subsection{Photon+jet production}

The production of a photon with associated jets in
the final state is another powerful and direct probe of the dynamics of
hard QCD interactions. As compared with the inclusive photon production, 
information about the accompanying jet allows to calculate 
parton fractions $x$ in the leading order approximation (see e.g. Ref.~\citen{Owens}).
Different $\Ptg$ and angular configurations between the photon and the jets can
be used to extend inclusive photon production measurements and 
simultaneously test the underlying dynamics of QCD hard-scattering subprocesses 
in different regions of $x$ and hard-scattering scales $Q^2$.

The triple differential cross section
$\mathrm{d^3}\sigma / \mathrm{d}\Ptg\mathrm{d}y^{\gamma}\mathrm{d}y^{\mathrm{jet}} $ for
the associated inclusive photon and jet production process $p\bar{p}\rightarrow \gamma +\mathrm{jet} +X$ 
is measured for events with central ($|y^\gamma|\lt 1.0$) and forward ($1.5\lt |y^\gamma|\lt 2.5$) photons
in four jet rapidity intervals ($|y^\text{jet}|\leq 0.8$, $0.8<|y^\text{jet}|\leq 1.6$,
$1.6<|y^\text{jet}|\leq 2.4$, and $2.4<|y^\text{jet}|\leq 3.2$), for
configurations with same and opposite signs of photon and jet rapidities \cite{gj_tcs}.
The pQCD NLO predictions describe data with central photons
in almost all jet rapidity regions except low $\Ptg$ ($<40$~\GeVc) and the opposite-sign rapidity events at high $\Ptg$
with very forward jets ($2.4<|y^{\mathrm{jet}}|<3.2$).
They also describe data with forward photons except for the same-sign rapidity events 
with $\Ptg>70$ \GeVc and $2.4<|y^{\mathrm{jet}}|\leq3.2$.
The measured cross sections typically have similar or smaller uncertainties than
the NLO PDF and scale uncertainties,
and can be used as inputs to global fits determining gluon and other PDFs.

\begin{figure}
\center
\includegraphics[width=6.2cm]{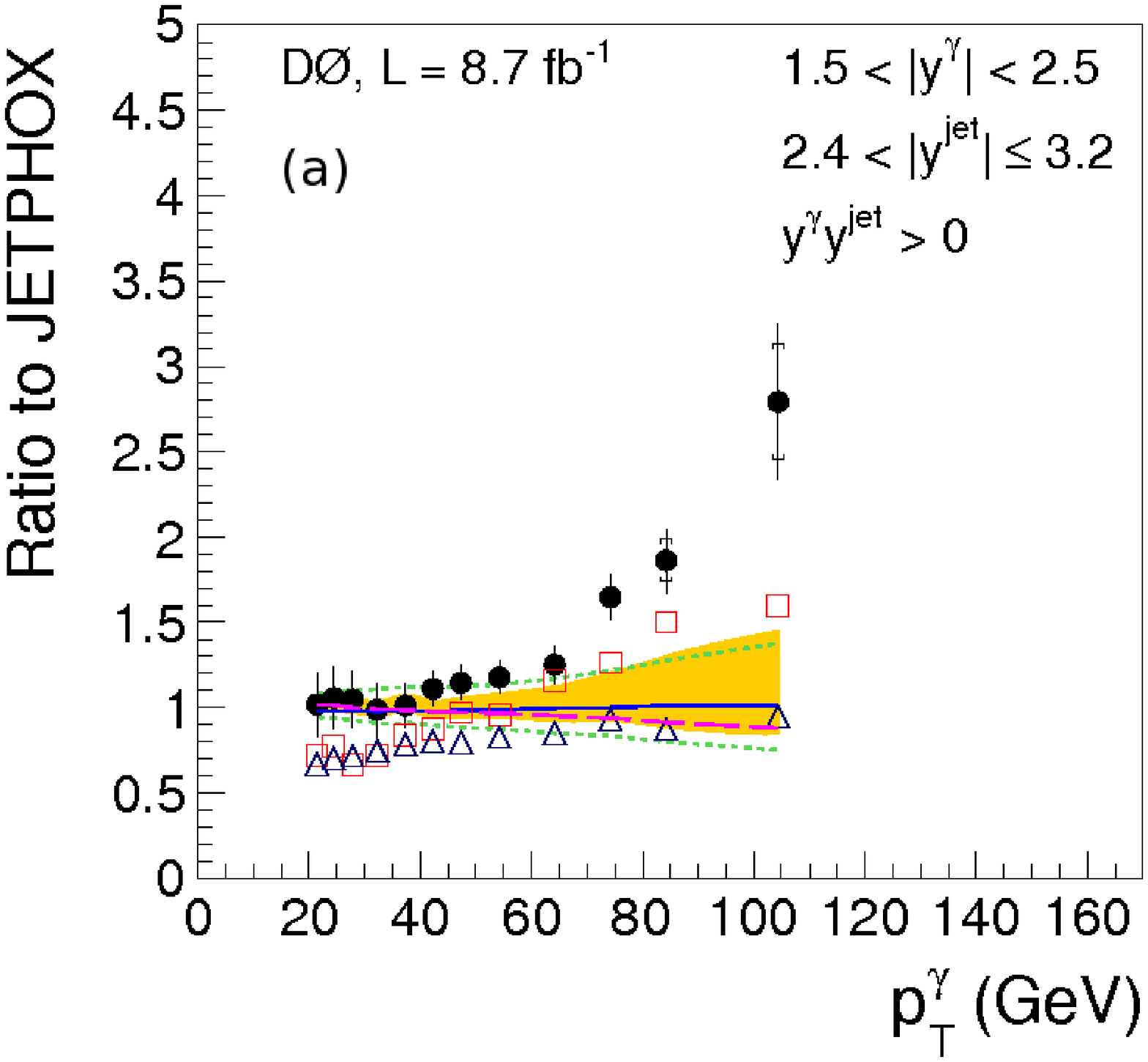} 
\includegraphics[width=6.2cm]{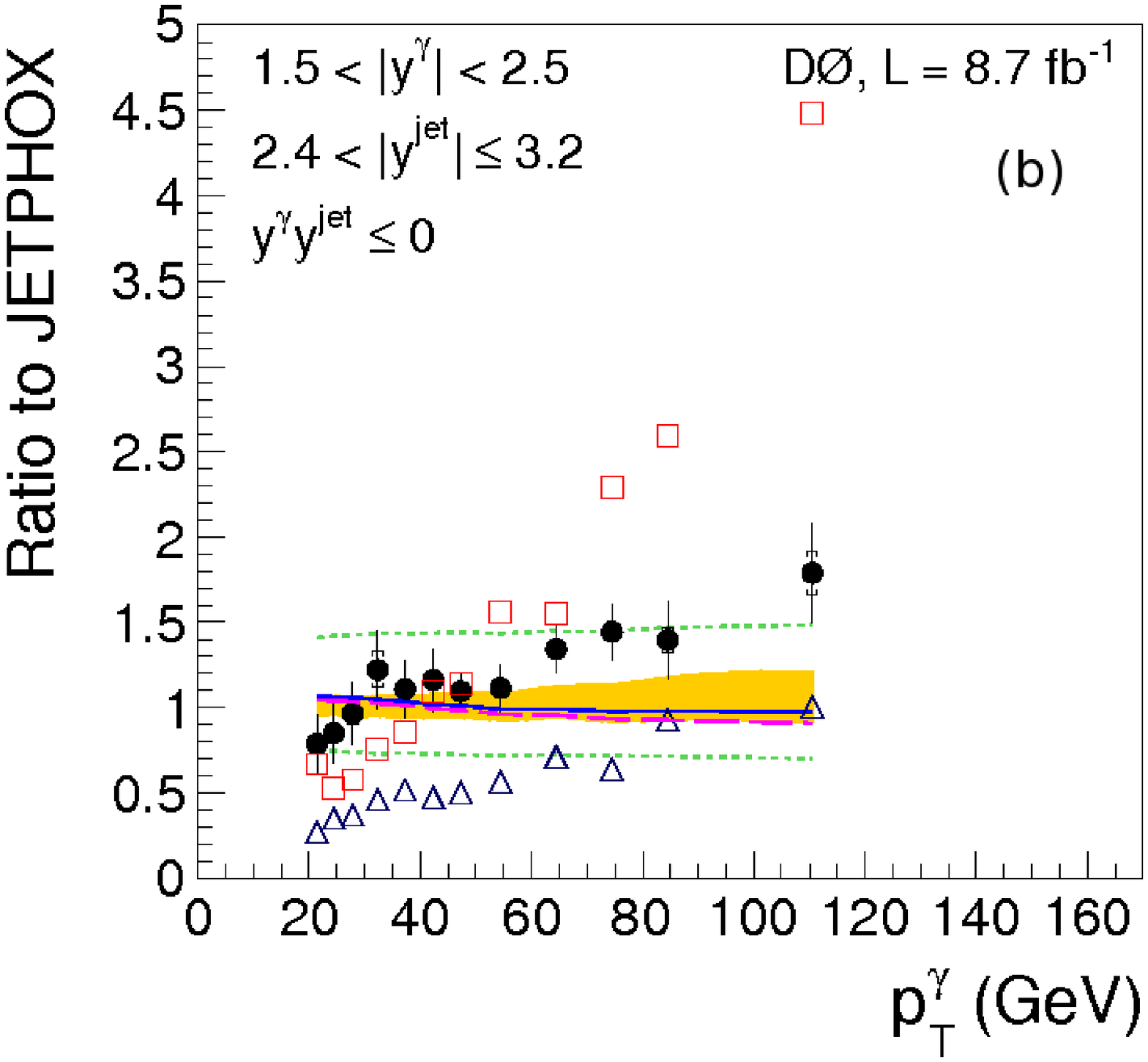}
 \caption{  
\label{Fig:tcs_d0}
Ratios of the measured differential cross sections of $\gamma+$jet production 
with forward photons and jet rapidity interval $2.4<|y^{\mathrm{jet}}|\leq3.2$
to the pQCD NLO prediction using {\sc jetphox} \cite{JETPHOX} with the CT10 PDF set and $\mu_{R}=\mu_{F}=\mu_f=\Ptg$.
Plots (a) and (b) correspond to the same and opposite signs of photons and jet rapidities.
The solid vertical line on the points shows the statistical and $p_T$-dependent systematic uncertainties added in quadrature,
while the internal line shows the statistical uncertainty.
The two dotted lines represent the effect of varying the
theoretical scales of {\sc jetphox} by a factor of two. The shaded region is the CT10 \cite{CT10} PDF uncertainty.
The dashed and dash-dotted lines show ratios of the {\sc jetphox} predictions with MSTW2008NLO \cite{MSTW}
and NNPDFv2.1 \cite{NNPDF} to CT10 PDF sets.
The predictions from {\sc sherpa} and {\sc pythia} are shown by the open squares and triangles, respectively.
}
\end{figure}

\subsection{Photon + heavy flavor jet production}
\label{sec:ghf}

Study of events with photons produced in association with a $b(c)$-quark jet
provides information about the $b(c)$-quark and gluon PDFs of the proton.
At high $p_T$'s, they are also sensitive to the events
with double $b(c)$ quarks produced in the annihilation process
$q\bar{q}\to \gamma g, g\to Q\bar{Q}$ ($Q=b,c$).
These events also provide a test for the models with intrinsic charm and beauty \cite{cteq6c,BHPS}.

The D0 and CDF experiments have measured the differential cross sections
of $\gamma+b$-jet and $\gamma+c$-jet productions as a function of $\Ptg$ at the Fermilab Tevatron $p\bar{p}$ collider 
\cite{gb_d0, gc_d0, gbc_cdf}.
The results cover the kinematic range $30<\Ptg<300$ \GeVc, 
$|y^\gamma|<1.0$, and $|y^{\rm jet}|<1.5$. 
In the same kinematic region, and in the same $\Ptg$ bins, D0 has also 
measured the $\sigma(\gamma+c)/\sigma(\gamma+b)$ cross section ratio.
None of the theoretical predictions considered (QCD NLO \cite{Tzvet}, 
$k_T$ factorization \cite{Zotov}, {\sc sherpa} and {\sc pythia}) 
give good description of the data in all $\Ptg$ bins. Such a description  
might be achieved by including higher-order corrections
into the QCD predictions. At $\Ptg\gtrsim 80$ \GeVc,
the observed difference from data may also be
caused by an underestimated contribution from gluon splitting $g\to c\bar{c}$ \cite{LEP,LHCb,Atlas}
in the annihilation process or by contribution from intrinsic charm \cite{cteq6c,BHPS}.

\begin{figure}
\includegraphics[width=0.48\linewidth]{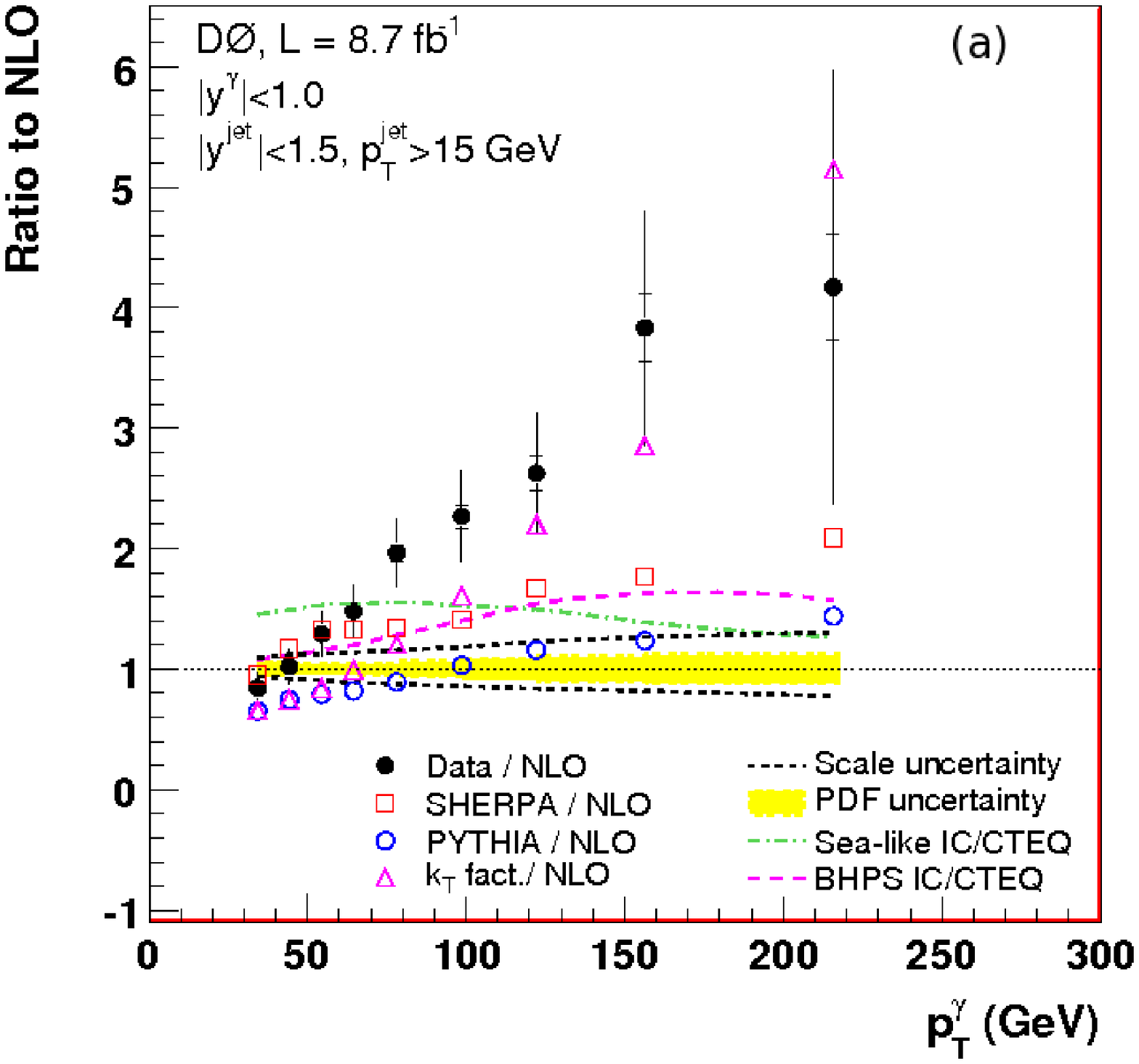}
\includegraphics[width=0.48\linewidth]{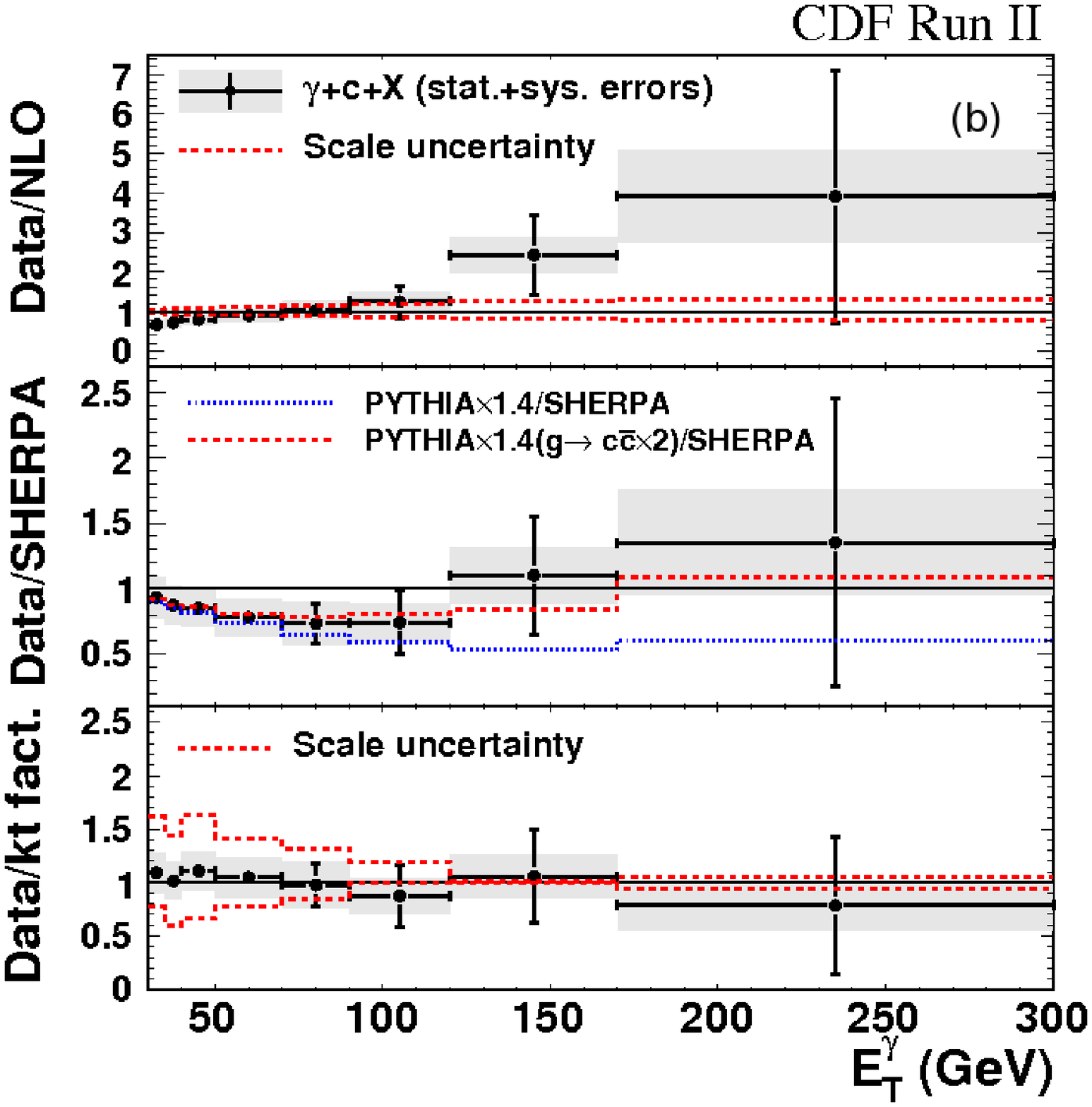}
~\\[-4mm]
\caption{%
The ratio of measured \gc-jet production cross sections to predictions. 
The  plot (a) is for D0 and the plot (b) is for CDF measurements.
The uncertainties on the data include both statistical (inner error bar) and total uncertainties (full error bar).
Also shown are the uncertainties on the theoretical QCD scales and the {\sc cteq}6.6M PDFs.
The ratio for intrinsic charm models \cite{cteq6c} are presented.
as well as the predictions given by $k_{\rm T}$-factorization \cite{Zotov}, {\sc sherpa}~\cite{Sherpa} and {\sc pythia}~\cite{PYT}.
The result of increased $q\bar{q}\to \gamma+g (g\to c\bar{c})$ rates by a factor 1.4 in {\sc pythia} predictions is also shown
(on the right).
}
\label{fig:xsectratio}
\end{figure}

\begin{figure}
\center
\includegraphics[width=0.60\linewidth]{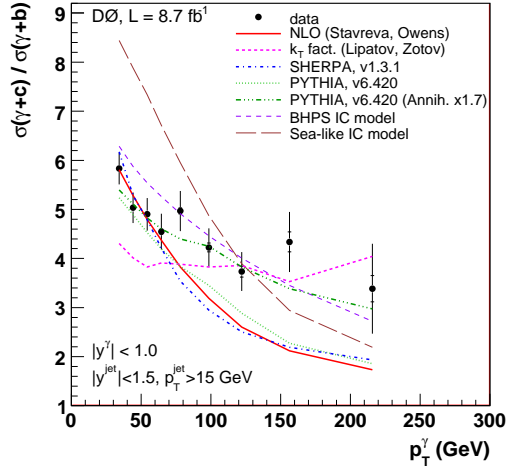}
~\\[-4mm]
\caption{%
The ratio of \gc-jet and \gb-jet production cross sections for data together with theoretical predictions as a function of $\Ptg$.
The uncertainties on the data include both statistical (inner error bar) and total uncertainties (full error bar).  
Predictions given by $k_{\rm T}$-factorization \cite{Zotov}, {\sc sherpa} 
\cite{Sherpa} and {\sc pythia} \cite{PYT} are also shown. The {\sc pythia} predictions with a contribution from the annihilation
process increased by a factor of 1.7 are shown as well.
The predictions for intrinsic charm models \cite{cteq6c} are also presented.  
}  
\label{fig:xsectratio2}  
\end{figure} 

\begin{figure}
\center
\includegraphics[width=0.60\linewidth]{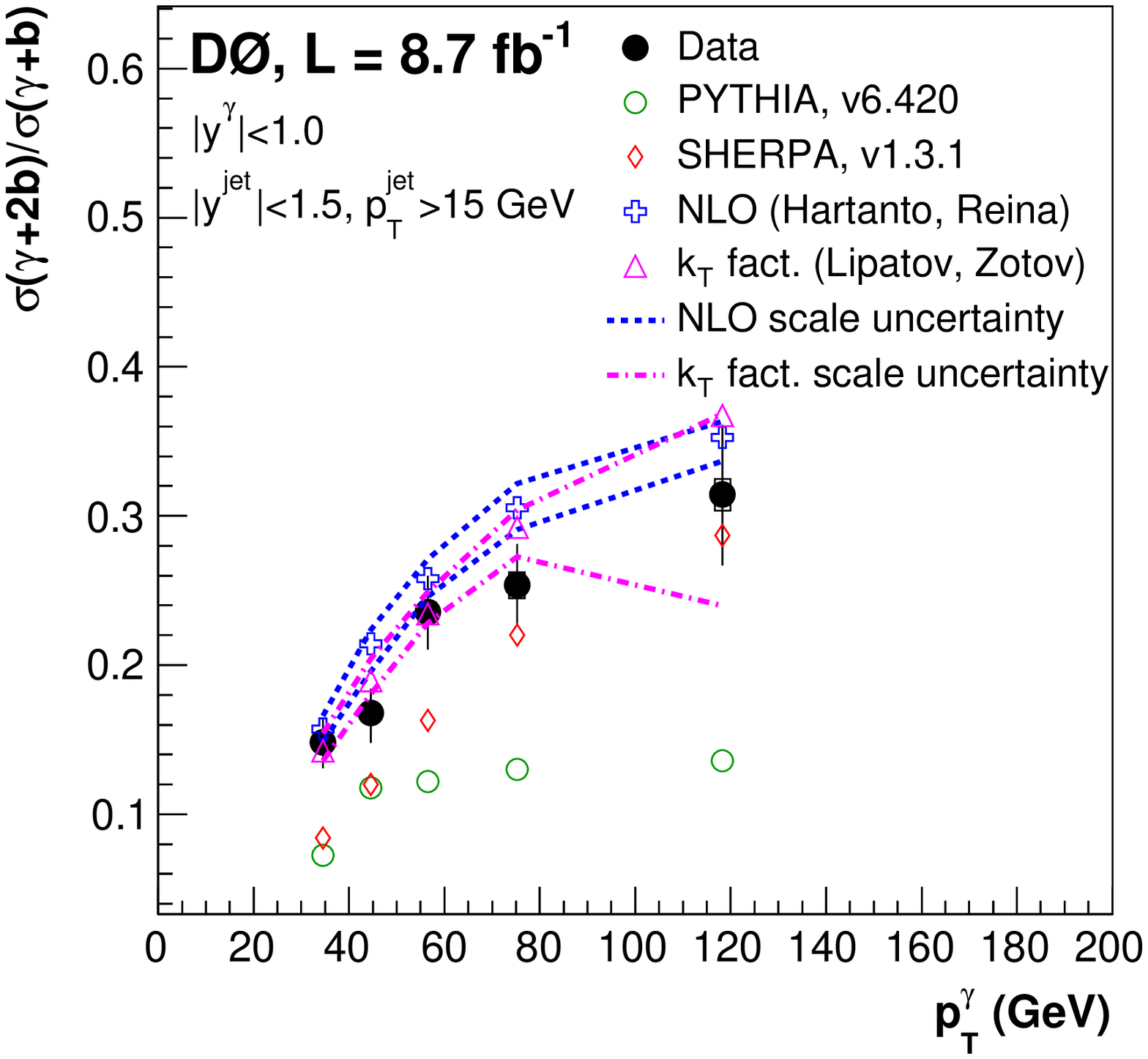}
~\\[-4mm]
\caption{%
The ratio of $\gamma+2~b$-jet to $\gamma+b$-jet production cross sections for data together 
with theoretical predictions as a function of $\Ptg$.
The uncertainties on the data points include both statistical (inner error bar) and
the full uncertainties (full error bar). 
The measurements are compared to the NLO QCD calculations \cite{NLO_gbb}.
The predictions from  {\sc sherpa} 
\cite{Sherpa}, {\sc pythia} \cite{PYT}  and $k_{\rm T}$-factorization \cite{Zotov}
are also shown along with the scale uncertainties on NLO and $k_{\rm T}$-factorization prediction.
}
\label{fig:gbb}
\end{figure}

Production of $\gamma+2~b$-jet events has been studied by D0 collaboration \cite{gbb_d0} differentially
in $\Ptg$ bins.  The ratio of differential production cross sections for 
$\gamma+2~b$-jet to $\gamma+b$-jet is also measured, see Fig.~\ref{fig:gbb}.
The ratio agrees with the predictions from NLO QCD and $k_{\rm T}$-factorization
approach within the theoretical and experimental uncertainties in the full studied
$\Ptg$ range while is not described by {\sc sherpa}  and {\sc pythia} generators.

\subsection{Diphoton production}

\begin{figure}[htbp]
\hspace*{-2mm}  \includegraphics[scale=0.21]{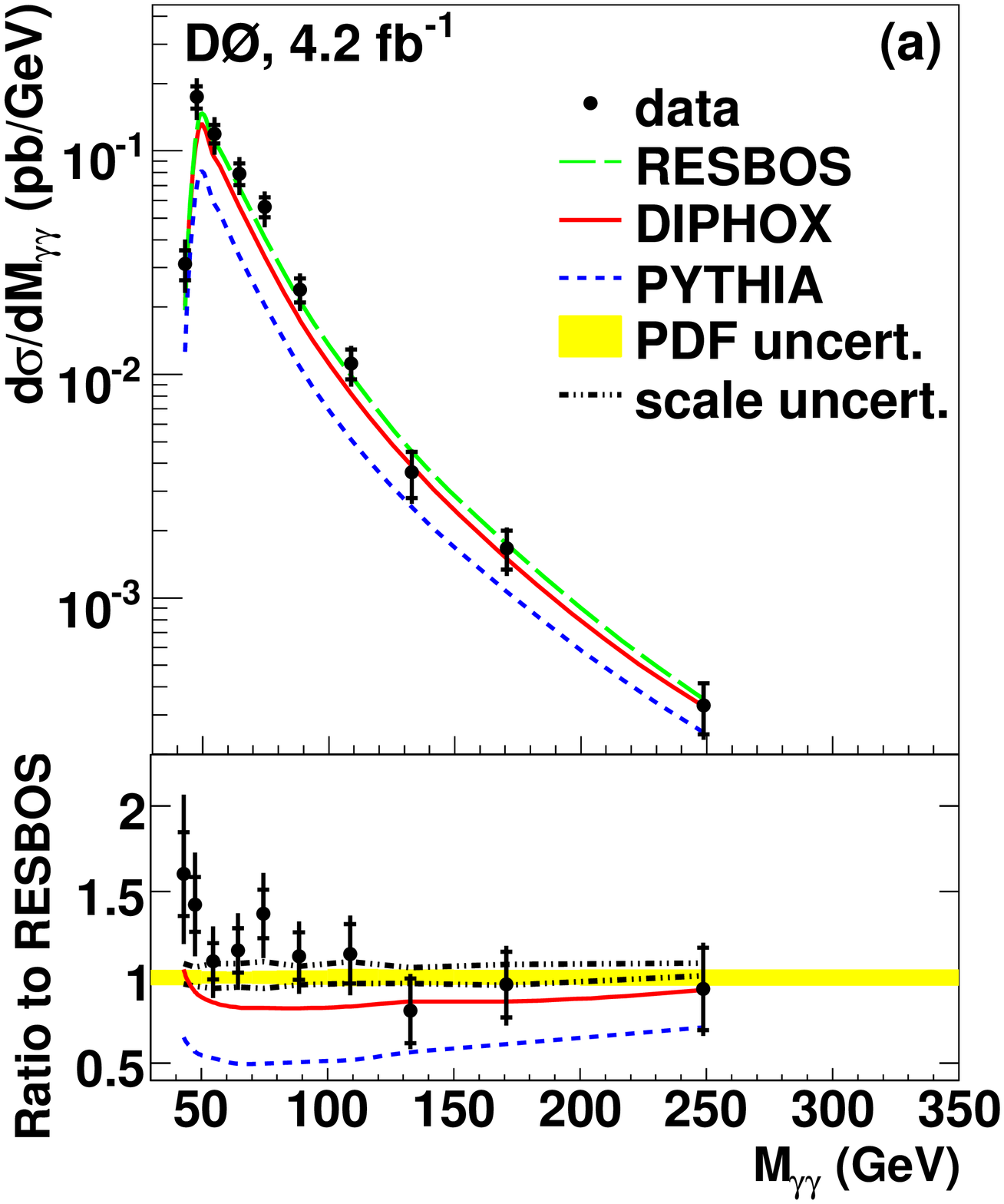}
\hspace*{-2mm}  \includegraphics[scale=0.21]{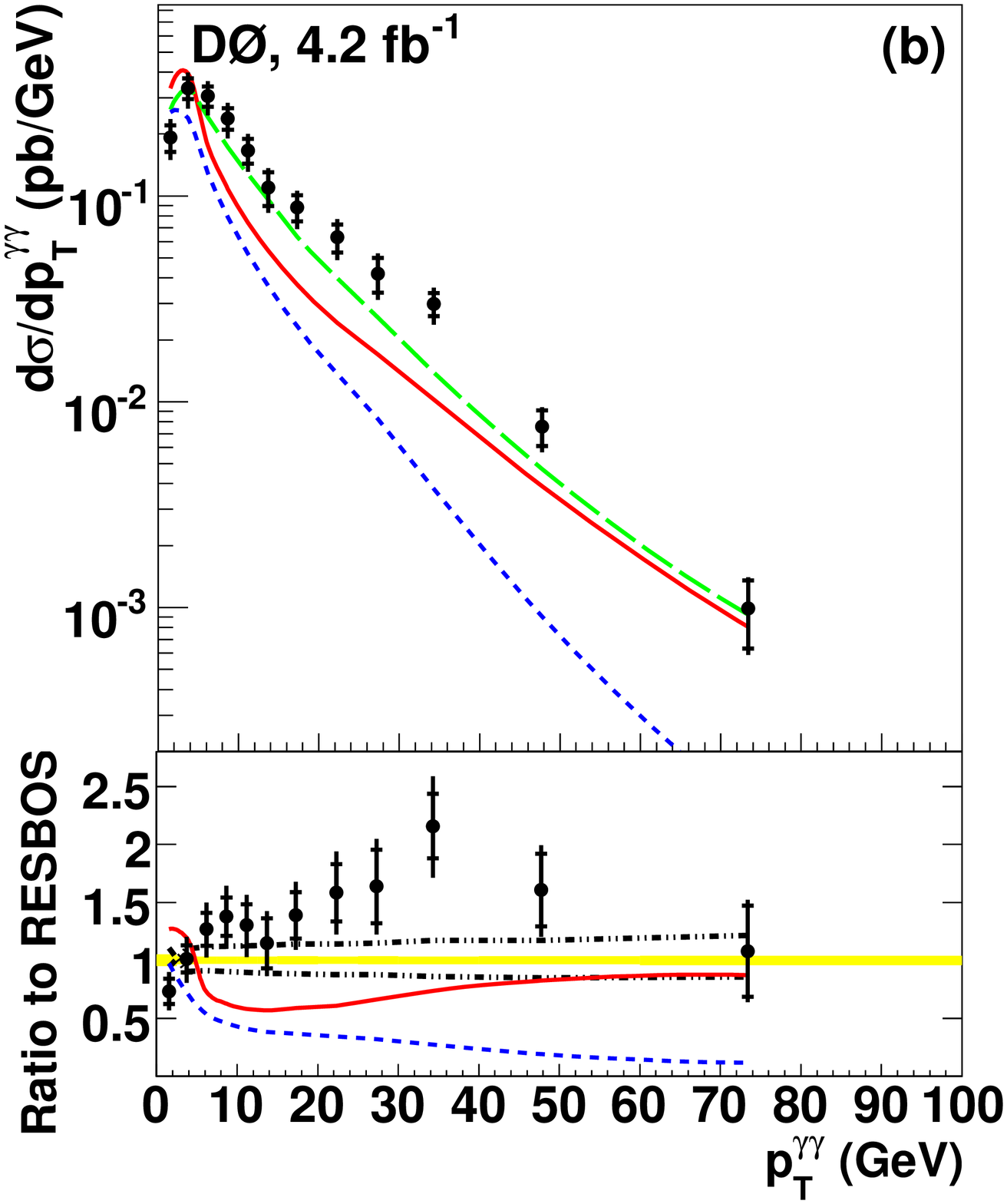}
\hspace*{-1mm}  \includegraphics[scale=0.21]{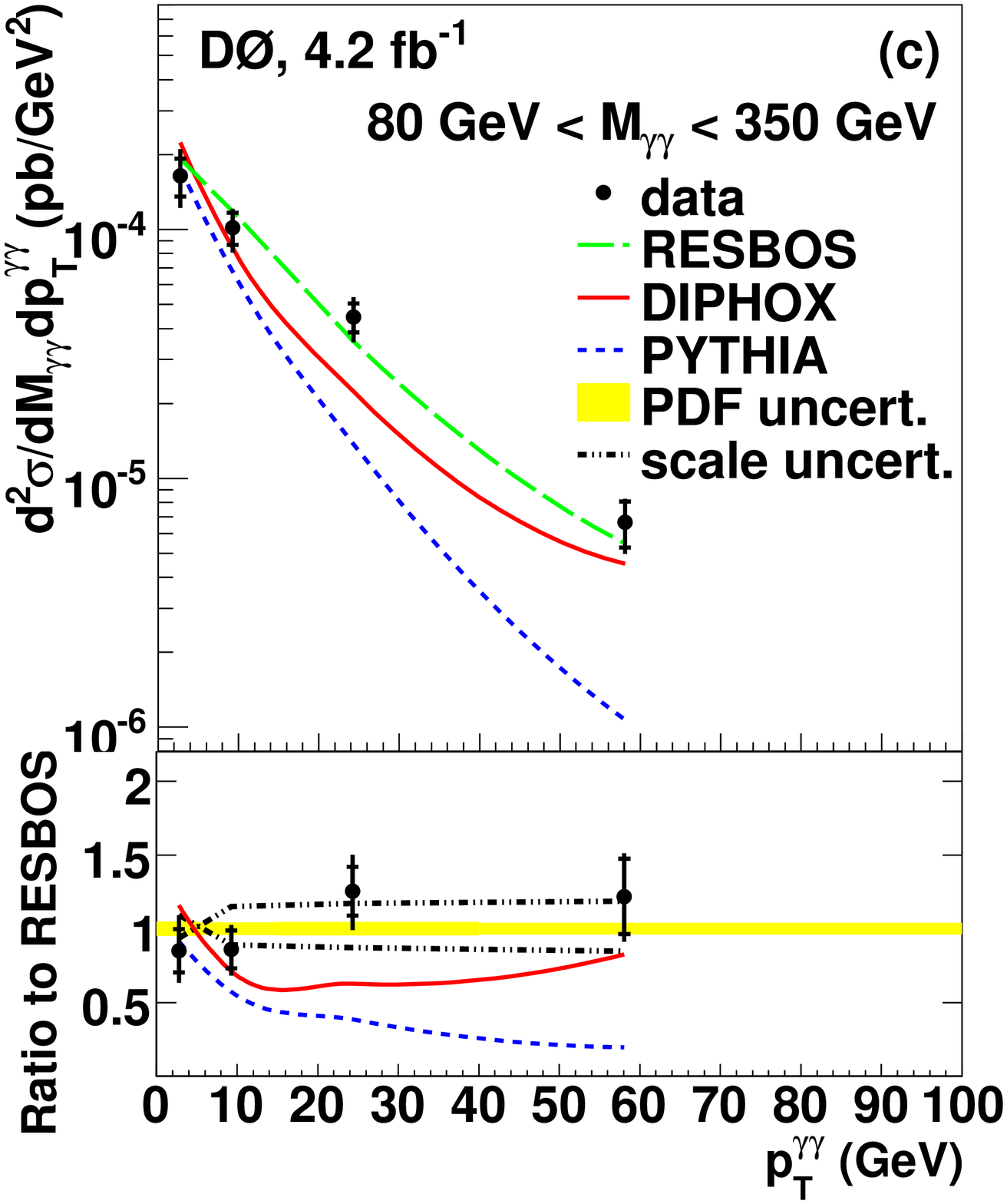}
\caption{The measured double differential diphoton production
cross sections as functions of \mgg (a), \ptgg for $30<$\mgg~$<50$~\GeVc (b), and \mgg~$>80$~\GeVc (c)
by the D0 experiment.}
\label{fig:diph_d0}
\end{figure}

\begin{figure}[htbp]
\hspace*{-2mm}  \includegraphics[scale=0.22]{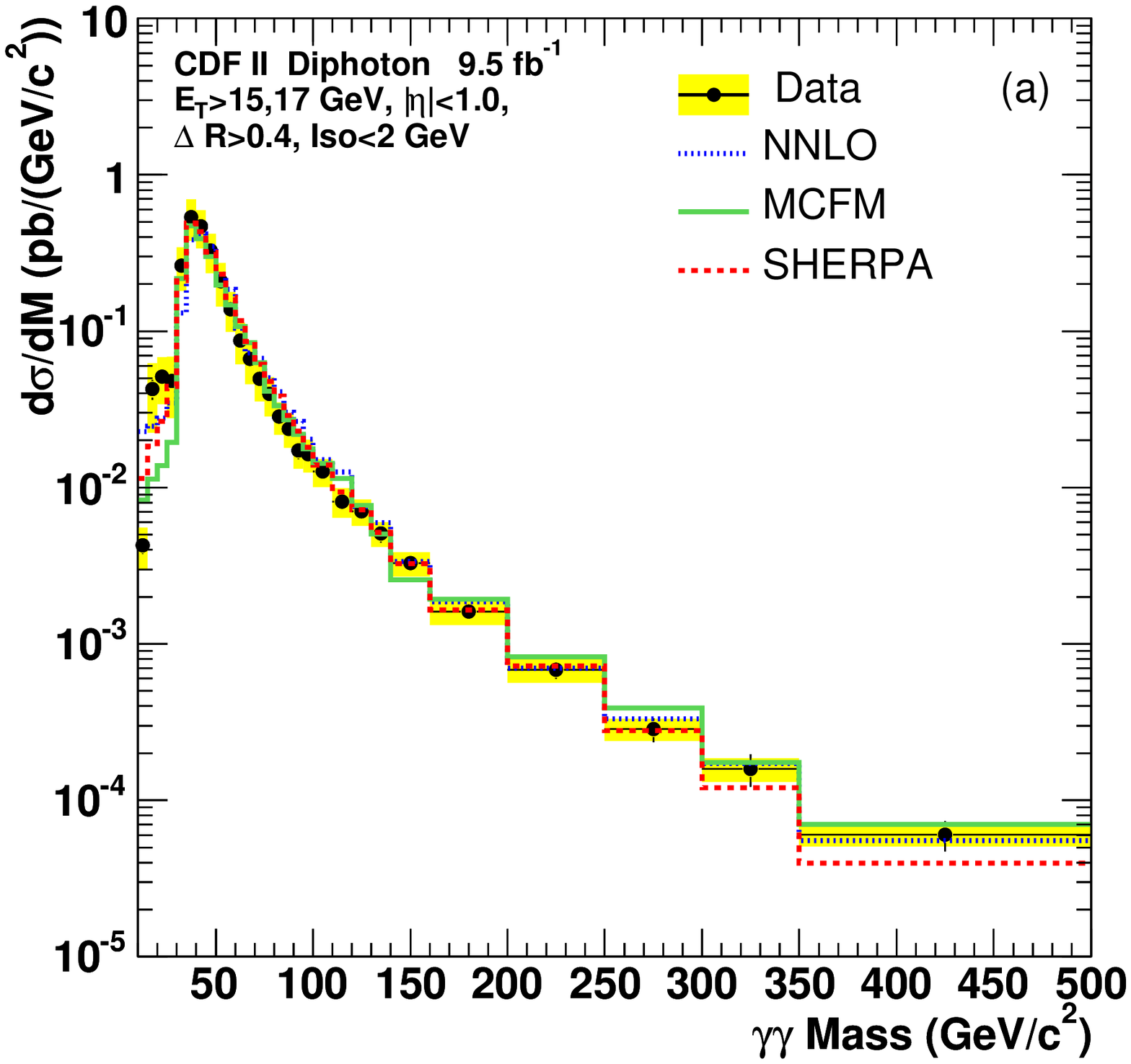}
\hspace*{-4mm}  \includegraphics[scale=0.22]{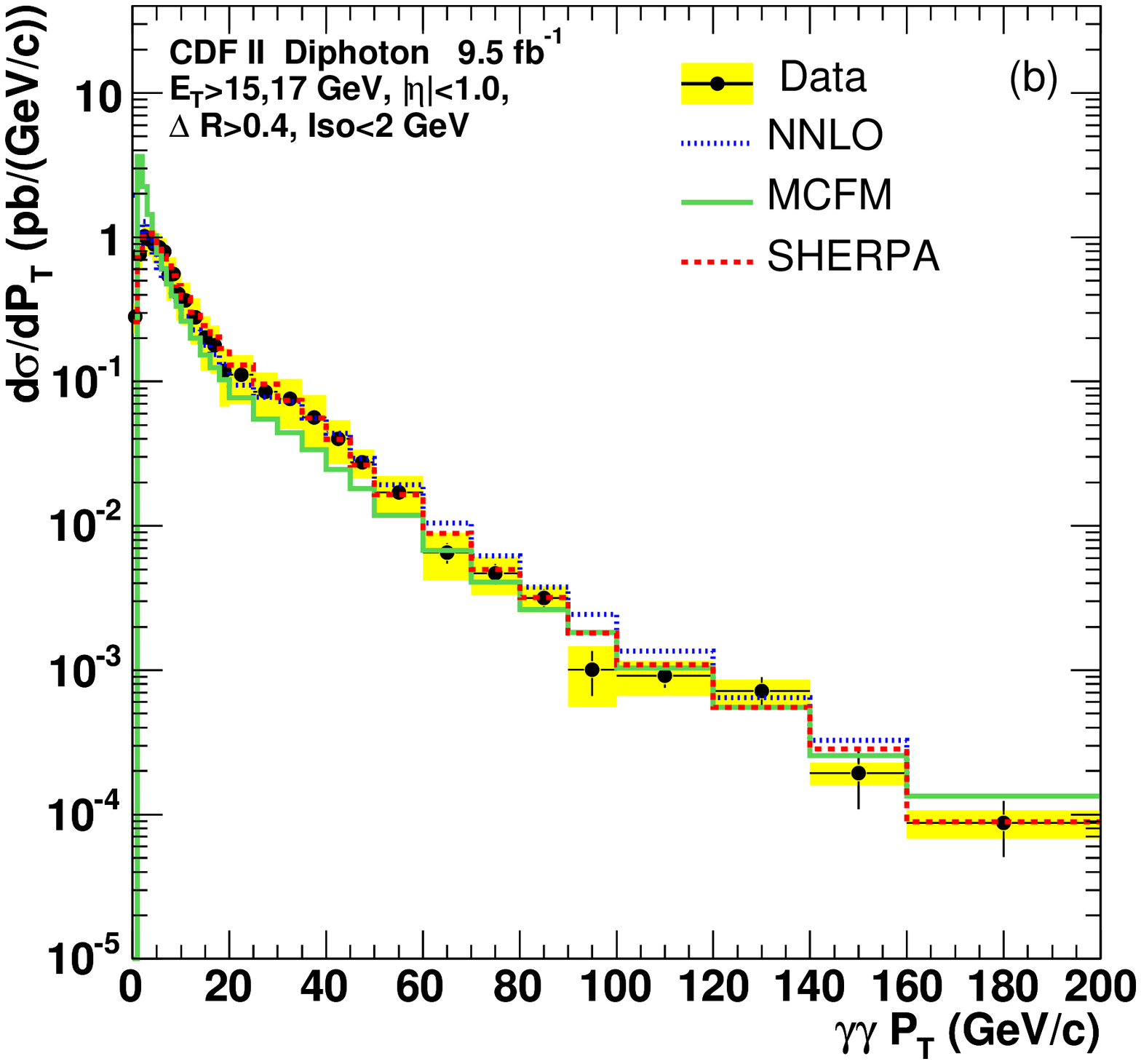}
\hspace*{-4mm}  \includegraphics[scale=0.22]{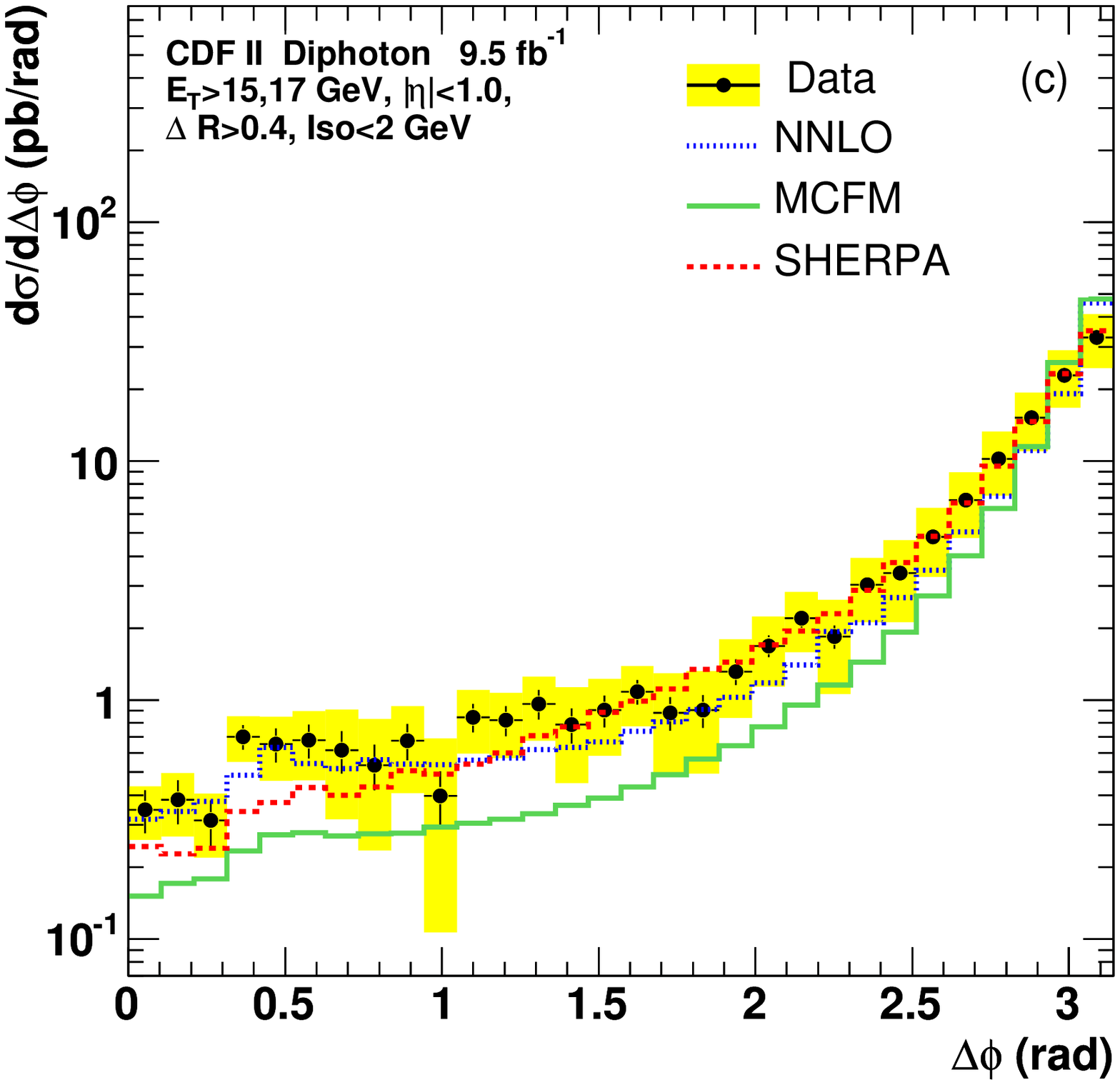}
\caption{The measured differential diphoton production cross sections as functions of
\mgg (a),  \ptgg (b) and \dphigg ~(c) by the CDF experiment.}
\label{fig:diph_cdf}
\end{figure}

In light of the Higgs boson search and other possible resonances decaying to a photon pair,
both collaborations performed a thorough study of diphoton production.
D0 measured the diphoton cross sections 
as a function of the diphoton
mass \mgg, the transverse momentum of the diphoton system \ptgg, the azimuthal angle
between the photons \dphigg, and the polar scattering angle of the photons.
The latter three cross sections are measured in the three \mgg ~bins, $30-50, 50-80$ and $80-350$ \GeVc.
The photons are considered with $|\eta|<0.9$, $p_{T,1}>21$, $p_{T,2}>20$ \GeVc and also requiring \ptgg$<$\mgg
~to reduce the contribution from the fragmentation photons~\cite{resbos}.
The measurements are compared to NLO QCD (provided by {\sc resbos} \cite{resbos} and {\sc diphox} \cite{diphox})
and {\sc pythia} \cite{PYT} predictions, see Fig.~\ref{fig:diph_d0}.
The results show that the largest discrepancies between data and NLO predictions
for each of the kinematic variables originate from the lowest \mgg~region
(\mgg~$<50$~\GeVc), 
where the contribution from $gg\to\gamma\gamma$ is expected to be largest~\cite{diphoton_d01}.
The discrepancies between data and the theory predictions are
reduced in the intermediate \mgg~region, and a quite satisfactory description
of all kinematic variables is achieved for the \mgg$>80$~\GeVc region, the
relevant region for the Higgs boson and new phenomena searches.
The CDF collaboration has also measured the diphoton production cross sections
functions of \mgg,  \ptgg and \dphigg \cite{diphoton_cdf2} . They are shown in Fig.~\ref{fig:diph_cdf}.
None of the models describe the data well in all kinematic regions, in particular
at low diphoton mass (\mgg$<60$ \GeVc), low \dphigg ($<1.7$ rad) and moderate
\ptgg ($20-50$ \GeVc).
Both experiments have also studied the diphoton production in separate kinematic regions,
with \dphigg$<\pi/2$ and \dphigg$>\pi/2$,
as well as for different \ptgg selections \cite{diphoton_d02,diphoton_cdf1}.

Measurements of the diphoton cross section done by CMS \cite{diphoton_cms} 
and ATLAS  \cite{diphoton_atlas} experiments provide a complementary information
to the extensive studies done by D0 and CDF experiments.

\section{W/Z+jets final states}
\label{sec:Vjets}

\subsection{W/Z+jet production}

The production of $W$ or $Z$ with accompanying hadronic jets 
provides quantitative tests of QCD through comparison 
of the rate of multijet production as a function of the strong coupling constant and 
comparison of various kinematic distributions with the theoretical predictions to probe the underlying matrix elements.
In addition, events with multiple jets in association with $W$ or $Z$ form a background 
for a variety of physics processes, including Higgs boson, top quark production and supersymmetry searches. 

In Run I  study of $W$ and $Z$ boson production in association with jets 
were initiated by measurement of ratio of $W$+1 jet to $W$+0 jet events~\cite{D0-Run1b-alphas-W+jets},
the measurement of the  cross section  and study of kinematic properties of direct single
$W$ boson production with jets~\cite{CDF-Run1b-W+jets}, study of jet properties in $Z$+jets~\cite{CDF-Run1a-Z+jets}\cite{CDF-Run1b-Z+jets}, and study of color coherence effects in $W$+jet events~\cite{D0-Run1b-W+jets}.

Large data sample in Run II allowed  
both CDF and  D0 experiments to conduct extensive studies of  $W$ and $Z$ boson production  in association with jets.

\begin{figure}[htbp]
\centerline{\includegraphics[width=0.50\columnwidth]{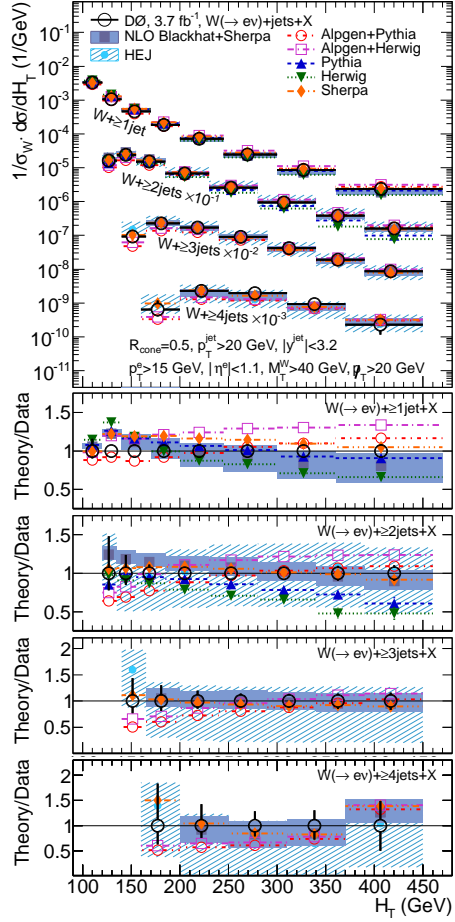}}
  \caption{Measurement of the distribution of the scalar sum of transverse energies if the $W$ boson and all jets and comparison to various theoretical predictions. Lower panels show theory/data comparisons for each of the $n$-jet multiplicity bin results separately.}
\label{D0-Wjets}
\end{figure}

The D0 collaboration published a comprehensive analysis of inclusive $W(\rightarrow e\nu)$+$n$-jet production for $n\geq$1,2,3,4 
using 3.7 fb$^{-1}$ of data~\cite{D0-W+jets}. 
Differential cross sections are presented as a function of many observables, such as  jet rapidities, lepton transverse momentum, 
leading dijet $p_T$ and invariant mass, etc. Many of the variables were studied for the first time in $W$+$n$-jet events,
e.g. a probability of the third jet emission as a function of dijet rapidity separation in inclusive $W+2$-jet events
(such a variable is important for understanding the Higgs boson via vector-boson fusion, and also sensitive to BFKL-like dynamics).
The data corrected for detector effects and the presence of backgrounds is compared 
to a variety of theoretical predictions. 
Fig.~\ref{D0-Wjets} shows the differential distributions of $W$+$n$-jet events as  functions of $H_T$, 
the scalar sum of the transverse energies of the $W$ boson and all $p_T>$20 \GeVc jets in the event. This variable is often used as the renormalization and factorizations scale for theoretical predictions for vector boson plus jets processes, so accurate predictions of $H_T$ are important. There is significant variation in the shapes of 
the $H_T$ spectrum from the various theoretical  predictions, {\sc pythia}, {\sc sherpa}, {\sc herwig}, {\sc alpgen} show discrepancies of 
the order of 25\% in one-jet bin and up to 50\% in 4-jet bin. This data is significantly more precise 
than theoretical predictions and can be used to improve the modeling.

\begin{figure}[htbp]
\centerline{\includegraphics[width=0.95\columnwidth]{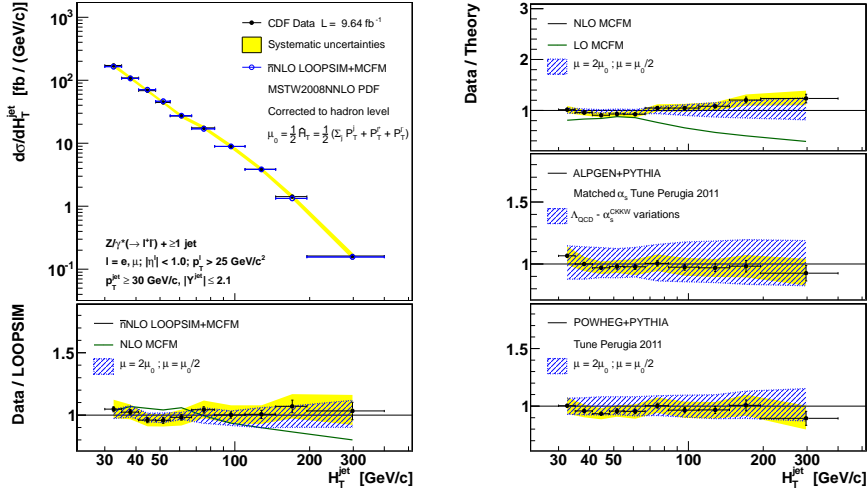}}
  \caption{%
Measurement of the $Z/\gamma^{\star}+\ge$1 jet differential cross section as a function of $H_T^{jet}$=$\Sigma p_T^{jet}$.
The lower and right panels show the data/theory ratio with respect to the theoretical predictions, with blue bands showing the scale uncertainty of each prediction, and yellow band corresponding to the experimental systematic uncertainty.
.}
\label{CDF-Zjets}
\end{figure}

The CDF experiment presented similarly extensive analysis of $Z/\gamma^{\star}(\rightarrow +e^+e^-, \mu^+\mu^-)$+jets 
production utilizing the 
full CDF dataset of 9.6 fb$^{-1}$~\cite{CDF-Z+jets}. The cross sections are unfolded to the particle level and combined. 
Results for various observables are compared with the most recent theoretical predictions. In addition, the effect 
of NLO electroweak virtual corrections~\cite{NLO-EW} on the $Z\gamma^{\star}$+jet production has been studied 
and included in the comparison with the measured cross section. Fig.~\ref{CDF-Zjets} shows measurement of the differential 
cross section as a function of $H_T^{jet}=\Sigma p_T^{jet}$ variable similar to one described previously. 
The approximate NNLO LOOPSIM+MCFM ($\bar{n}$NLO) prediction~\cite{nNLO}   used with NNLO PDF and 3-loop running $\alpha_S$ 
provides better modeling of the data distribution and shows a significantly reduced scale uncertainty.

\subsection{W/Z+heavy flavor jet production}

The measurement of the $W$ boson production in association with a $b$-quark jet provides
an important test of QCD, as it is sensitive to heavy-flavor quarks in the initial
state. $W$+$b$-jet production
is a large background to searches for the Higgs boson
in $WH$ production with a decay of $H\rightarrow  bb$, 
to measurements of top quark properties in single 
and pair production, and to searches for physics
beyond the Standard Model. 
The CDF collaboration published results for the cross section for jets from $b$ quarks produced with $W$
 boson using 1.9 fb$^{-1}$ of data~\cite{CDF-W+b}. The events were selected by identifying electron or muon decays of $W$ and containing one or two jets with $E_T>$20 \GeVc and $|\eta|<$2.0. The measured $b$-jet production cross section   of $\sigma\times B(W\rightarrow\ell\nu)$=2.74$\pm$0.27(stat)$\pm$0.42(syst) pb is 
higher than theoretical predictions   based on NLO calculations of 1.22$\pm$0.14(syst)pb. 

The D0 collaboration published results for the same process based on a data sample of 6.1 fb$^{-1}$~\cite{D0-W+b}. The combined results for electron and muon channels,
 defined using a common phase space for $p_T^{e,\mu}>$20 \GeVc, $|\eta^{\mu}|<$1.7 ($|\eta^e|<$1.1 or 1.5$<|\eta^e|<$2.5), $p_T^{\nu}>$25 \GeVc, 
$p_T^{\rm b-jet}>$20 \GeVc, $|\eta^{\rm b-jet}|<$1.1, are $\sigma(W\rightarrow\ell\nu)+b+X)$=1.05$\pm$0.12(stat+syst) pb 
for $|\eta^{\ell}|<$1.7. The result is in  agreement with NLO predictions using MCFM v6.1~\cite{MCFM} based on CTEQ6M PDF~\cite{cteq6} 1.34$^{+0.41}_{-0.34}$ pb as well as with predictions from the 
{\sc sherpa} and {\sc madgraph}~\cite{MADGRAPH} Monte Carlo event generators.

 The study of associated production of a $W$ boson and a charm  quark at hadron colliders
provides direct access to the strange-quark content
of the proton at an energy scale of the order of the W-boson mass. This
sensitivity is due to the dominance of strange quark-gluon fusion. At leading order the production of $W$ boson with single charm 
in $p\bar{p}$ collisions is described by the scattering of a gluon with a $d$, $s$ or $b$ quark; 
however at the Tevatron the large $d$ quark PDF in the proton is compensated by the small quark-mixing 
CKM matrix element $|V_{cd}|$, while contribution from $gb\rightarrow W c$ is heavily 
suppressed by $|V_{cb}|$ and $b$ quark PDF. 
The CDF collaboration presented the first observation of the production of  
$W$ boson  with a single charm quark jet in $p\bar{p}$ collisions at $\sqrt{s}$=1.96 TeV~\cite{CDF-W+c}. The charm quark is identified through 
the semileptonic decay of the charm hadron into an electron or muon, {\it soft lepton}, so charm jets are required 
to have an electron or muon candidate within the jet, so called {\it soft lepton tagging}, while the $W$ boson is identified through its leptonic 
decay by looking for an isolated electron or muon carrying large transverse energy $E_T$ and large missing $\not\!\!{E}_T$ in the event. 
Events are classified based on whether the charge of the lepton from $W$ boson and 
the charge of the soft lepton are of opposite signs or the same sign. 
The $Wc$ signal is observed with a significance of 5.7 standard deviations. 
The production cross section for $p_{T_c}>$20 \GeVc and $|\eta_c|<$1.5 is $\sigma_{Wc}\times B(W\rightarrow \ell\nu)$=13.5$^{+3.4}_{-3.1}$pb 
and is in agreement with theoretical predictions. 

Measurements of the production for a $Z$ boson in association with $b$ jets were published by the CDF and D0 collaborations.
 Both results provide good agreement with the theoretical predictions. The D0 experiment utilized 4.2 fb$^{-1}$~\cite{D0-Z+b-4.2} of 
data for $Z\rightarrow+\ell\ell$ events with a jet with 
$p_T>$20 \GeVc and pseudorapidity of $|\eta|\leq$2.5 to measure the ratio of $Z$+b-jet to $Z$+jet cross sections of 0.0193$\pm$0.0027. 
The CDF results~\cite{CDF-Z+b} correspond to the ratio of integrated $Z+b$ jet cross sections to the inclusive $Z$ production for jets with $E_T\geq$20 \GeVc and $|\eta| <$1.5 and is 3.32$\pm$0.53(stat)$\pm$0.42(syst)$\times$10$^{-3}$. The predictions from Monte Carlo generators 
and NLO QCD calculations are consistent with this result.

\begin{figure}[htbp]
\centerline{\includegraphics[width=0.50\columnwidth]{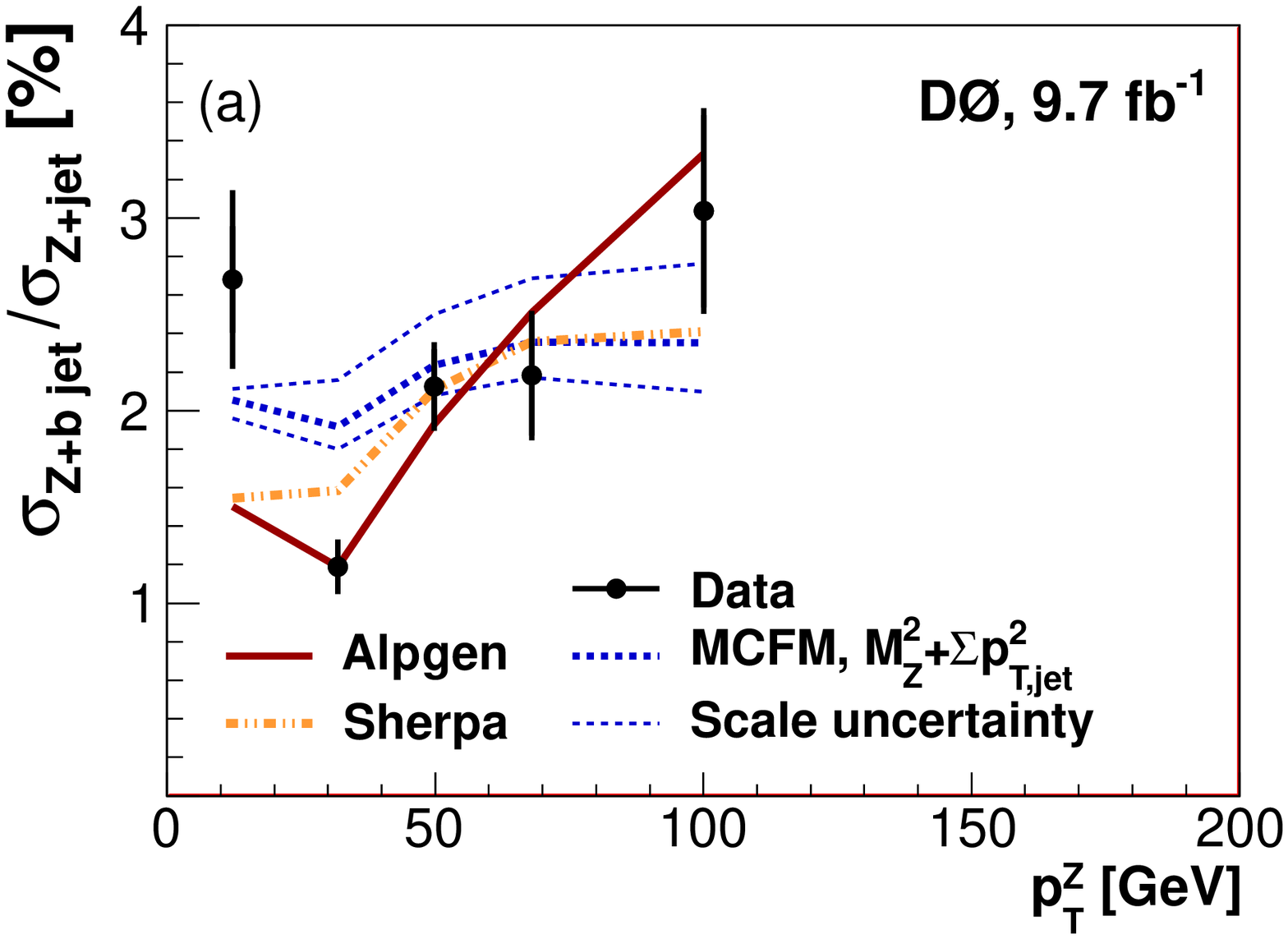}
     \includegraphics[width=0.50\columnwidth]{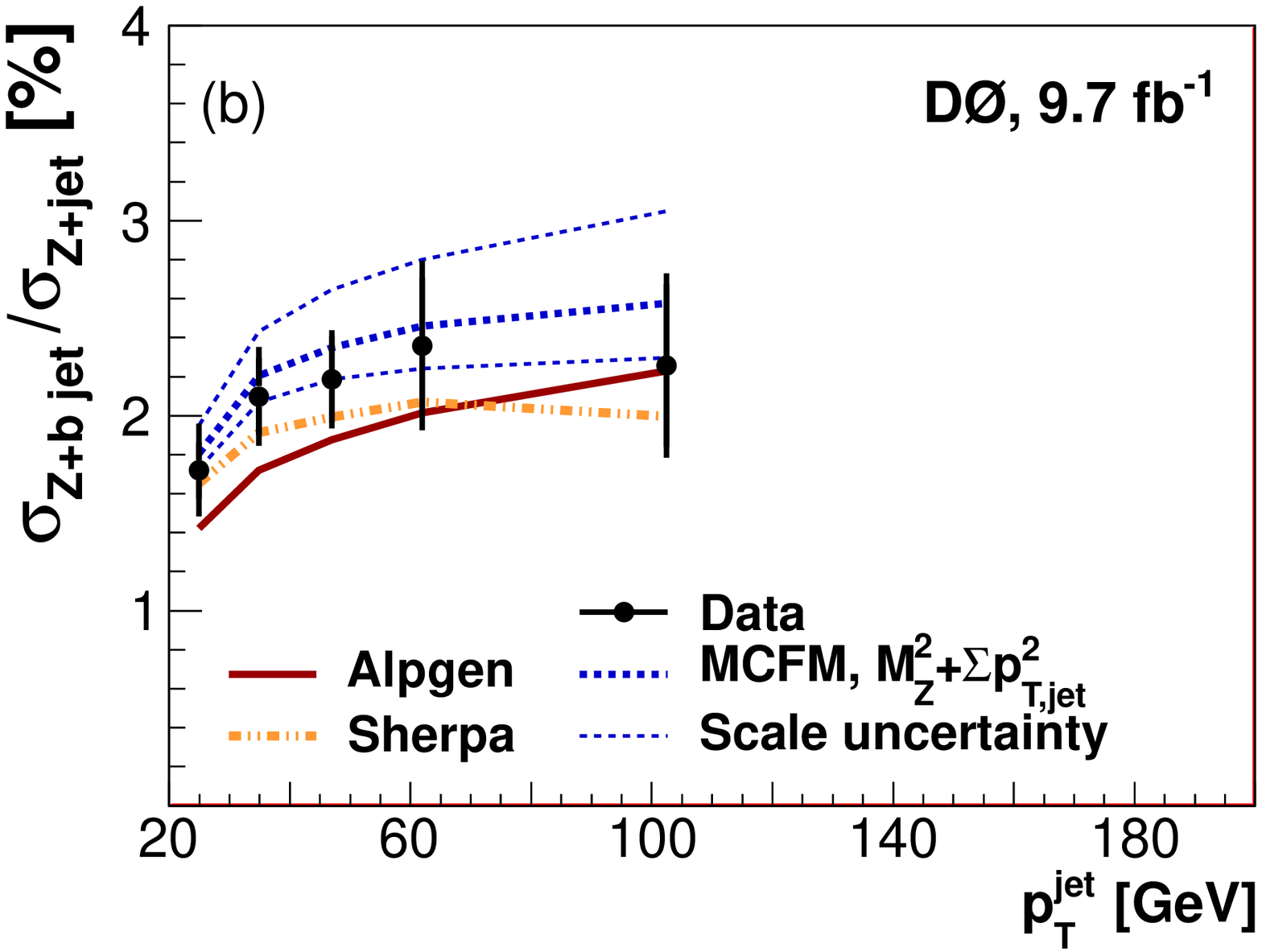}}
  \caption{%
Ratio of the differential cross section for $p_T^Z$ (a) and $p_T^{jet}$ (b). The error bars
 include statistical and systematic uncertainties added in quadrature. the scale uncertainty band represent the variation of the renormalization and factorization scales by a factor of 2.}
\label{D0-Z-bjets}
\end{figure}

The D0 collaboration extended the study of $Z+b$-jet production by utilizing the full D0 data set of 9.7 fb$^{-1}$~\cite{D0-Z+b-9.7}. The ratios of the 
differential cross sections as a function of $p_T^Z$ (a) and $p_T^{jet}$ (b) are presented in Fig.~\ref{D0-Z-bjets} compared with MCFM, {\sc alpgen}, {\sc sherpa} predictions. None of the predictions used provide a  consistent description of the  variables. 

\begin{figure}[htbp]
\centerline{\includegraphics[width=0.50\columnwidth]{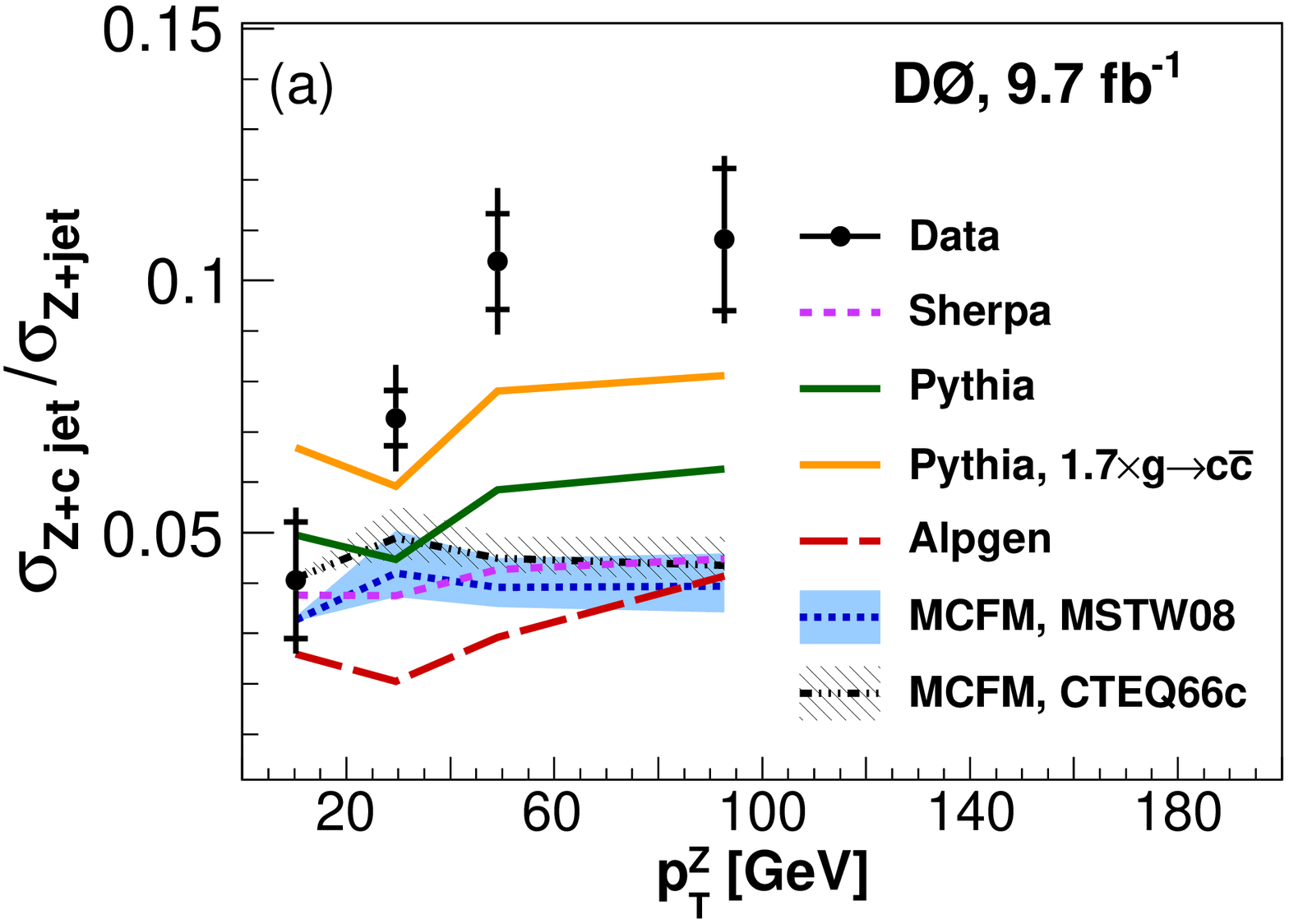}
     \includegraphics[width=0.50\columnwidth]{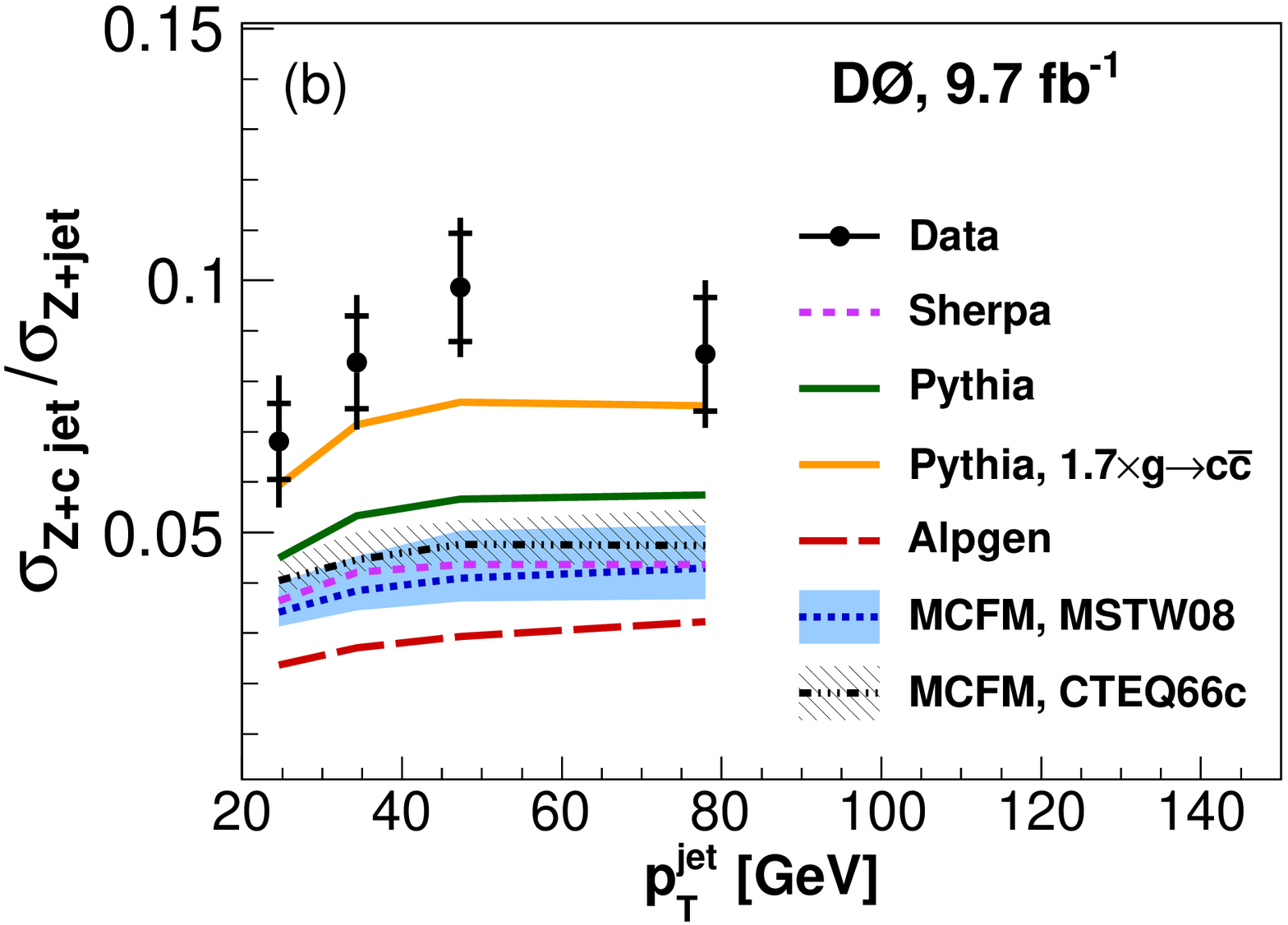}}
  \caption{%
Ratios of the differential cross sections of $Z+c$-jet to $Z+$jet as a function of (a) $p_T^Z$ and (b) $p_T^{jet}$. The uncertainties if the data include statistical (inner error bars) and full uncertainties (entire error bars).}
\label{D0-Z-cjets}
\end{figure}

The D0 collaboration reported the first measurement
of associated charm jet production with a $Z$ boson~\cite{D0-Z+c}.
Results are presented as  measurements of the ratio
of cross sections for the  $Z+c$ jet to $Z$ + jet production as
well as the $Z+c$ jet to $Z+b$ jet production in events with at
least one jet to benefit from the cancellation of some systematic uncertainties. 
This analysis is based on the complete Run II D0 data set of 9.7 fb$^{-1}$.   
The ratios of differential cross sections as a function of
$p_T^{jet}$
 and $p_T^Z$
are compared to various predictions in Fig.~\ref{D0-Z-cjets}.
On average, the NLO predictions significantly underestimate
the data. Perugia-0 tune with CTEQ6L1 PDF set are used for {\sc pythia} comparison.  Improvement in predictions can be achieved 
by enhancing the default rate of $g\rightarrow c\bar{c}$ in {\sc pythia} by
a factor of 1.7, motivated by the $\gamma+c$ jet production
measurements at the Tevatron discussed in Section \ref{sec:ghf}.

\section{Soft QCD}
\label{sec:softQCD}

The theory of strong interactions, QCD, is very successful in describing 
processes where a hard scale is present, either given by a large transverse momentum, $p_T$, or by a large mass of the exchanged particles,
 or a highly virtual particle. 
In these types of processes the strong coupling constant, $\alpha_S$, is small enough to allow for perturbative calculations to be valid.
 Soft interactions, which are usually understood as the interactions of hadrons at a relatively small scale, or low $p_T$, 
although making up the bulk of the hadronic cross section, lack precise theoretical predictions in the absence of the hard energy scale needed for the perturbative QCD calculations to converge. 
The fundamental importance to improve our understanding of soft strong interactions can be demonstrated by the lack of reliable 
predictions for such important quantities as the total hadronic cross section,  cross-sections for elastic scattering of hadrons, 
or the mass and  size of the proton.  
From the practical point of view, the  majority of collisions produced at the colliders belong to the category of ``soft processes'' 
and thus are very important to the modeling of the background activity. 

Hadron-hadron collisions can be divided into several categories.
Elastic scattering is a 2-to-2 color singlet exchange process in which two outgoing particles 
are the same as the two incoming particles. This process is described by the single variable $t$, squared four-momentum transfer.
 Single (double-) diffraction corresponds 
to the color singlet exchange between the initial hadrons, 
where for single (double) diffraction one (both) of the incoming particles 
 is (are) excited into a high mass color singlet state, with the mass $M_X$ ($M_X$and $M_Y$), which then decays.  
This process can be be described in terms of the variables, $t$, 
and either mass $M_X$ ($M_X$ and $M_Y$), 
or the fractional energy loss of the intact proton (and antiproton) $\xi=M_X^2/s$ ($\xi_1=M_X^2/s$ and $\xi_2=M_Y^2/s$).
The  non-diffractive production includes all processes not described by the elastic and diffractive channels. In this case, particle production is taking place through all available rapidity space. 

\subsection{Non-diffractive production}

\subsubsection {Minimum Bias Studies}

The minimum bias final state observables represent a complicated mix of different physics effects ranging 
from purely soft to very ``hard'' ones. 
The term {\it minimum bias}  is a generic term which refers to events that are selected with very minimal trigger, 
to ensure that they are as inclusive as possible, and so as a result the definition of minimum bias differs from experiment to experiment. 
The majority of minimum bias events are ``soft'' and thus processes under these conditions are notoriously difficult to model. 
While the understanding of softer physics is interesting in its own right, a detailed understanding of minimum bias interactions is extremely important in very high luminosity environments where a large number of such interactions happen in the same bunch crossing. 
At the CDF experiment studies of minimum bias events were initiated at Run I when
inclusive charged particle distributions at $\sqrt{s}$=1800 \GeVc were measured~\cite{CDF-mb-RunI} , as well as studies of  different variables, such as the multiplicity, transverse momentum $p_T$, average $p_T$ for “soft” and “hard” interactions at $\sqrt{s}$ of 630 and 1800 \GeVc were performed. 
At Run II the CDF collaboration continued minimum bias studies by providing first measurement of the event 
transverse energy sum differential cross section representing an attempt at describing the full final state including neutral particles, by studying particle transverse momentum as a function of the event particle multiplicity, and significantly 
extending the range of the inclusive charged particle transverse momentum differential cross section while improving precision. 
The analysis was based on 506 pb$^{-1}$ data sample collected with CDF minimum bias trigger 
implemented by means of two sets of Cherenkov counters placed
on both sides of the detector and requiring a coincidence of both signals. The resulting 
MB sample contains most of the inelastic cross section with a small contamination of single- and
double-diffractive. 
\begin{figure}[htbp]
\centerline{\includegraphics[width=0.8\columnwidth]{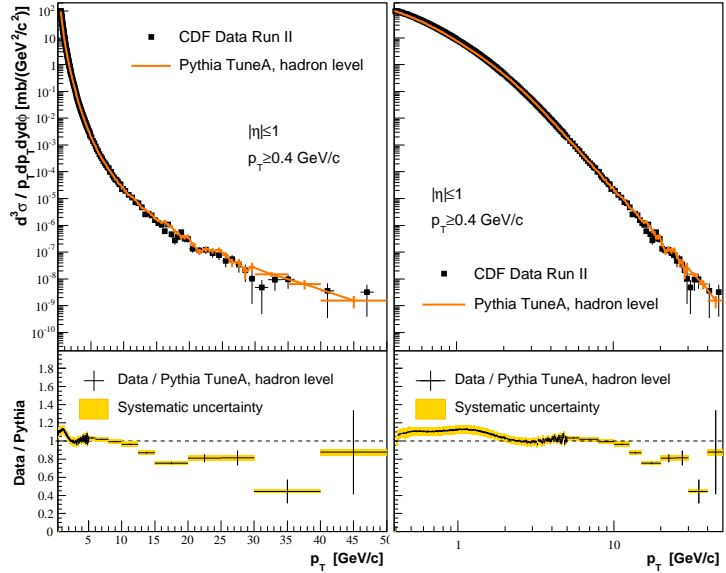}}
\caption{%
Comparison of the track $p_T$ differential cross section with {\sc pythia} Tune A prediction at hadron level. 
The data error bars describe the statistical uncertainty  on the data and the statistical uncertainty on the total correction.}
\label{CDF-minbias}
\end{figure}
Fig.~\ref{CDF-minbias} shows a comparison of track $p_T$ differential cross section with {\sc pythia} Tune A prediction at hadron level. Data and Monte Carlo prediction show good agreement.

\subsubsection{Underlying Event Studies}

\begin{figure}[htbp]
\includegraphics[width=0.4\columnwidth]{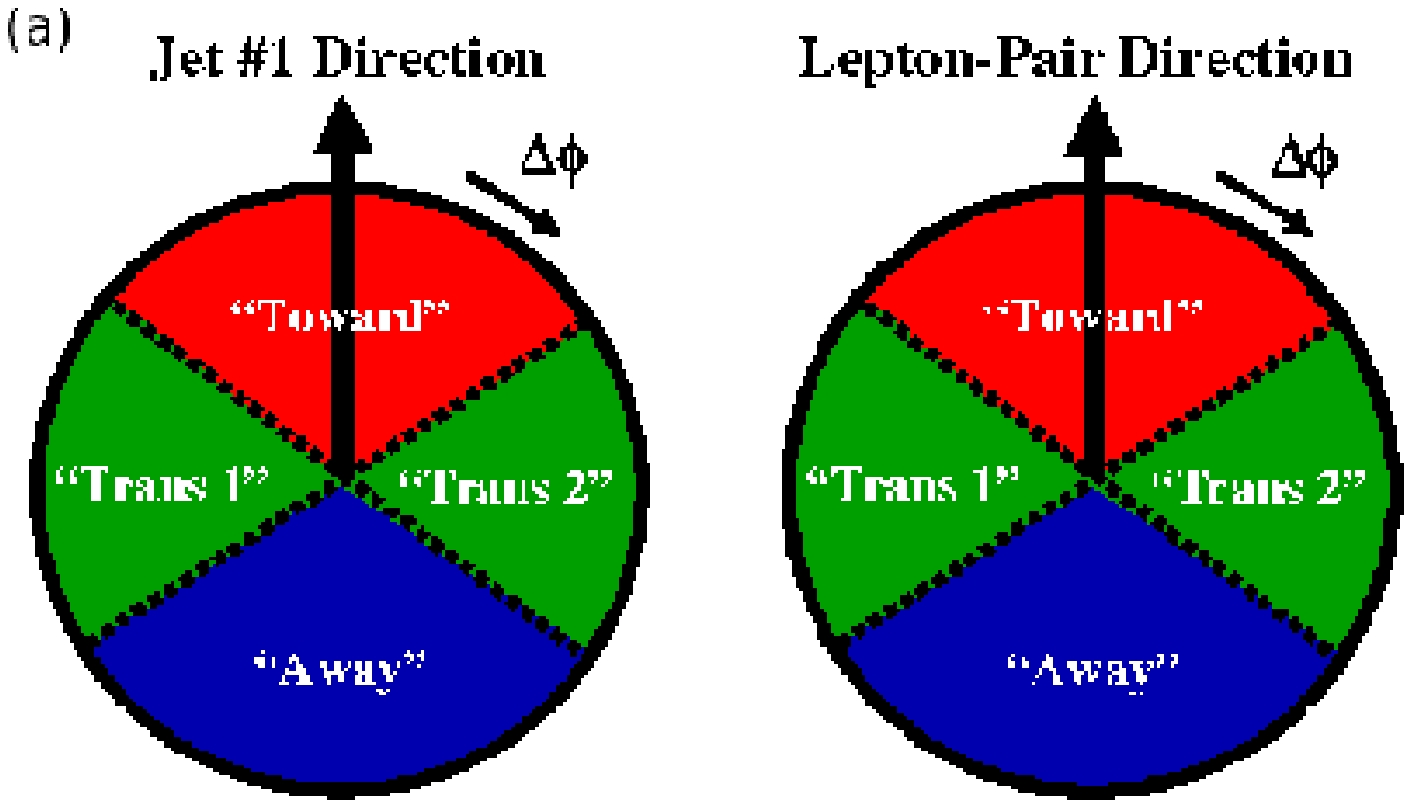}
\includegraphics[width=0.55\columnwidth]{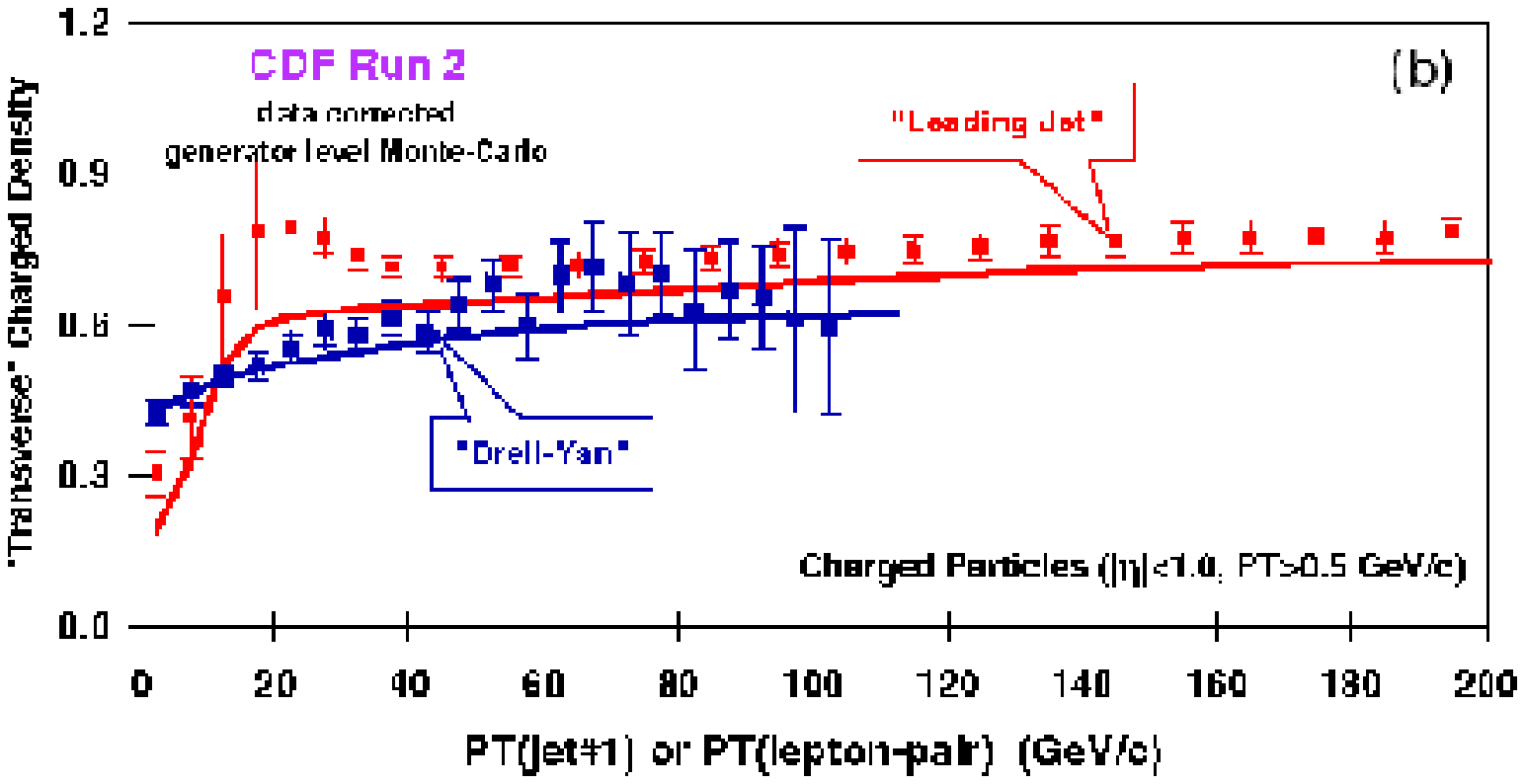}
  \caption{%
(a) Schematic division of different regions in azimuthal angle $\Delta\phi$ relative to the direction of the 
leading jet in the event or the direction of the lepton pair in Drell-Yan production; (b) the density of charged particles in the transverse region for leading jet and Drell-Yan events compared with {\sc pythia} Tune A and {\sc pythia} Tune AW.}
\label{CDF-UE}
\end{figure}

The existence of Monte Carlo models that accurately simulate  QCD hard-scattering events is essential for all {\em new} physics searches
 at hadron-hadron colliders. To achieve a given accuracy one should be able  to have not only a good model 
of the hard scattering part of the process, but also of the {\it underlying event} corresponding to all final state particles produced beyond those associated with the hardest scattering, an unavoidable background to most collider observables.
 The sources of the underlying event are  beam-beam remnants (BBR) and activity from multiple parton interactions (MPI).
The CDF pioneered a method providing a comprehensive set of measurements subjecting 
to the rigorous scrutiny particle production associated with the underlying event in a model-independent way.
Run II studies of the  underlying event were extended to the comparison of Drell-Yan production and leading jet topologies~\cite{CDF-UE-runII}. 
For Drell-Yan production, the final state includes a lepton-antilepton pair, and there is no colored final state radiation, thus providing a clean way to study the underlying event (UE).
The methodology of the presented study is similar to previous CDF UE studies~\cite{CDF-UE-runI}, by considering the {\em toward}, {\em away}, and {\em transverse} regions defined by the azimuthal angle $\Delta\phi$  relative to  the direction of the leading jet in the event, or  the direction 
of the lepton-pair in Drell-Yan production ($\Delta\phi=\phi-\phi_{jet_1/pair}$),  see Fig.~\ref{CDF-UE}(a). 
We study charged particles with $p_T>$0.5 \GeVc and $|\eta|<$1 in the above-mentioned regions. 
For high-$p_T$ jet production the leading jet in the event, reconstructed with the Midpoint algorithm, and with  $|\eta_{jet}|<$2 was required. 
For Drell-Yan production the requirement of the invariant mass of the lepton-pair to be in the mass region of the Z-boson, 70$<M_{pair}<$110 \GeVcc, with $|\eta_{pair}|<$6 was placed. 
For leading jet events, the toward and away regions are characterized by large contributions from the outgoing high energy jets, whereas the transverse region is perpendicular to the plane of the hard scattering and is sensitive to the underlying event. For Drell-Yan events, while the 
away region receives large contributions from the balancing jet, both the toward and  transverse regions are sensitive to the underlying event. Many observables were studied for all three regions of interest. Here we will describe just one, the charged particle density, $dN/d\eta d\phi$
 in the transverse region for both the  leading jet and Drell-Yan topologies, see Fig.~\ref{CDF-UE}(b).   
The underlying event observable is found to be reasonably flat with the increasing lepton pair transverse momentum and  quite similar for both topologies. The small ``bump'' for low-$p_T$ values for leading jet distribution reflects the fact that there are many low $p_T$ jets and  for this $p_T<$30 \GeVc values 
the leading jet is not always the jet resulting from the hard two-to-two scattering. {\sc pythia} Tune A for leading jet events, and {\sc pythia} Tune AW for Drell-Yan events provide reasonable agreement with the experimental data. 

\subsubsection{Particle Production}

\begin{figure}[htbp]
\centerline{\includegraphics[width=0.45\columnwidth]{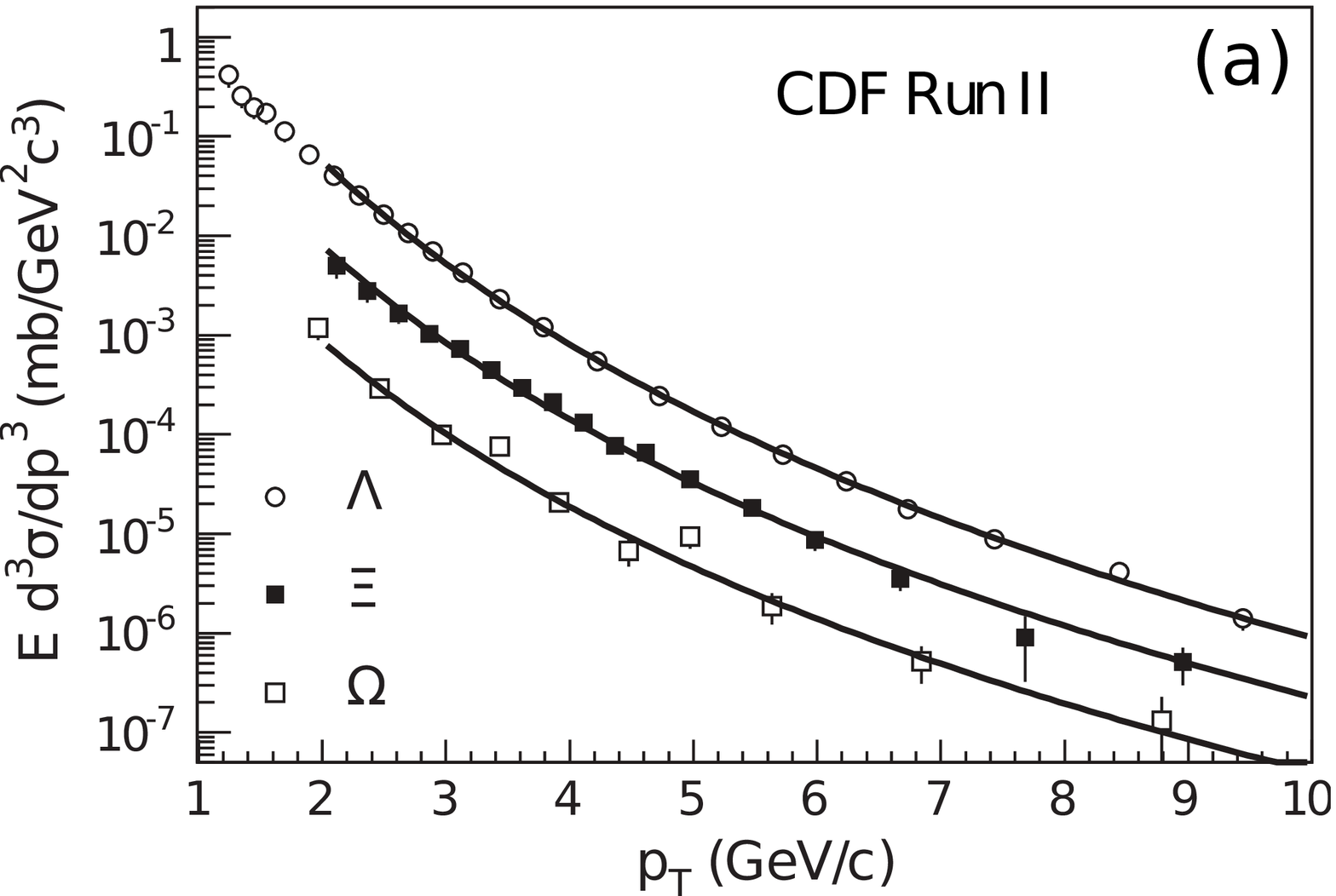}
\includegraphics[width=0.45\columnwidth]{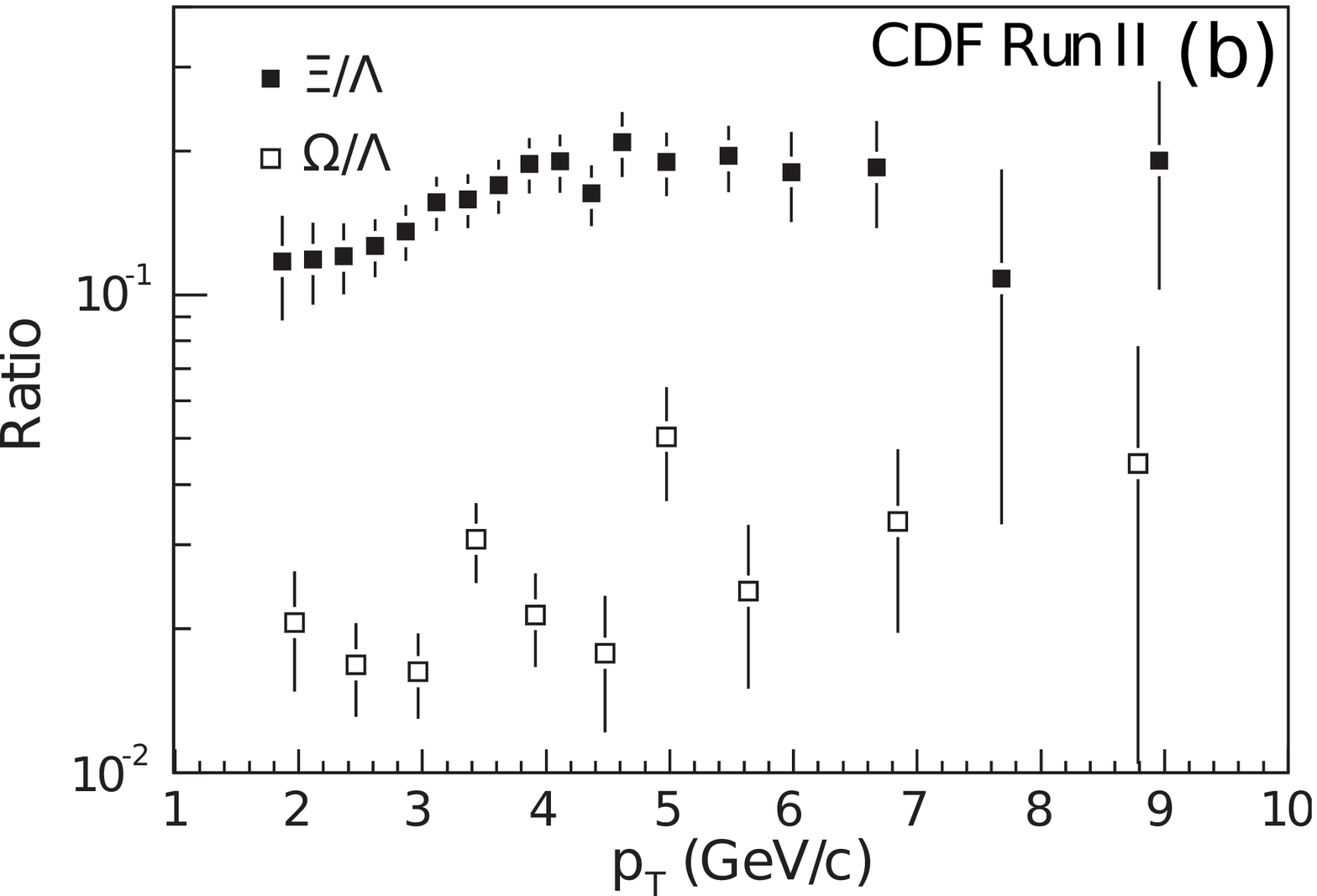}}
  \caption{
The inclusive invariant $p_T$ differential cross section distributions for $\Lambda^0$, $\Xi^-$, and $\Omega^-$ for 
$|\eta|<$1 uncertainties for data points include all statistical and systematic uncertainties except one associated with normalization uncertainty due to the  minimum bias trigger cross section. The solid curves are from fits to a power law function; (b) the ratios of $\Xi^-/\Lambda^0$ and $\Omega^-/\Lambda^0$ as a function of $p_T$.}
\label{CDF-hyperons}
\end{figure}

The measurements of the production of particles with different quark flavors and number of quarks 
is an essential step in understanding hadron production. 
Since the strange quark is heavier than the up and down quarks, strange hadron production is usually suppressed, with an 
amount of suppression used for refining the phenomenological models and parameters of the  Monte Carlo models. At the same time, the enhanced production of the strange particle has been frequently suggested as a manifestation of the formation of quark-gluon plasma. 
The CDF collaboration presented measurements of $\Lambda^0$, $\bar{\Lambda^0}$, $\Xi^\pm$, and $\Omega^\pm$ hyperons under minimum bias conditions\cite{CDF-hyperons} and $K_S^0$, $K^{\star\pm}(892)$ and $\phi^0$ in minimum bias events and $K_S^0$ and $\Lambda^0$ in jets~\cite{CDF-Kshorts}. 
All particles were reconstructed in the central region with $|\eta|<$1.0, and for minimum bias produced particles with $p_T$ up to 10 \GeVc and particles in jets with $p_T$ up to 20 \GeVc. 
From the ratio of cross sections, see Fig.~\ref{CDF-hyperons}(b), it is clear that cross sections depend on the number of strange quarks, however very similar $p_T$ slopes for distributions on Fig.~\ref{CDF-hyperons}(a) indicate a universality in particle production as $p_T$ increases.
%
Results of ~\cite{CDF-Kshorts} also demonstrate that the ratio of $\Lambda^0$ to $K_S^0$ as a function of $p_T$ in minimum bias events becomes similar to the fairly constant ratio in jets at $p_T\sim$5 \GeVc. This confirms the earlier observation from CDF underlying event studies 
that particles with $p_T\geq$5 \GeVc in minimum bias events are from ``soft'' jets and that the $p_T$ slope of particles in jets is insensitive to light quark flavor and to the number of valence quarks. 
These results are providing an important contribution for tuning of Monte Carlo models.

\subsubsection{Fragmentation Studies}

\begin{figure}[htbp]
\centerline{\includegraphics[width=0.5\columnwidth]{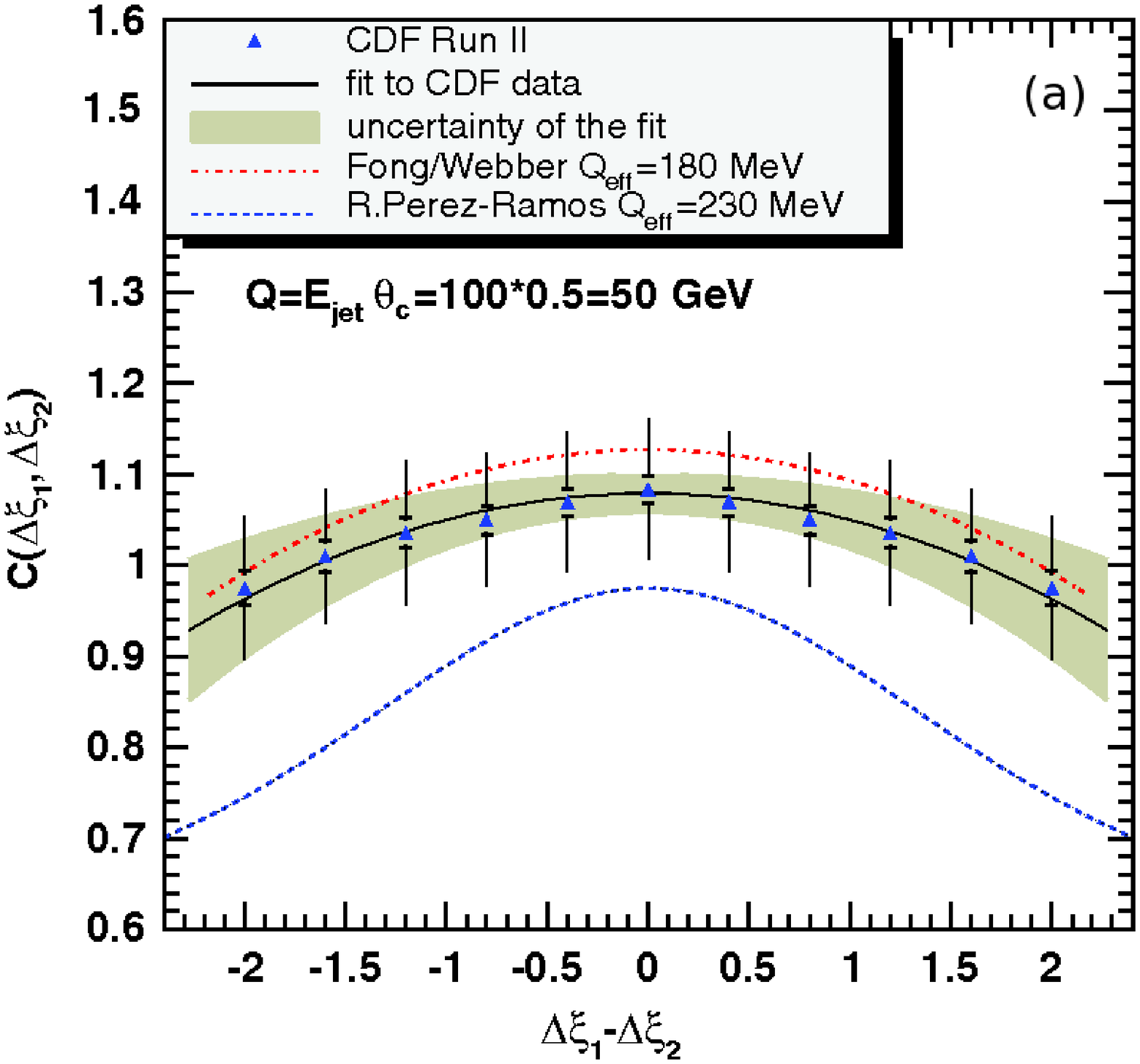}
\includegraphics[width=0.5\columnwidth]{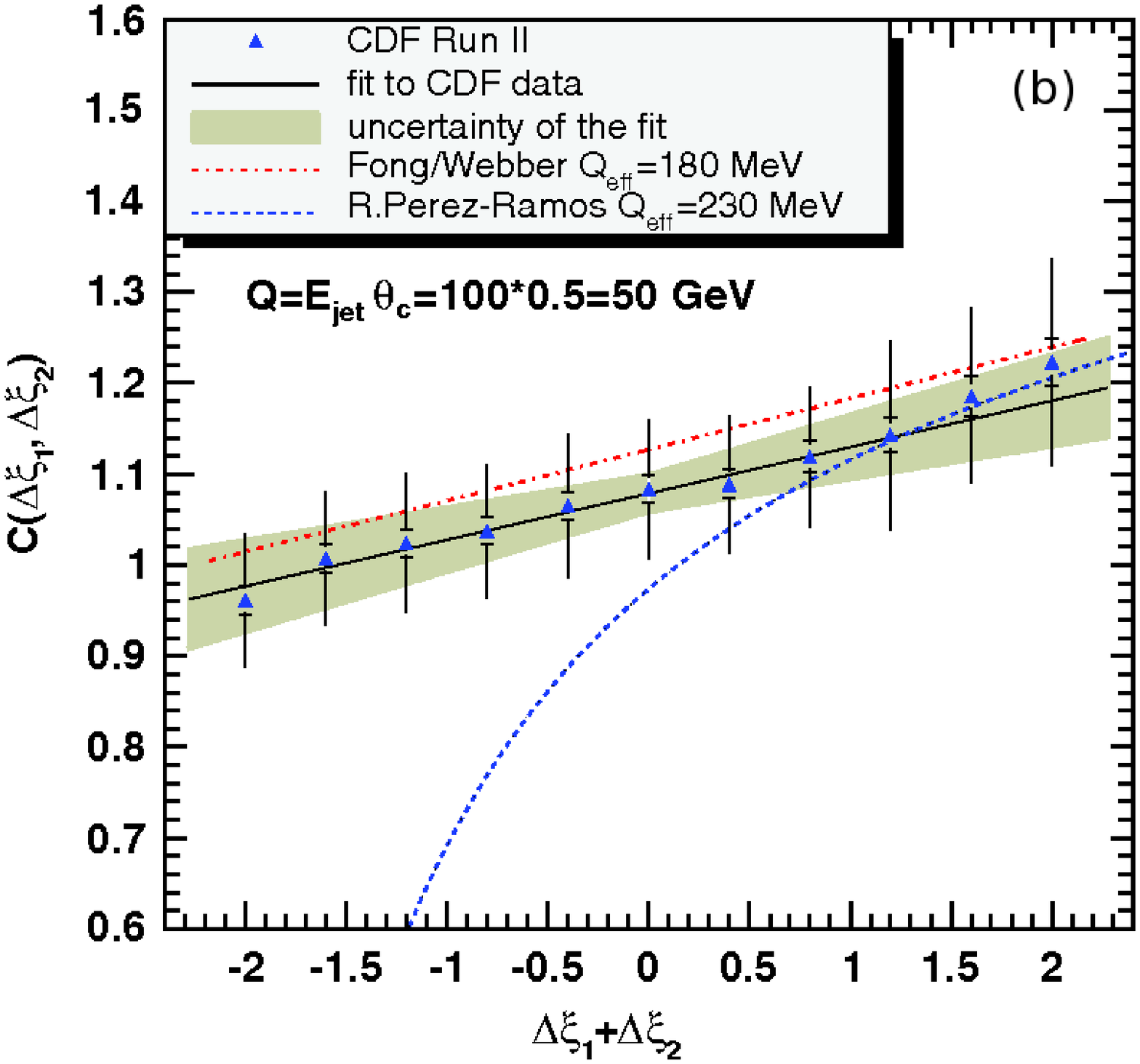}}
\caption{ Central diagonal profiles $\Delta\xi_1$=-$\Delta\xi_2$ (a) and $\Delta\xi_1$=$\Delta\xi_2$ (b) of two-particle momentum correlations in jets in the restricted cone of size $\theta_c$=0.5 radians for dijet mass bin with $Q$=50 \GeVc.The correlation in data 
is compared to that of theory.}
\label{CDF-2momentumcorrelation}
\end{figure}

The transition from partons to hadrons, or {\it hadronization},  is not understood from perturbative QCD 
and has to be described by a phenomenological model. Detailed studies of jet fragmentation allow us 
to understand the relative roles 
of the perturbative and non-perturbative stages of jet formation and to probe boundaries of parton shower and hadronization.
 The characteristics of soft particle production, such as particle multiplicities, 
inclusive distributions and correlation functions can be  described by analytical predictions
of the next-to-leading log approximation (NLLA)\cite{NLLA} describing parton shower formation supplemented with the 
local parton-hadron duality approach~\cite{LPHD} prescribing that hadronization process takes place locally and thus applies perturbative predictions at the partonic level directly to hadronic distributions. 
Past studies of inclusive particle distributions at $e^+e^-$ experiments~\cite{jet-fragmentation-e+e-} and CDF\cite{CDF-runi-jetfragmentation} have given strong support to this 
theoretical framework. 
In Run II the CDF collaboration extended studies to the measurements of the two-particle momentum correlations in jets as a function of 
jet energy~\cite{CDF-two-momentum}. The correlation function is introduced as $\xi=\ln{E_{jet}/p_{hadron}}$ and is defined as a ration of two- and on-particle inclusive momentum distributions: $C(\Delta\xi_1,\Delta\xi_2)={D(\xi_1,\xi_2)}/({D(\xi_1)D(\xi_2)})$, where both inclusive distributions
 $D(\xi)=\ln({d\eta}/{d\xi})$ and $D(\xi_1,\xi_2)$ are normalized to unity. 
The results are obtained for charged particle within a restricted cone with an opening angle
 of $\theta_c$=0.5 radians around the jet axis for events with dijet masses between 66 and 563 \GeVcc with underlying event contributions subtracted using the complimentary cones technique. 
The characteristic features of the theoretical predictions are follows: the correlation should be stronger for partons with equal momenta, 
or $\Delta\xi_1$=$\Delta\xi_2$, and the strength of this effect should increase for lower momentum partons. 
Figure~\ref{CDF-2momentumcorrelation} shows overall good agreement between the data and 
theoretical predictions based on Fong-Webber calculation ~\cite{Fong-Webber} 
that provided the predictions at the level of NLLA precisions, the modified leading log approximation (MLLA)~\cite{MLLA} referred in 
Fig.~\ref{CDF-2momentumcorrelation} as R.Perez-Ramos approach, is an approach similar to NLLA but including higher-order terms ($\alpha_S^n\ln^{2n-2}{E_{jet}}$ and higher). 
The data follows the theoretical trends and shows an enhanced probability of finding two particles with the same value of momenta, 
indicated by the parabolic shape of the $\Delta\xi_1$=-$\Delta\xi_2$ central diagonal profile with maximum at 
$\Delta\xi_1$=$\Delta\xi_2$=0, with the effect becoming larger for particles with 
lower momenta, represented by the positive slope of the  
$\Delta\xi_1$=$\Delta\xi_2$ central diagonal profile. 

The measurement of the transverse momenta of particles 
in jets with respect to the jet axis, $k_T$~\cite{CDF-kT} allows probing for  softer particle spectra 
than from the previously discussed  observables. The CDF measurement is based on 1 fb$^{-1}$ of data in events with dijet masses between 66 and 
737 \GeVcc. The shape of $k_T$ distribution is compared to the theoretical predictions from MLLA and 
NMLLA, next-to-leading log approximation~\cite{NMLLA}, 
as well as for {\sc pythia} Tune A Monte Carlo
generator. 
The NMLLA results for $Q_{eff}$=230 MeV provide an excellent description of the data 
over the entire range of particle $k_T$ and dijet masses used in this measurement.
Predictions of Monte Carlo generators for final stable particles are in agreement 
with the results obtained from data. The good qualitative agreement between NMLLA predictions  
and charged hadrons from {\sc pythia} Tune A  is due to the tunings of the hadronization parameters in 
{\sc pythia} Tune A, discussed previously, while distribution from {\sc pythia} at the parton level shows significant deviations.

\subsubsection{Event Shapes}

\begin{figure}[htbp]
\begin{center}
\centerline{\includegraphics[width=0.41\columnwidth]{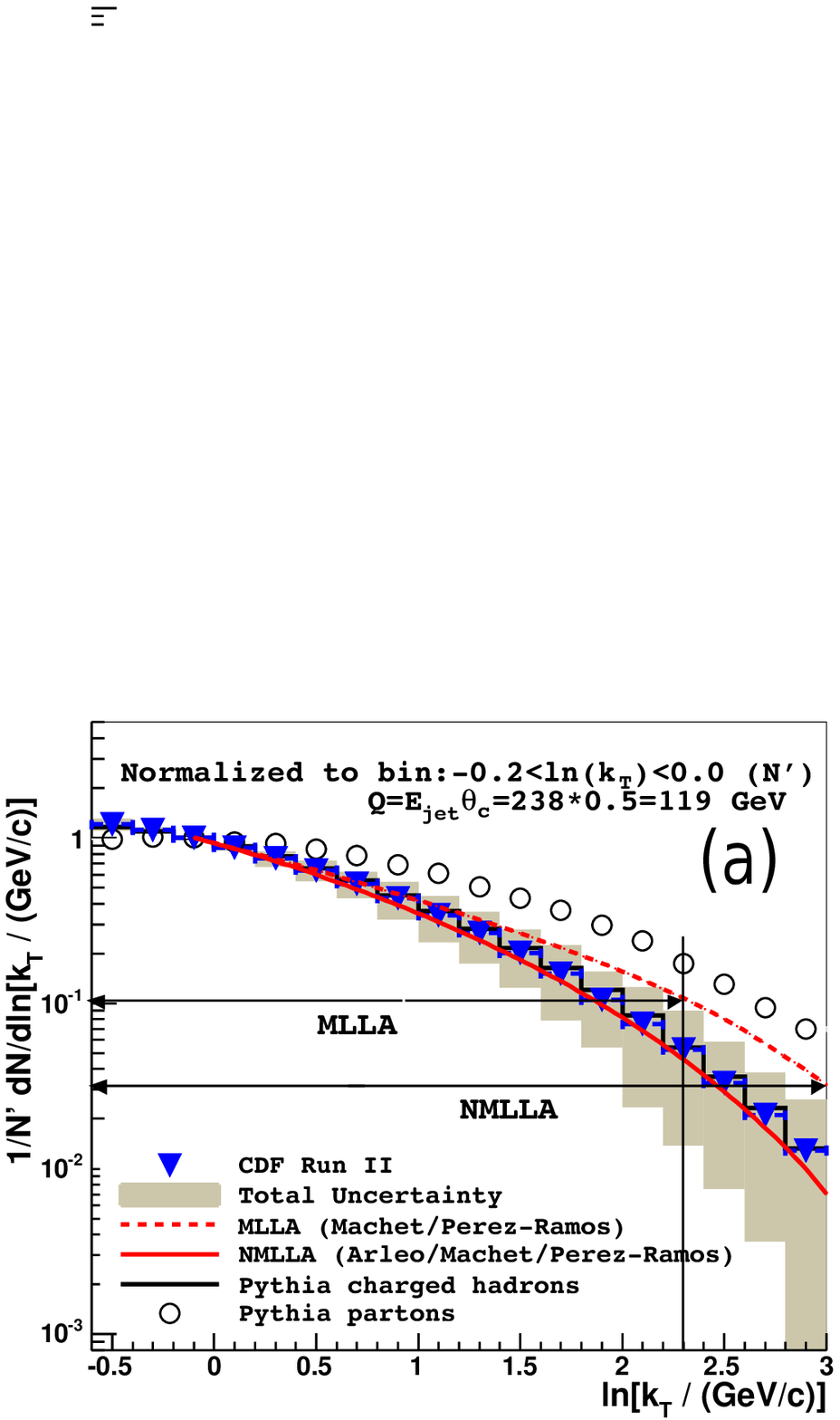}
\includegraphics[width=0.45\columnwidth]{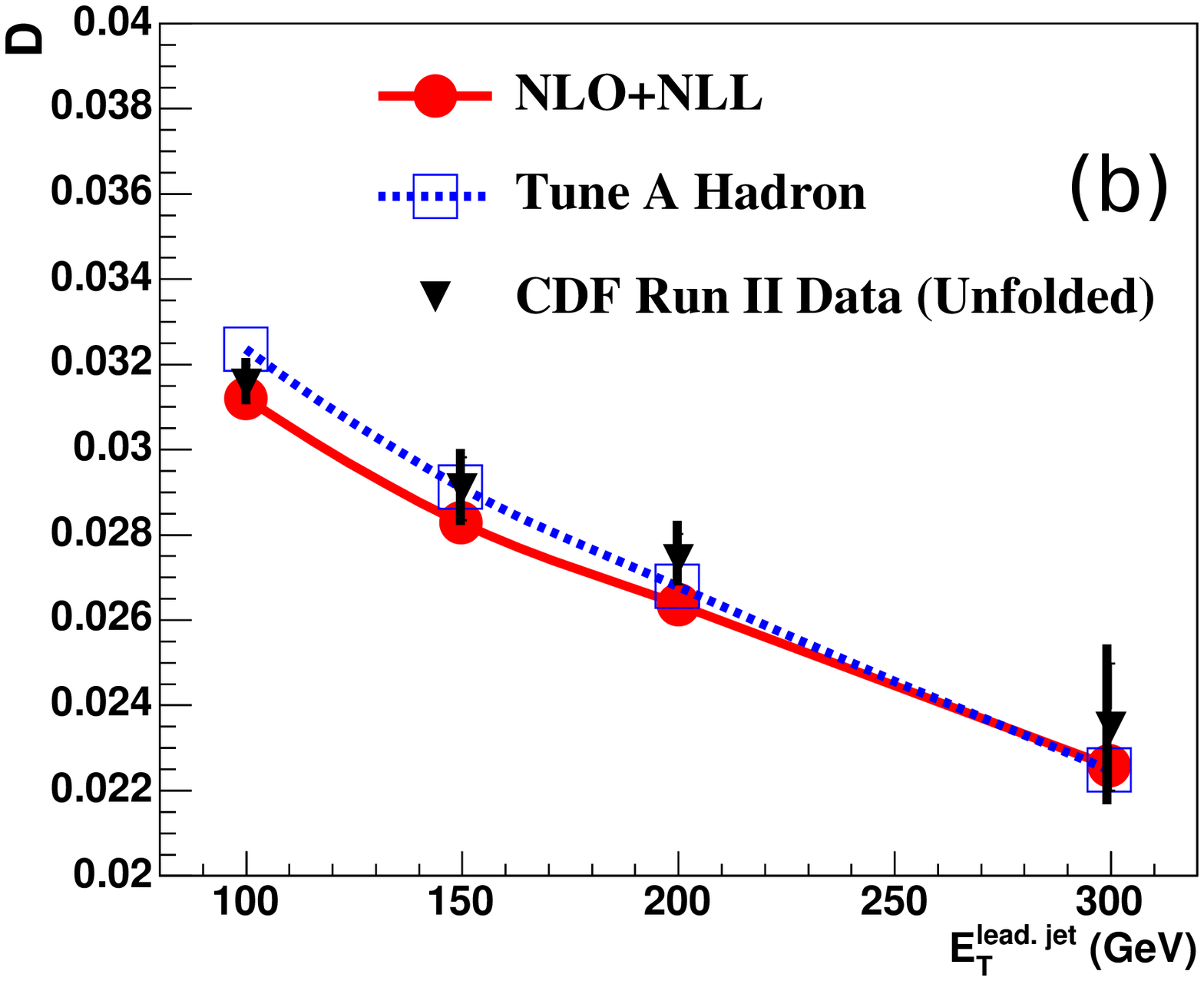}}
\caption{ (a) measured $k_T$ distribution of particles in the restricted cone of size $\theta_c$=0.5 
around the jet axis in dijet mass bin of $Q$=119 \GeVc. The data are compared to the analytical predictions MLLA and NMLLA and to the predictions of the {\sc pythia} Tune A for partons and charged hadrons. Ranges of validity for MLLA and NMLLA predictions are shown by arrows; (b) the CDF corrected results for the dependence of the thrust differential on the transverse energy
 of the leading jet. The experimental results are compared with a parton-level NLO+NLL calculation and with {\sc pythia} Tune A at the hadron level. The error bars correspond to statistical and systematic uncertainties added in quadrature.}\label{CDF-kt-eventshapes}
\end{center}
\end{figure}

Event shapes describe geometric properties of the energy flow in  the 
QCD final states by encoding information about the energy flow of an event in a continuous fashion. 
By having  sensitivity to both perturbative and non-perturbative aspects of QCD they can be an important 
addition to the jet fragmentation studies. 
Event shapes have been studied extensively in $e^+e^-$ and DIS experiments~\cite{event-shapes-review}. 
However at hadron colliders they have received far less attention,
primarily due to the difficulties in the theoretical description associated with the environment. 
From a
theoretical point of view, a description over the full range of an event shape observable
at a hadron collider requires not only perturbative QCD calculations 
 but also the inclusion of a phenomenological model of the
underlying event. 
The CDF collaboration performed studies~\cite{CDF-eventshapes} of transverse thrust and thrust minor, 
both defined in the plane perpendicular to the beam direction  
to reduce the conflict between requirements of calculations for variables to be ``global'' 
and reality of the limited detector coverage of any collider experiment. 
By  using energies from unclustered calorimeter towers to measure the variables one can be 
free from the arbitrariness associated with jet definition.  
The transverse-thrust variable $\tau$ is defined to vanish in the limit of two back-to-back objects, and for the 
isotropic event  $\tau$ = 1-2/$\pi$,  
while the transverse-thrust minor  $T_{min}$ is a measure of the out-of-plane transverse momentum 
and varies from zero for an event in the event plane to $2/\pi$ for a cylindrically symmetric event. Both $\tau$ and $T_{min}$  are sensitive to the modeling of the underlying event 
and agree with the distributions obtained from the {\sc pythia} Tune A. 
These observables can be used to improve the modeling of the underlying event.   
In addition to these variables, a new variable, {\it thrust differential}, 
constructed to be less sensitive to the underlying event and hadronization effects, 
was introduced by taking the weighted difference of the mean values of the thrust and thrust minor over the event sample. 
 The evolution of this quantity as a function of the leading jet energies allows
  to have meaningful comparison between data and the theoretical predictions.
As can be seen from  Fig.~\ref{CDF-kt-eventshapes}(b), both the 
{\sc pythia} Tune A and resummed next-to-leading-logarithm (NLL) parton-level predictions that were matched to fixed-order results at 
next-to-leading-order (NLO), referred to as NLO+NLL calculations~\cite{NLO-NLL}, describe the data quite well. 
This study illustrates the need to include underlying event contributions when comparing data with pQCD in hadron-hadron collisions.

\subsection{Elastic Scattering}

Elastic scattering $p\bar{p}\rightarrow p\bar{p}$ is a very important
process that  probes the structure of the proton. It is characterized
by different $t$-dependencies, starting with the lowest values of $t$:
the Coulomb region where elastic scattering is dominated by photon
exchange, the nuclear/Coulomb interference region; the ``single pomeron
exchange region'', where $d\sigma/dt$ is proportional to $e^{-bt}$,
followed by a region with 
a local diffractive minimum which moves to lower $|t|$ values
as $\sqrt{s}$ increases, so called {\it shrinkage}, and a high $|t|$ region described
by perturbative QCD. 

\begin{figure}[htbp]
\centerline{\includegraphics[width=0.70\columnwidth]{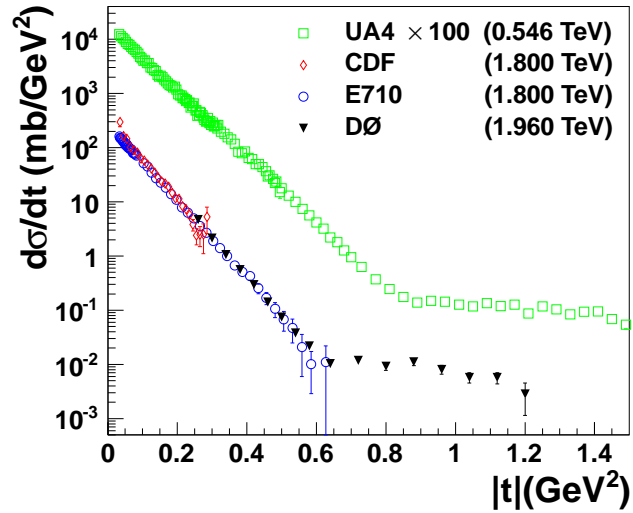}}
  \caption{ The $d\sigma/dt$ cross section as a function of $t$ compared to the results from the CDF, E710 and the  UA4 experiments.}
\label{elasticscatt-D0}
\end{figure}

The D0 collaboration extended the $|t|$ range previously measured
by CDF (0.025$<|t|<$0.29 \GeVcc)~\cite{CDF-elastic}
to 0.26$<|t|<$1.2 \GeVcc~\cite{D0-elastic}. 
The elastically scattered protons and antiprotons were tagged with the
forward proton detector spectrometer system. The data sample
corresponds to an integrated luminosity of 31 nb$^{-1}$ and was
collected with dedicated beam conditions. 
Fig.~\ref{elasticscatt-D0} shows the measured ${d\sigma}/{dt}$
differential cross section. The uncertainties correspond to the total
experimental systematic uncertainties not including the  14\% normalization
uncertainty. The distribution shows a change in the $|t|$
dependence. The fit to the $d\sigma/dt$ in the region of $|t|$
from 0.25-0.6 \GeVc$^2$ with an exponential function $Ae^{-b|t|}$
provides a logarithmic slope parameter
$b=16.86\pm 0.10(stat)\pm 0.20(syst)$ in agreement with previous
measurements from the CDF and E710 experiments~\cite{E710}. Comparison in shape to
data from UA4 collected at the $\sqrt{s}$=546 GeV~\cite{UA4} confirms the presence of
the {\it shrinkage} as diffractive minima move toward lower $|t|$ values.

\subsection{Diffractive processes}

Diffractive reactions, which  constitute a substantial fraction of the total
cross section in hadron-hadron scattering, can be described in terms
of the {\it pomeron} ($\Pomeron$) exchange, a hypothetical object with the quantum numbers
of the vacuum. The experimental signatures of the diffraction consist in
particular kinematic configurations of the final states: the presence of
 non-exponentially suppressed large rapidity gaps and/or the presence of
 the intact leading particles. The diffractive processes became an
important tool in understanding many interesting aspects of QCD such
as low-$x$ structure of the proton, the behavior of QCD in the high density regime.

Significant progress in understanding diffraction has been made at the Tevatron $p\bar{p}$ colliders. 
The CDF and D0  collaborations contributed extensively\cite{CDF-RunI-SD-DJ}-\cite{CDF-D0-Diffraction-RunI},
 by studying a wide variety of diffractive processes at three different center-of-mass energies: 630 GeV, 1800 GeV - Run I of Tevatron, and 1960 GeV - Run II. Some important results include the observation of  QCD factorization breakdown in hard single diffractive processes, the discovery of large rapidity gaps between two jets, and the study of diffractive structure function  in double pomeron exchange dijet events.

\subsubsection{Hard Single Diffraction}

The signature of single diffractive (SD) dissociation at the Tevatron is either a forward rapidity gap along the direction of one of the initial particles, or a presence of leading particle, antiproton, with $\xi <$0.1. The process $\bar{p}p\rightarrow\bar{p}X$, which can be described by assuming that a pomeron is emitted by the incident antiproton and undergoes a hard scattering with the proton, is an ideal reaction to study the partonic content of the pomeron, and the diffractive structure function. 
The high energies of the Tevatron collider allows the study of diffraction in terms of  perturbative QCD, i.e. in 
the presence of a hard scale. These types of diffractive processes are called {\em hard diffraction} and 
were extensively studied in Run I~\cite{CDF-D0-Diffraction-RunI}. 

\begin{figure}[htbp]
\centerline{\includegraphics[width=0.50\columnwidth]{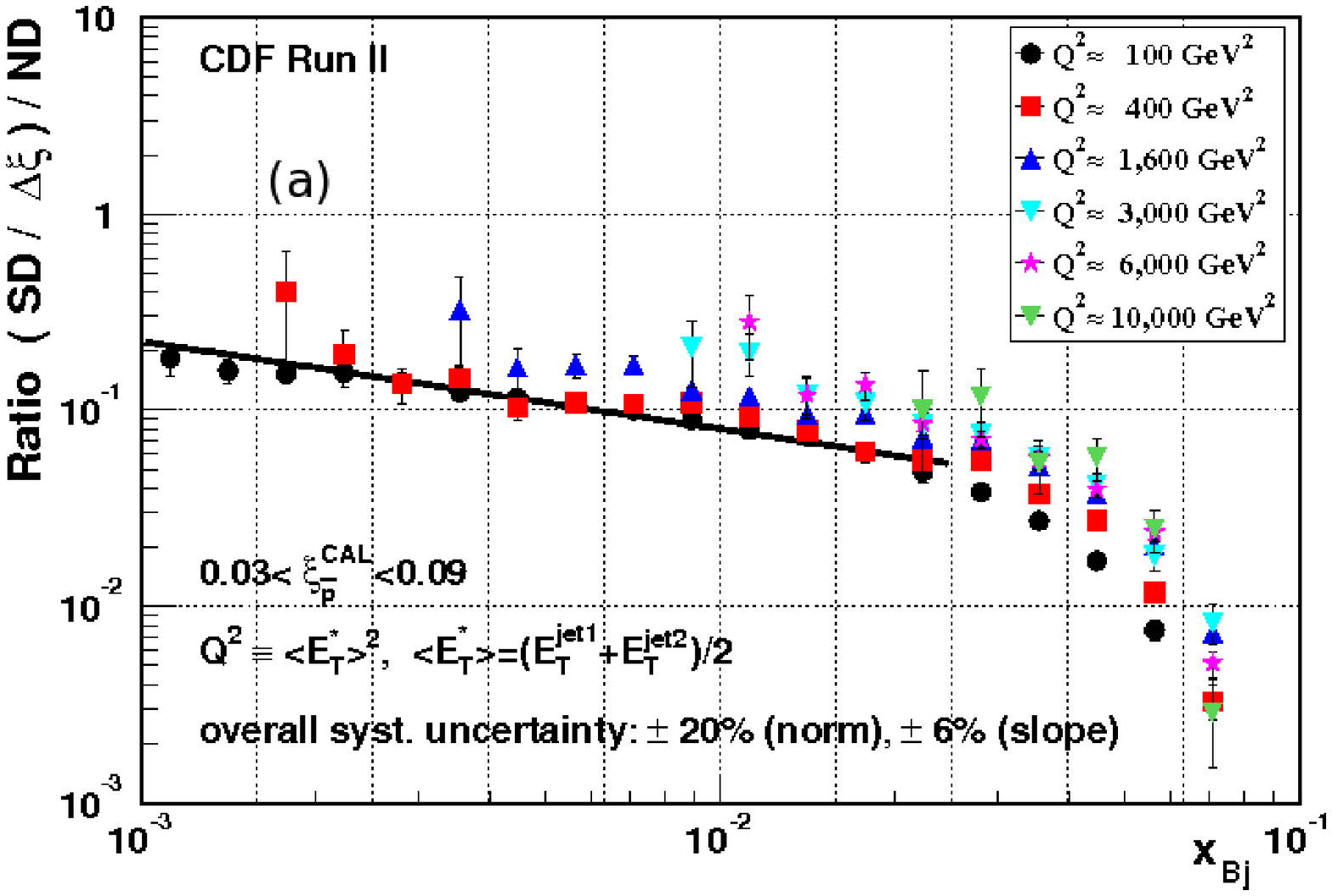}
     \includegraphics[width=0.50\columnwidth]{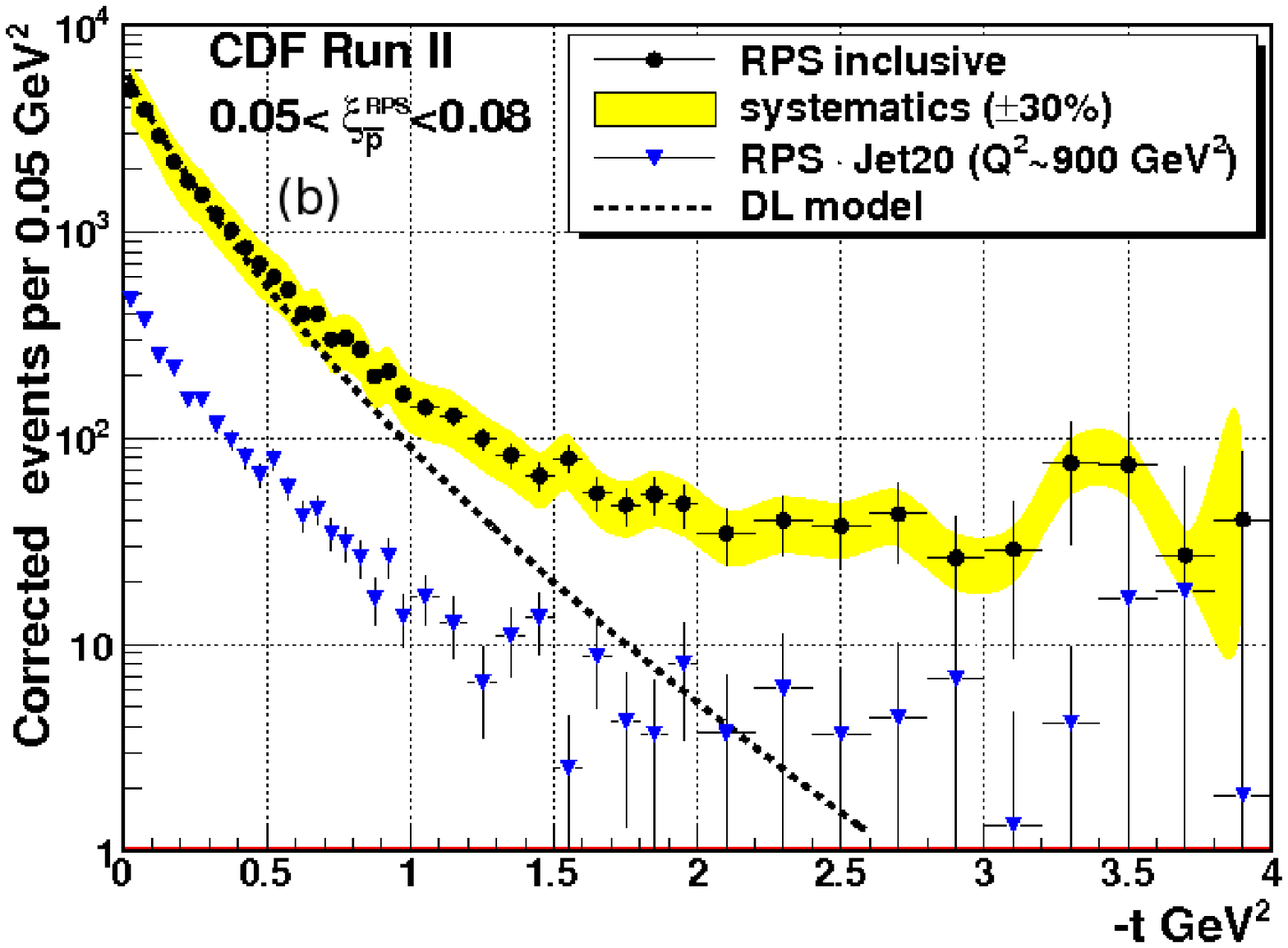}}
  \caption{ (a) The ratio of diffractive to non-diffractive dijet event
rates as a function of $x_{Bj}$ (momentum fraction of parton in the
antiproton) for different values of $Q^2$ equal to the square of the mean dijet transverse energy $E_T$; (b) measured $t$ distribution 
for two type of  events (circles) soft single diffractive inclusive
events and hard single  diffractive (triangles) events triggered by presence of at least one jet with $E_T>$20 GeV. The
curve represents the distribution expected for soft SD in the
DL (Donnachie-Landshoff) model.}
\label{diffdijets}
\end{figure}

One of the important questions in hard  diffraction is whether these type of processes obey QCD factorization, 
or in other words, whether the  pomeron has a universal process independent diffractive parton structure function.
Results from Run I\cite{CDF-RunI-SD-DJ}-\cite{CDF-RunI-SD-Jpsi} 
show that the rate of single diffractive relative to non-diffractive processes is lower by on order of magnitude than  expectations from diffractive PDFs determined at HERA   $ep$ collider. This presents breakdown of QCD factorization in hard diffraction between Tevatron and HERA. This suppression was further studied by investigating diffractive structure function in diffractive dijet production in Run I~\cite{CDF-RunI-SD} and continuing these studies in Run II~\cite{CDF-SD-dijet} by comparing two samples of dijets events,
 diffractive (SD) triggered by the presence of an intact antiproton detected in the 
Roman Pot Spectrometer (RPS), and non-diffractive (ND). 
By taking the ratio of SD dijet rates to ND, 
which in a good approximation is the ratio of the diffractive to  the known proton structure function, 
the diffractive structure function can be extracted. The dependence of diffractive structure 
function on the average value of mean dijet $E_T$, $Q^2$ was studied. Fig.~\ref{diffdijets}(a) shows the ratio of the single diffractive dijet event rate to those of non-diffractive dijet events  as a function of $x_{BJ}$, the Bjorken-x of the struck parton of the antiproton. In the range of 100$<Q^2<$100000 GeV$^2$ no significant $Q^2$ dependence is observed. 

In addition, the $t$-distribution was measured for both soft and hard single diffractive processes, see Fig.~\ref{diffdijets}(b).
The slope of the distribution shows no dependence on the $Q^2$ of the process; for both soft and hard  samples it is very similar.
 For low-$t$ values, experimental results are well described by the curve based  on predictions from the 
Donnachie-Landshoff model~\cite{Donnachie-Landshoff} for soft diffractive processes. High $t$ 
values can be used to search for so-called ``diffraction minimum'' similar  to the one observed for elastic scattering processes. 
However the experimental data shows flat behavior and not enough discriminating resolution power.

Diffractive $W/Z$ production is an important process for probing the quark content of the pomeron, 
since to leading order, the $W/Z$ is produced through a quark, while  gluon associated production is 
suppressed by a factor of $\alpha_S$ and can be identified by an additional  jet.  
CDF studied diffractive $W$ production in Run I~\cite{CDF-SD-W-RunI} by using the rapidity gap signature of  diffractive events. 
In Run II, events were selected with the ``intact leading antiproton'' signature, where $\bar{p}$ is detected in the Roman Pot Spectrometers (RPS).
The RPS allows very precise measurement of the fractional momentum loss of $\bar{p}$  ($\xi$),  
eliminating the problem of {\em gap survival probability}.  
The  novel feature of the analysis, the determination of the full kinematics of the $W\rightarrow l\nu$ decay, is 
made possible by obtaining the neutrino ${E_T}^{\nu}$ from the missing $E_T$, ${E\!\!\!\!/}_T$,  and $\eta_{\nu}$
from the formula ${\xi}^{RPS}-{\xi}^{cal}=({E\!\!\!\!/}_T/\sqrt{s})\exp(-{\eta}_{\nu})$,  
where $\xi^{RPS}$ is the true $\xi$ measured in RPS and ${\xi}^{cal}=\sum_{i(towers)}{(E_T^i/\sqrt{s})\exp(-\eta^i)}$.
The fractions of diffractive $W$ and $Z$ events are measured to 
be 
$[0.97\pm0.05(stat.)\pm0.11(syst.)]$\% and $[0.85\pm0.20(stat.)\pm 0.11(syst.)]$\% 
for the kinematic range 0.03$<\xi<$0.10 and $|t|<$1 \GeVc.
  The measured diffractive $W$ fraction is consistent with the Run I CDF 
result~\cite{CDF-SD-W-RunI} when corrected for the $\xi$ and $t$ range.

\subsubsection{Central Exclusive production}

Central exclusive production, defined as the class of reactions $p+\bar{p}\rightarrow p+X+\bar{p}$, where the colliding particles emerge intact and a produced state, $X$, is fully measured, has been the subject of much interest recently, particularly at large $\sqrt{s}$
 where    the rapidity range $\Delta y_{total}=2\times\ln{\sqrt{s}/m_p}$=15.3 at the Tevatron allows the possibility of  
large rapidity gaps  produced between state $X$ and proton and antiproton.   
There are three production mechanisms responsible for this processes: $\gamma\gamma\rightarrow X$,  $\gamma\Pomeron\rightarrow X$, and $\Pomeron\Pomeron\rightarrow X$, the so called Double Pomeron Exchange (DPE). The first two processes were observed at CDF for the first time and will
 be discussed later.

CDF made an observation of exclusive dijet production~\cite{CDF-DPE-excldijet}
 by studying events triggered by the intact leading antiproton on one side and a large rapidity gap in the proton direction.
 Although the $\xi_{\bar p}$ variable can be measured directly from the RPS information, $\xi_p$ can be calculated by summing information from all the observed particles in the detector, $\xi_p=(1/\sqrt{s})\sum{E^i_T e^{\eta^{\imath}}}$.

\begin{figure}[htbp]
\centerline{\includegraphics[width=0.70\columnwidth]{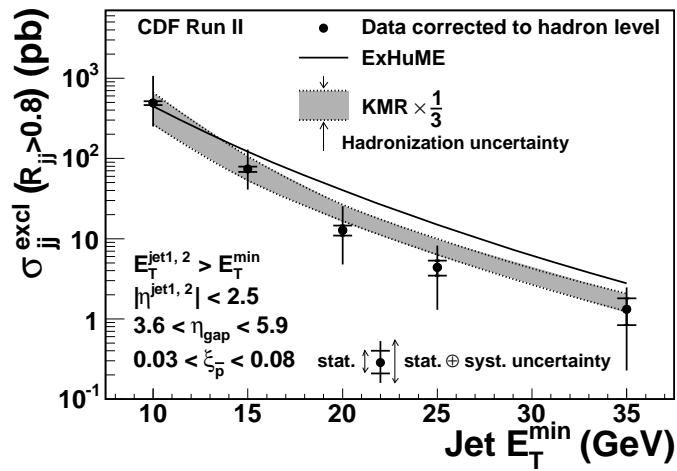}}
 \caption{ Exclusive cross section for events with two jets of $E_{T^{jet}}>$ 10  GeV 
with $R_{jj}>$0.8 compared with ExHuME~\cite{EXHUME}, event generator based  on perturbative calculations of Ref.~\cite{Khoze-excldijet}, (solid curve) and with the LO analytical calculation KMR.}
\label{excldijets-CDF}
\end{figure}

The exclusive dijet production was first studied by CDF in Run I data and a limit of 
$\sigma_{excl}<$3.7 nb (95\% CL) was placed~\cite{CDF-DPE-excldijet-RunI}.
This study was continued in Run II when the observation of the exclusive dijet production was reported~\cite{CDF-DPE-excldijet}.
The exclusive signal is extracted using the dijet mass
fraction  method: 
the
ratio $R_{jj}\equiv M_{jj}/M_X$ of the dijet mass $M_{jj}$ to the total mass
$M_X$ of the final state is formed and
used to discriminate between the signal of exclusive dijets,
defined as $R_{jj}>$0.8, and the background of
inclusive DPE  dijets, expected to have a continuous distribution
concentrated at lower $R_{jj}$ values. 
The measured cross sections, see  Fig.~\ref{excldijets-CDF}, are consistent with KMR predictions by 
 Khoze et al. ~\cite{Khoze-excldijet}. 
The D0 collaboration extended the study of exclusive dijets into the highest mass states by presenting evidence~\cite{D0-excldijets}  
for diffractive exclusive dijet production with an invariant dijet mass $M_{jj}$ greater than 100 GeV.

Another central exclusive process proceeding through the double pomeron exchange mechanism is exclusive
diphoton production $p\bar{p}\rightarrow p\gamma\gamma\bar{p}$. CDF has performed a search for exclusive $\gamma\gamma$  
in combination with the search for CEP $e^+e^-$~\cite{CDF-exclusive-e+e-}. The analysis
techniques were identical - to require absolutely empty detectors by
triggering on very large forward rapidity gaps on both sides, except for
two photon (electron) candidates. In contrast to exclusive diphoton
search, where the theoretical predictions vary significantly  for
different models, central  exclusive electron-positron production is a
QED process with a well known cross section, so while reporting an observation of
 exclusive electron-positron production in hadron-hadron collisions the 
The CDF collaboration was able to validate the method for exclusive diphoton searches. 
The initial search for exclusive $\gamma\gamma$ production~\cite{CDF-exclusive-diphoton-limit} using 532
pb$^{-1}$ of data resulted in finding 3 candidate events and placing a
limit. 
New analysis~\cite{CDF-exclusive-diphoton-observation} utilized 1.11 fb$^{-1}$ of data by selecting events
with two electromagnetic showers, each with $E_T>$2.5 GeV and
pseudorapidity $|\eta|<$1.0 while requiring no other particles
detected in the rest of the detector in pseudorapidity range from -7.4 to +7.4. 
The observed 43 candidate events have the kinematic properties expected for exclusive $\gamma\gamma$ production, see Fig.~\ref{CDF-excldiphoton}. This result constitutes the first observation of exclusive diphoton production in hadron-hadron collisions. The corresponding cross section of $\sigma_{\gamma\gamma,excl}=2.48^{+0.40}_{-0.35}(stat)^{+0.40}_{-0.51}(syst)$ pb is in agreement with the theoretical predictions, where dependence on low-x gluon density contributes to significant uncertainty due to the choice of PDFs.

CDF II also studied dimuon production~\cite{CDF-excl-dimuon}, when the event signature requires two oppositely 
charged central muons, and either no other particles (large forward rapidity gaps), or one additional photon detected. 
 Within the kinematic region $|\eta_{\mu}|<$0.6 and $M_{\mu\mu}\in[3.0,4.0]$ \GeVcc, there are 402 events with no 
electromagnetic shower, see the $M_{\mu\mu}$ spectrum in Fig.~\ref{CDF-excldiphoton}(a). 

The $J/\psi$ and $\psi(2S)$ are prominent, so the exclusive vector meson production expected for the elastic photoproduction 
$\gamma+p\rightarrow J/\psi(\psi(2S))+p$ is observed for the first time in hadron-hadron collisions. The obtained cross sections 
${d\sigma}/{dy}_{y=0}(J/\psi)$=3.92$\pm$ 0.25(stat)$\pm$ 0.52(syst) nb and for $\psi(2S)$ 0.53$\pm$0.09(stat)$\pm$0.10(syst) nb 
agree with the predictions~\cite{Jpsi-predictions}, while  ratio R=$\sigma(\psi(2S))/\sigma(J/\psi)$=0.14$\pm$0.05 is in agreement with the HERA 
value~\cite{HERA-Jpsi} of 0.166$\pm$0.012 at similar $\sqrt{\gamma p}$.
\begin{figure}[htbp]
\centerline{
\includegraphics[width=0.3\columnwidth]{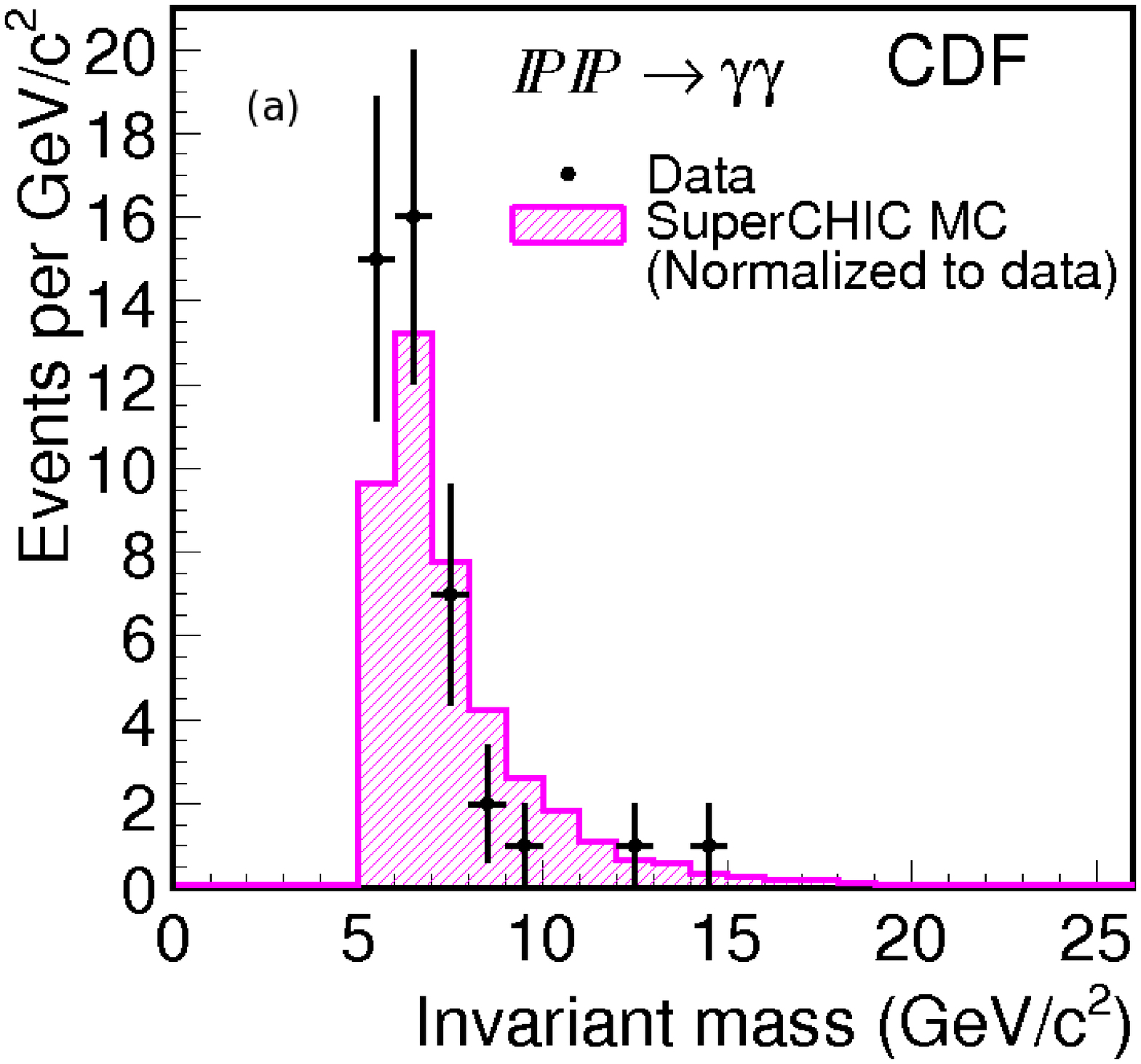}
\includegraphics[width=0.3\columnwidth]{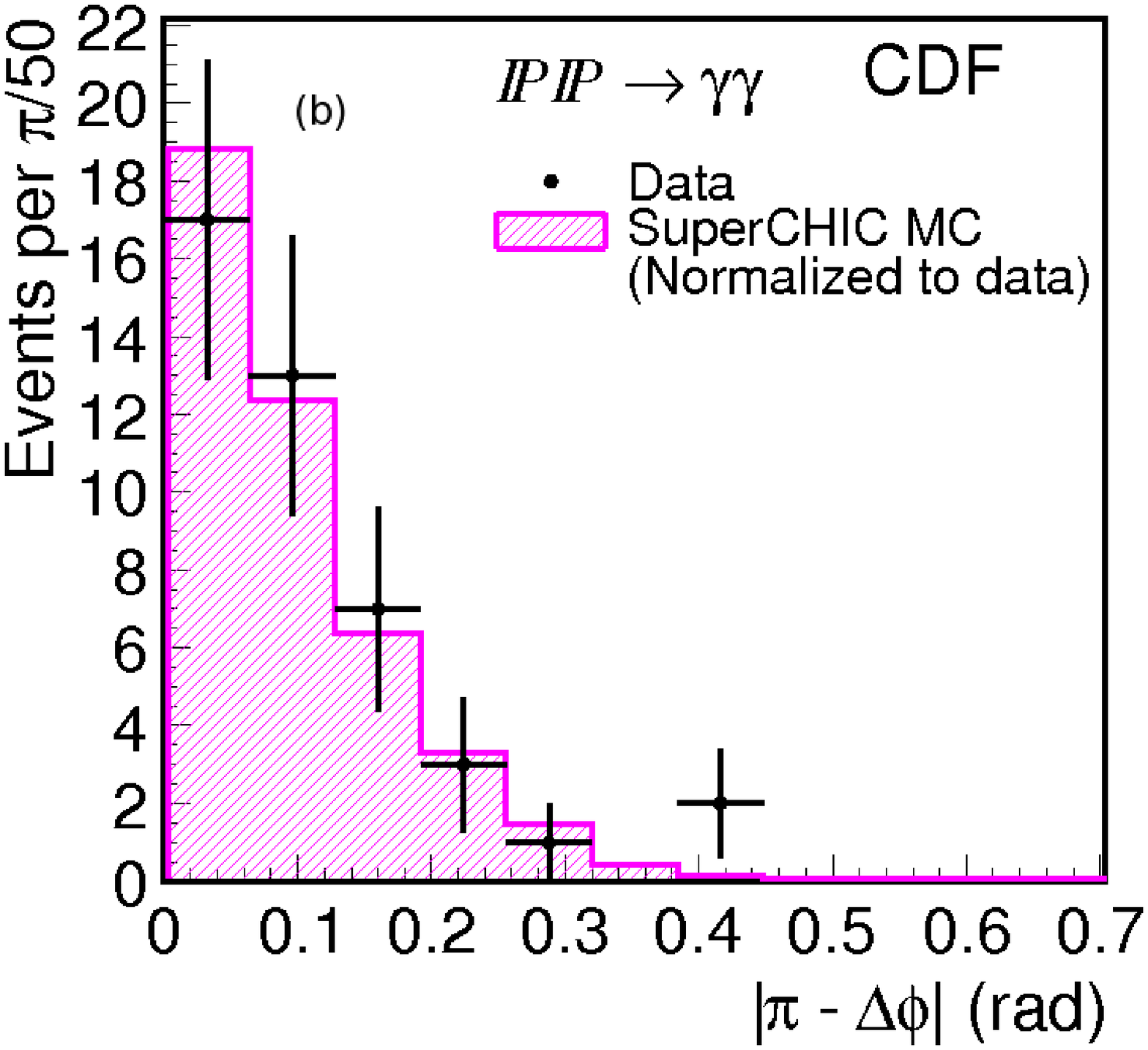}
\includegraphics[width=0.31\columnwidth]{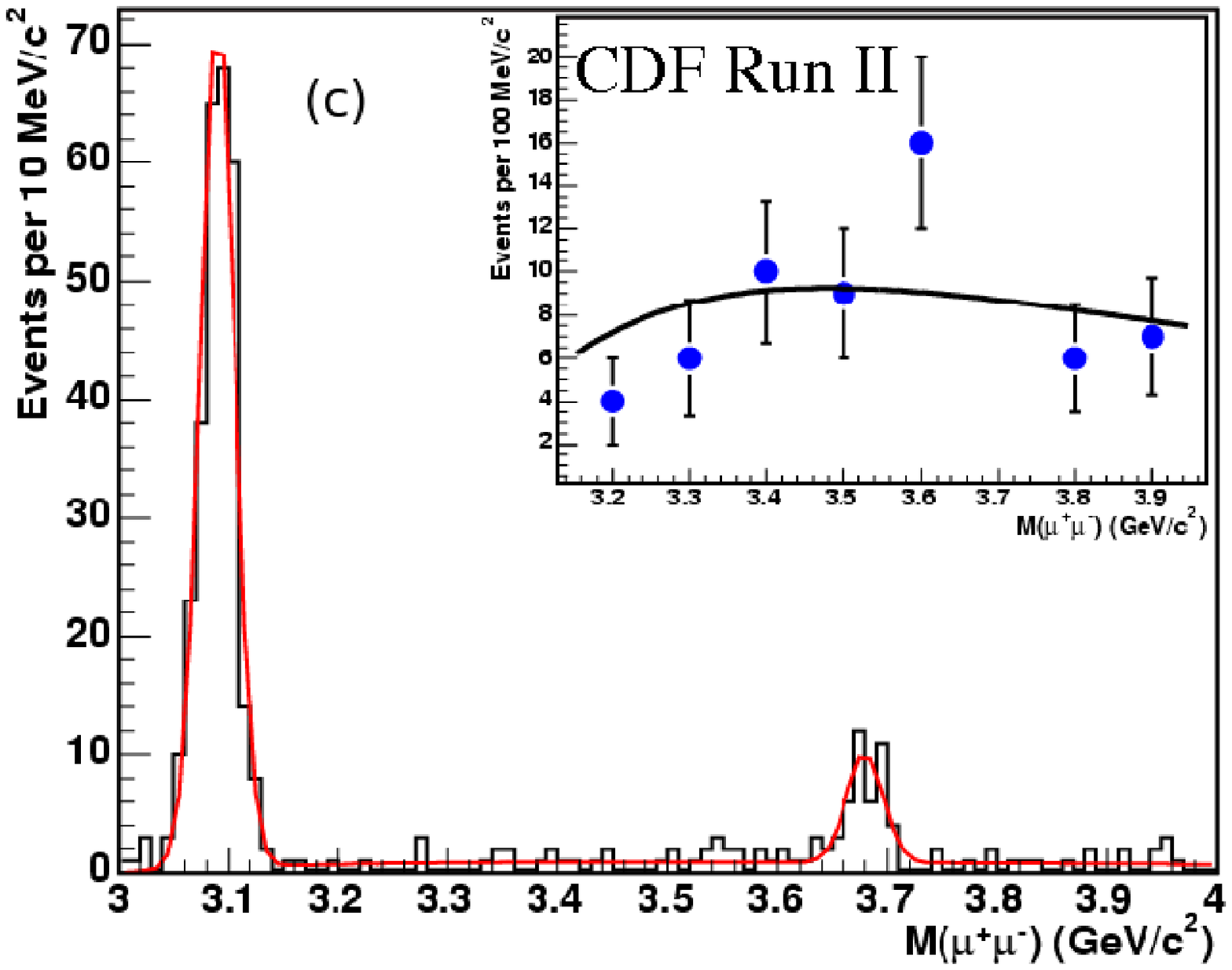}
\label{CDF-excldiphoton}}
 \caption{(a) Invariant mass of exclusive photon pair events compared to MC; (b) azimuthal angle difference (from back to back) between the exclusive photons compared to MC; (c) mass, $M_{\mu\mu}$ of exclusive dimuon events, with no EM shower, (histogram) together with a fit to two Gaussians for the $J/\psi$  and  $\psi(2S)$, and a QED continuum. All three shapes are predetermined, with only the normalizations floating. Inset: Data above the $J/\psi$  and excluding 3.65 $<M_{\mu\mu}<$ 3.75 \GeVcc ( $\psi(2S)$) with the fit to the QED spectrum times acceptance (statistical uncertainties only).}
\label{CDF-excldiphoton}
\end{figure}
By requiring one EM shower with $E_T^{EM}>$80 MeV in addition to the requirement 
mentioned above, we are able to measure $\chi_{c0}$ production. Allowing the EM tower 
causes a large increase (+ 66 events) in the $J/\psi$ peak and minor change (+1 event) in 
the $\psi(2S)$ peak. 
 After correcting for background, efficiencies, and the branching fraction, 
we observe $\chi_{c0}\rightarrow J/\psi +\gamma$ production for the first time in hadron-hadron collisions 
and obtain a cross section for exclusive $\chi_{c0}$ production of 75$\pm$10(stat)$\pm$10(syst) nb~\cite{}, 
which is compatible with the theoretical predictions ~\cite{chi_c_theory}.

In addition CDF performed a search for exclusive $Z$ boson production.
No exclusive Z$\rightarrow\l^+l^-$ candidates were observed.  The first upper limit on the 
 exclusive $Z$ cross section in hadron collisions, $\sigma_{excl}(Z)<$
0.96 pb at 95\% confidence level was placed.

\subsection{Study of double parton interactions}

The CDF and D0 collaborations comprehensively studied the phenomenon of MPI events in a series of 
Run I  and Run II measurements.
The events with double parton (DP) scattering 
provide insight into the spatial distribution of partons in the colliding hadrons. 
They can be also a background to many rare processes, especially with multijet final state. 

In Run I, CDF collaboration studied DP event using four-jet \cite{mpi_cdf1_run1} and 
$\gamma+3$ \cite{mpi_cdf2_run1} events. The observed fraction of DP events is found
to be much higher in the $\gamma+3$ final state (about 57\%) than in the four-jet events
(about 6\%). Both these analyses measured the so-called effective cross section, $\sigma_{\rm eff}$, 
that characterizes rates of the DP events, e.g.
$\sigma^{\gamma j, jj}_{\rm DP} = \sigma^{\gamma j}\sigma^{jj} /\sigma_{\rm eff}$.
This parameter is tightly related with the parton spatial distribution (see e.g. \cite{Threl}).

D0 has studied the DP events in $\gamma+3$ jet final state \cite{mpi1_d0},
in which two pairs of partons undergo two hard interactions in a single $p\bar{p}$ collision.
D0 measured $\sigma_{\rm eff}$, 
and found it to be $\sigma_{\rm eff}=16.4\pm 0.3({\rm stat})\pm2.3({\rm syst})$.
It is in agreement with the previous CDF result \cite{mpi_cdf2_run1},
$\sigma_{\rm eff}=14.5\pm 1.7({\rm stat})^{+1.7}_{-2.3}({\rm syst})$, 
as well as with $\sigma_{\rm eff}=12.1\pm 10.7^{+10.7}_{-5.4}$ 
\cite{mpi_cdf2_run1}.

\begin{figure}[htbp]
\hspace*{-2mm}  \includegraphics[scale=0.3]{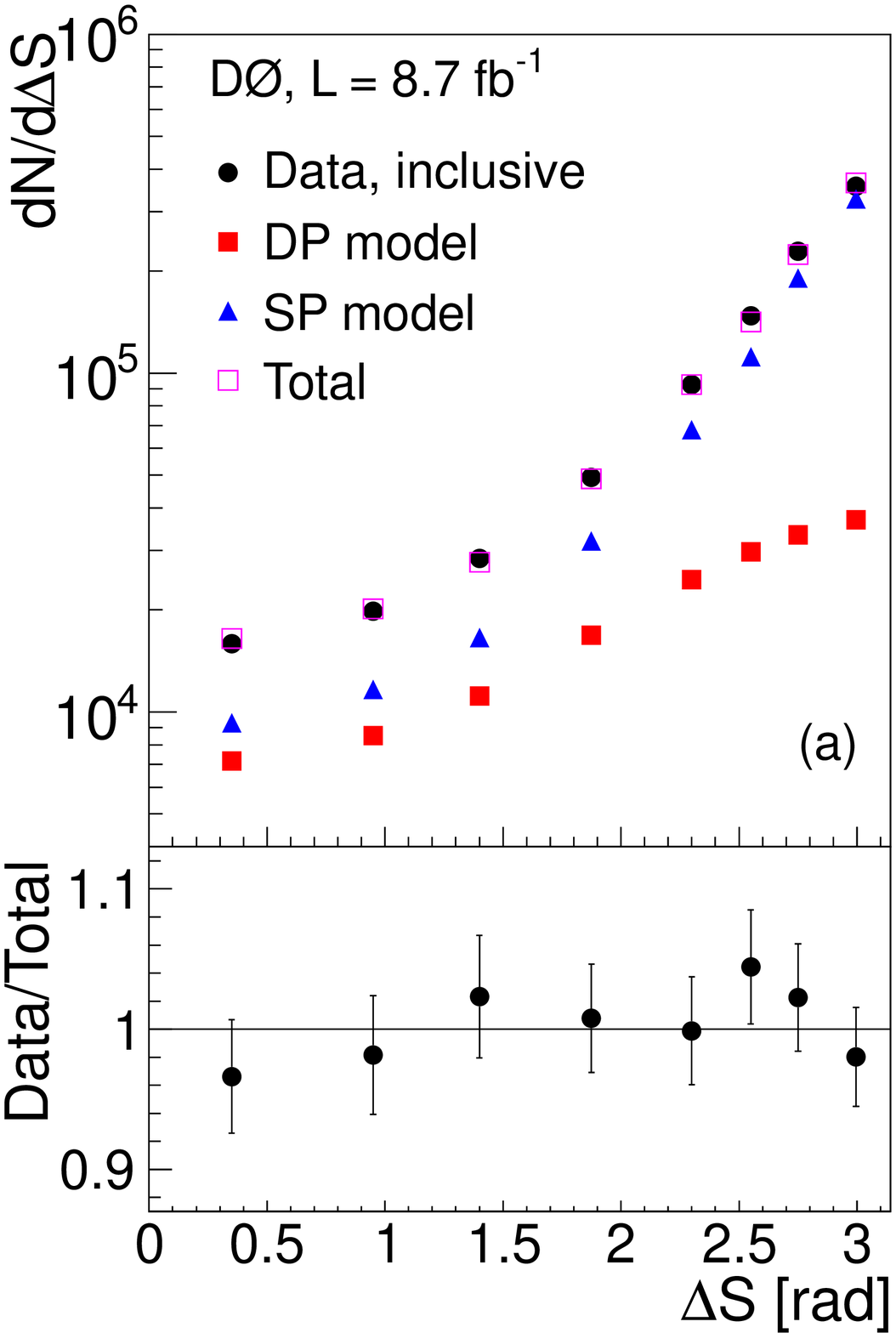}
\hspace*{-3mm}  \includegraphics[scale=0.33]{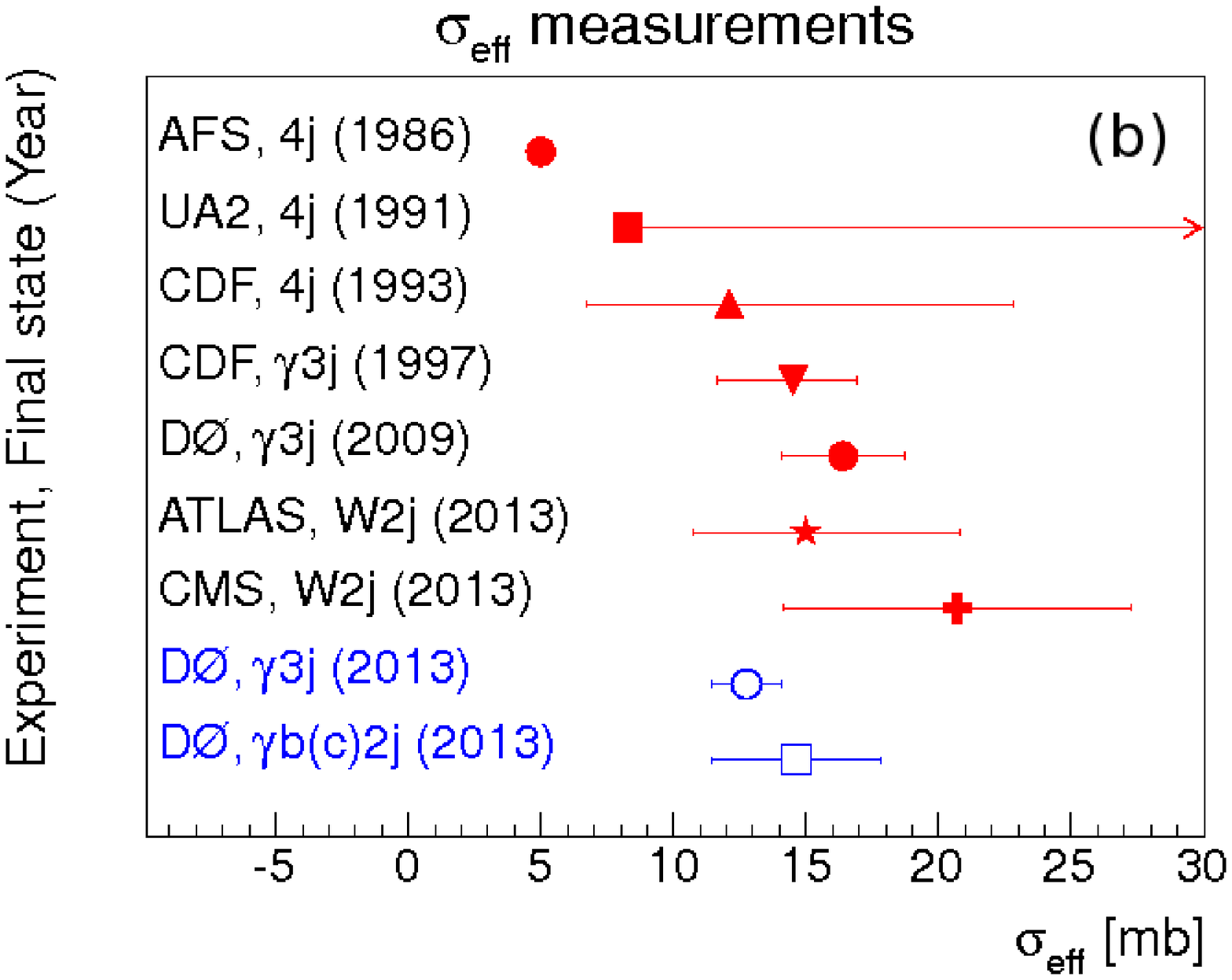}
\caption{ (a)The $\Delta S$ distribution in the data, DP and SP models,
and the sum of the DP and SP contributions weighted with their fractions (``Total'').
(b) Existing  measurements of effective cross section, $\sigma_{\rm eff}$,
compared with result presented here
(AFS: no uncertainty is reported; UA2: only a lower limit is provided).}
\label{fig:ds1_d0}
\end{figure}

D0 collaboration has also tested a dependence of
the effective cross section on the initial quark flavor using 
the $\gamma+3$-jet and \gpHFjj events with $\Ptg>26$ GeV with inclusive and heavy flavor leading jet \cite{mpi2_d0}.
The effective cross sections 
are found to be $\sigma_{\rm eff}^{\rm incl} = 12.7 \pm 0.2\thinspace({\rm stat}) \pm 1.3\thinspace({\rm syst})$ mb
and
$\sigma_{\rm eff}^{\rm HF} = 14.6 \pm 0.6\thinspace({\rm stat}) \pm 3.2\thinspace({\rm syst})$ mb
for the two event types, respectively.
This is the first measurement of  $\sigma_{\rm eff}$ with heavy flavor jets in the final state.
Due to the significant dominance of the Compton-like process 
$qg\to q\gamma$,
one can conclude that there is no evidence for a dependence of $\sigma_{\rm eff}$ on the initial parton flavor.
The  plot (a) of Fig.~\ref{fig:ds1_d0} shows the $\Delta S$ distribution in the data, DP and single parton (SP) models,
and the sum of the DP and SP contributions weighted with their fractions.
Here the variable $\Delta S$ is defined as an azimuthal angle between the $p_T$ vectors of two object pairs
($\gamma$+jet and jet+jet) in \gpTHRj events. The found DP fractions vary within about $17-20\%$.
The plot (b) summarizes the published world measurements (AFS, UA2, CDF, D0, ATLAS 
and CMS experiments) of the effective cross section.

\begin{figure}[htbp]
\hspace*{0mm}  \includegraphics[scale=0.29]{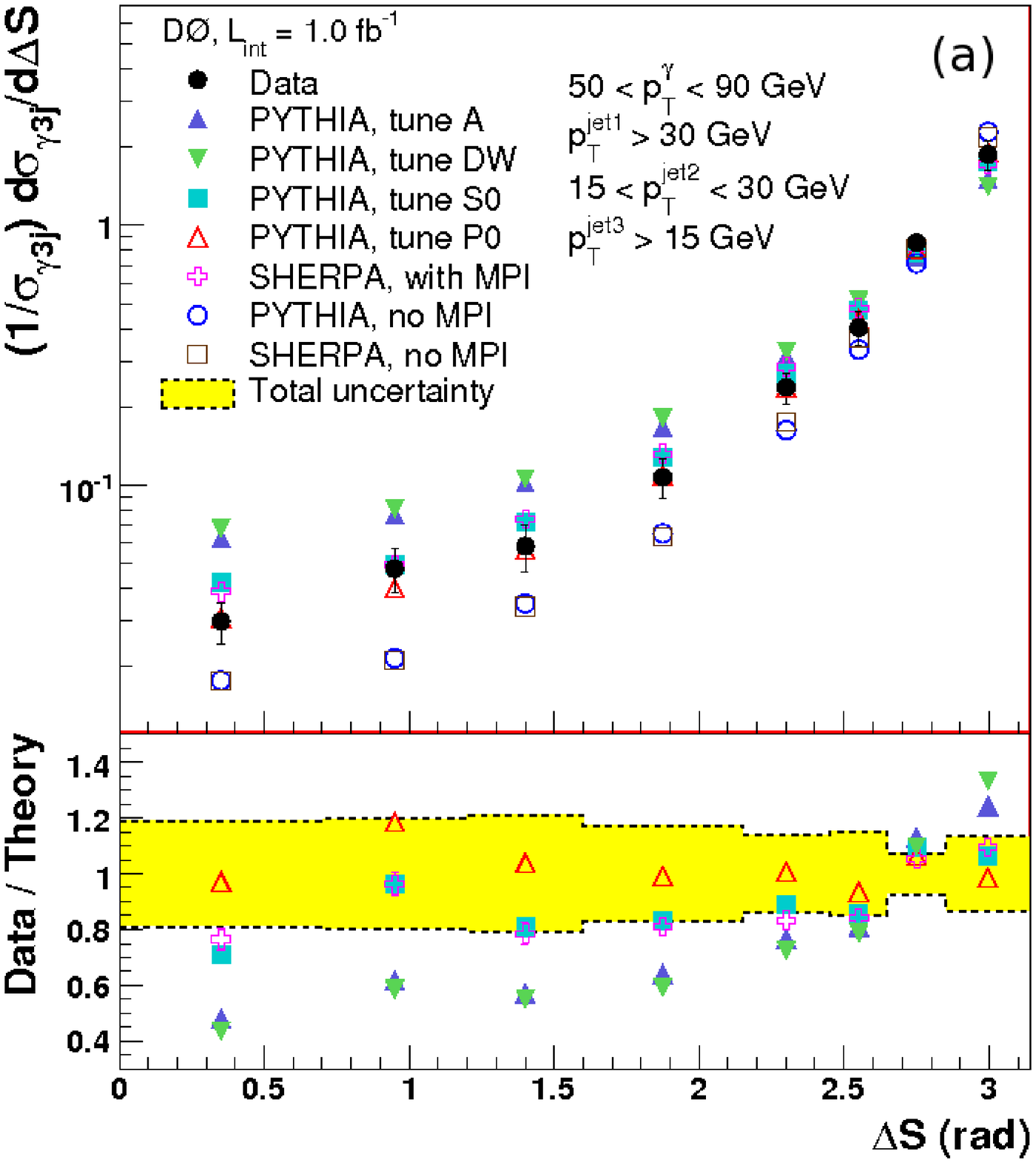}
\hspace*{0mm}  \includegraphics[scale=0.29]{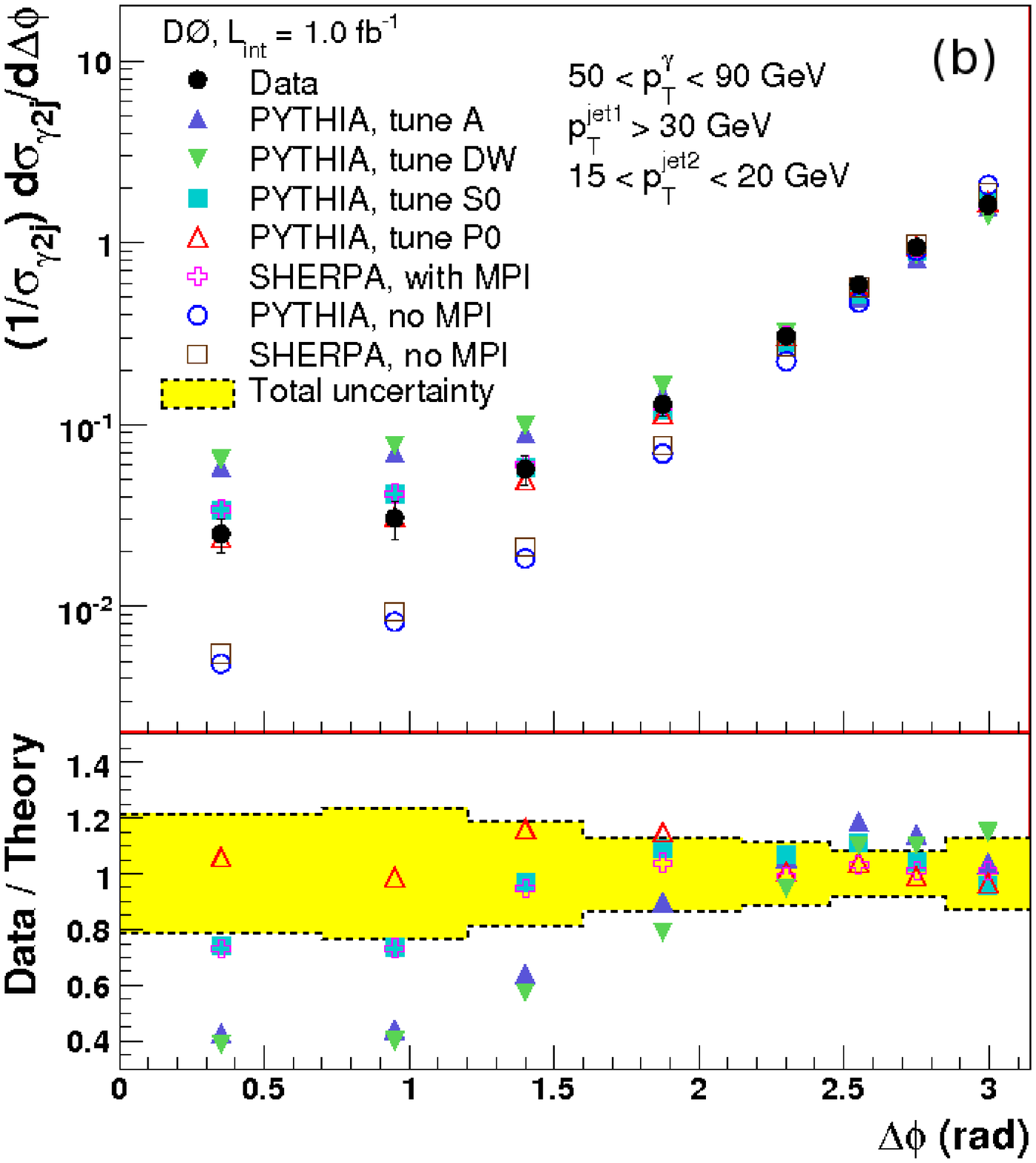}
\caption{ (a) Normalized differential cross section in the $\gamma + 3$-jet + X
events, $(1/\sigma_{\gamma 3j})\sigma_{\gamma 3j} /d\Delta S$, in data compared to Monte Carlo models
and the ratio of data over theory, only for models including
MPI, in the range $15 < p_T^{jet2} < 30$ GeV. 
(b) Normalized differential cross section in $\gamma + 2$-jet + X
events, $(1/\sigma_{\gamma 2j})\sigma_{\gamma 2j} /d\Delta\phi$, in data compared to Monte Carlo models
and the ratio of data over theory, only for models including
MPI, in the range $15 < p_T^{jet2} < 20$ GeV.} 
\label{fig:ds2_d0}
\end{figure}

To tune MPI models, D0 also measured differential cross sections for the $\Delta S$ variable 
in the three $p_T$ bins of the 2nd jet $p_T$ \cite{mpi2_d0}.
Comparison of data with a few MPI and two ``no MPI'' models are shown in Fig.~\ref{fig:ds2_d0}.
One can see that data clearly contain DP events and favor more Perugia MPI tunes \cite{PYT}.

D0 collaboration has also studied recently the double $J/\psi$ events 
produced due to DP interactions \cite{mpi4_d0}
with a dominated $gg\to J/\psi J/\psi$ production mechanism in each of the DP scatterings.
Using DP events with a fraction of $f_{\rm DP} = 0.30\pm 0.10$, the effective cross section
has been estimated as 
$\sigma_{\rm eff}^{JJ} = 5.0 \pm 0.5\thinspace({\rm stat}) \pm 2.7\thinspace({\rm syst})$ mb.

\section{Summary and Conclusions}
\label{sec:summary}

In this short review, we presented main measurements and studies
of the QCD processes performed by CDF and D0 experiments in Run II,
for the data taking period from April 2002 to September 2011
(some Run I results are also briefly mentioned).
These processes can be roughly split into those which
are typically described in the framework of perturbative QCD,
and those which are treated using phenomenological QCD models.

The measurements of the first type with jet final state
are used to tune the gluon distribution at 
parton momentum fractions $x\gtrsim0.2$, test running 
of $\as$ for momentum transfers up to 400~GeV and extract
a precise integrated value $\alpha_s(m_Z)=0.1161^{+0.0041}_{-0.0048}$,
impose limits on some new phenomena which were expected at high energies
(e.g. excited quark, axigluon, technicolor models, 
$W'$, and $Z'$ productions). 
Measurements with photons provided 
a valuable input for tuning gluon PDFs at low $x$, 
soft-gluon resummations, and contributions from 
the parton-to-photon fragmentation processes, check 
a wide variety of approaches used
to predict the differential $\gamma+$(heavy flavor) jet and diphoton
cross sections.
Numerous studies of $W/Z$+jets productions allowed
extensive tests and tuning pQCD NLO and Monte Carlo  
event generators, which have been used to predict
backgrounds for Higgs boson production and searches
for new phenomena at the Tevatron and LHC.

A broad physics program dedicated to studying 
soft strong interactions resulted in a bulk of interesting
results. A series of minimum bias events studies allows
tuning the non-perturbative QCD models which are used
to describe underlying events and hadronization effects.
Studies of MPI phenomenon at high $p_T$ regime
constrain existing MPI models, parton spatial densities
inside a nucleon, and tune Monte Carlo event generators.
Measurements of diffractive and elastic cross sections
test low-$x$ structure of the proton and 
constrain many phenomenological models.

The obtained results affected a variety of models
describing strong interactions between partons,
may suggest directions of following studies,
and compose a valuable legacy for other ongoing and 
future experiments and input for theoretical models.

\section*{Acknowledgments}
Authors would like to than Anwar Bhatti and Robert Hirosky for useful comments and discussions.

We thank the Fermilab staff and technical staffs of the participating institutions
for their vital contributions. We acknowledge support from the DOE and NSF
(USA), ARC (Australia), CNPq, FAPERJ, FAPESP and FUNDUNESP (Brazil),
NSERC (Canada), NSC, CAS and CNSF (China), Colciencias (Colombia), MSMT
and GACR (Czech Republic), the Academy of Finland, CEA and CNRS/IN2P3
(France), BMBF and DFG (Germany), DAE and DST (India), SFI (Ireland), INFN
(Italy), MEXT (Japan), the KoreanWorld Class University Program and NRF (Korea), CONACyT (Mexico), FOM (Netherlands), MON, NRC KI and RFBR (Russia),
the Slovak R\&D Agency, the Ministerio de Ciencia e Innovacion, and Programa
Consolider–Ingenio 2010 (Spain), The Swedish Research Council (Sweden), SNSF
(Switzerland), STFC and the Royal Society (United Kingdom), the A.P. Sloan Foundation
(USA), and the EU community Marie Curie Fellowship contract 302103.


\begin{thebibliography}{10}
\bibitem{parp67} The {\sc pythia} parameter \texttt{PARP(67)} was
    increased from the current default of $1.0$ to $4.0$ which was
    the default before version 6.138.
\end{thebibliography}
\end{document}